\documentclass{report}
\usepackage{suthesis-2e}
\usepackage{cancel}
\usepackage[utf8]{inputenc}
\usepackage{array,amsfonts,amsmath,amssymb,amsthm, mathtools}
\usepackage{enumerate}
\DeclareMathOperator*{\argmax}{arg\,max}

\newcommand{\E}{\mathbb{E}}
\newcommand{\F}{\mathcal{F}}
\newcommand{\pp}{\mathbb{P}}

\newtheorem{theorem}{Theorem}

\usepackage{bbm}
\usepackage{bm}
\usepackage[ruled,vlined]{algorithm2e}
\usepackage[T1]{fontenc}
\usepackage{natbib}
\usepackage{graphicx}
\usepackage{algorithmic}

\dept{Statistics}
\tablespagefalse
\figurespagefalse

\begin{document}

\title{Adaptive Experiments and A Rigorous Framework for Type I \\ Error Verification and Computational Experiment Design}

\author{Michael Sklar - 
June 2021, with minor corrections as of}

\principaladviser{Tze Lai}
\firstreader{Ying Lu}
\secondreader{Philip Lavori}
 
\beforepreface
\prefacesection{Abstract}

What wouldn't we give for faster access to life-saving drugs, cancer cures, or pandemic-ending vaccines? In recent decades, modern statistics has found something to trade: at the price of additional complexity and the loss of Gaussian behavior of our estimators, we can get faster, more robust, more flexible, and more efficient experiments through the use of adaptive designs. This thesis covers breakthroughs in several areas of adaptive experiment design: (i) (Chapter 2) Novel clinical trial designs and statistical methods in the era of precision medicine. (ii) (Chapter 3) Multi-armed bandit theory, with applications to learning healthcare systems and clinical trials. (iii) (Chapter 4) Bandit and covariate processes, with finite and non-denumerable set of arms. (iv) (Chapter 5) A rigorous framework for simulation-based verification of adaptive design properties.

\prefacesection{Acknowledgments}

This thesis would not exist without the influence and investment of an enormous collection of people: My family and friends, who bring so much joy to my life - especially my parents, who have always been behind me to the greatest possible degree; and my girlfriend Debashri, who has been a constant source of love, support, and companionship throughout my PhD. (She has also contributed significantly to this thesis by putting up with my strange working hours.) My mentors and teachers, whose skill, patience, and generosity I strive to emulate: in particular, Tze Lai, whose boundless energy and eternal optimism in the face of uncertainty propelled me through my PhD; Phil Lavori, whose practical wisdom and philosophy of `putting another brick in the wall' guided me through the real world of clinical trials; and my grandfather Larry Shepp, who taught me my first theorems and showed me how powerful math can be. I also owe much to the undergraduate professors and mentors who drew me toward statistics and brought me under their wing after my grandfather passed away, including Ed George, Larry Brown, James Troendle, Elchanan Mossel, and Mark Low. Since coming to Stanford, I have enjoyed working alongside and learning from many wonderful colleagues and collaborators. My life and research are much richer for time spent with David Azriel, Adam Kapelner, Abba Krieger, Ying Lu, Mei-Chiung Shih, Nikolas Weissmueller, Huanhong Xu, Ben Stenhaug, Balasubramanian Narasimhan, and my coworkers on consulting projects and at the Stanford Data Science for Social Good program. Plus Nikos Ignatiadis, who has been an exceptional colleague, editor, roommate, and friend. Finally, I point to the towering structures our society has built for all of us to stand upon, supported by the contributions of many people whom I will never know.

\afterpreface

\chapter[Adaptive Experiments and the Need for Frameworks]{Adaptive Experiments and the Need for Robust Frameworks and Methods}

FDA's guidance on adaptive design \citep{food2019adaptive} defines an adaptive clinical trial design as ``one that allows for prospectively planned modifications to one or more aspects of the design based on accumulating data from subjects in the trial." Adaptive experiments can achieve multiple key goals, such as stopping a trial early for success or futility, reducing the expected sample size by estimating the necessary enrollment or information during the trial, treatment selection, subgroup selection, or allocating more patients to the superior treatment. They offer the ambitious designer a combinatorial explosion of choice. With increasing acceptance of adaptive designs and improving ability to execute them, scientists can look forward to more flexibility and capability to answer multiple questions in one shot; institutions, to resource savings, accelerated timelines, or care improvements; and trial designers to craftsmanship - each trial a special snowflake.

For a regulator, however, the last point poses a critical question: how to manage an explosion of design heterogeneity without either containing it and limiting its benefits, or allocating a rapidly growing budget of attention and resources. This challenge is not unique to the FDA; we may also conceive of a ``regulator" abstractly as a player in an applicant - gatekeeper game, where the applicant designs and executes a statistical experiment, and the gatekeeper must either approve or reject the result. Thus, the regulator may be the FDA or other medical regulators; a medical insurer who must make a reimbursement decision; a scientific journal editor weighing evidence to accept a paper; an internet company using A/B testing to decide how and whether to change a website design; or a business planning to purchase advertisements based on a vendor's claims. For better or worse, in confirmatory applications, where gatekeeping institutions are incentivized to avoid false positives and deter bluffing, this issue is often handled by requiring standardized design constraints and metrics such as Type I Error, leaving many other design features to the applicant. The robustness, efficiency, and accessibility of these requirements will ultimately affect not only which applications succeed, but also the energy which applicants must expend to comply with (or subvert) the regulator's requirements; and therefore, in the long term, which types of applicants and institutions can succeed under competitive pressures.

To the regulator, then, this landscape demands predictable, verifiable, hard-to-game, but otherwise mild requirements for design and analysis. Here, one may think of Type I Error requirements, FWER and FDR in multiple testing, or Bayesian estimation (particularly in the case when a prior and model can be agreed upon, but a prespecified design cannot). To the research statistician, it demands robust, cheap-to-implement, compliant, and broadly effective methodologies. For example, techniques such as bootstrapping, studentization, Gaussian approximations, hybrid resampling, uniform confidence bounds from concentration inequalities, always valid p-values, and simulation. For the design and decision rules, particularly for complex applications such as precision-guided drug development, basket protocols, Point-of-Care trials, or SMAR trials, one may think of Bayesian decision theory, closed testing, simulation, and multi-arm bandit approaches such as Thompson sampling, epsilon-greedy randomization, bandit processes, and arm elimination. We hope the results presented in this thesis will help guide the reader toward such appropriate and powerful tools, and push forward their boundaries of capability on multiple dimensions.

A roadmap of the developments in this thesis, following the chronological order of the associated projects: (i) (Chapter 2) Novel clinical trial designs and statistical methods in the era of precision medicine. (ii) (Chapter 3) Multi-armed bandit theory, with applications to learning healthcare systems and clinical trials. (iii) (Chapter 4) Bandit and covariate processes, with finite and non-denumerable set of arms. (iv) (Chapter 5) A rigorous framework for simulation-based verification of adaptive design properties.  Each of the subsequent chapters contains its own introduction section, background literature review, and conclusion section with further discussion.

\chapter[Novel Trial Designs and Methods for Precision Medicine]{Novel Clinical Trial Designs and Statistical Methods in the Era of Precision Medicine}

\section{Introduction } 
The first topic of this chapter is concerned with ``Adaptive Subgroup Selection in Confirmatory Clinical Trials"  addressed in Section 2.2.4, where we describe master protocols for precision-guided drug development and efficacy/safety testing. The second topic of this chapter is "Group Sequential and Adaptive Designs of Confirmatory Trials of New Treatments" as addressed in Section 2.2.2, and Section 2.2.3 considers related advances in statistical analysis of trials for regulatory submission. Section 2.2.1 provides the literature review and background of efficient adaptive designs, and in the remainder of this section we discuss the background of master protocols.

\subsection{Targeted Therapies in Oncology and FDA's Drug Guidance}

Precision medicine considers the ``individual variability in genes, environment, and lifestyle” of a patient to better prevent or treat illness \citep{NIHprecision, garrido2018proposal}. In his State of the Union Address in January 2015, President Barack Obama launched a precision medicine initiative, that was to focus first on the improvement of cancer therapies. Experts agreed that oncology was “the clear choice,” owing to recent advances in diagnostic technology, computational capability, and scientific understanding of cancers, which remain a leading cause of morbidity and mortality worldwide \citep{collins2015new}. Targeted therapies and immuno-oncology (IO) agents were among the forerunners of transformative new medicines, and have heavily utilized innovative statistical methods to meet the clinical development challenges inherent to personalized medicines \citep{de2010translating, snyder2014genetic}.

Targeted therapies have established their benefit over conventional cytotoxic therapy across multiple tumors \citep{hodi2010improved, borghaei2015nivolumab, postow2015immune}. However, a large unmet medical need remains for most malignancies. Patients are seeking better options urgently. 
The comprehensive evaluation of new investigational targeted therapies in oncology, in a timely and resource efficient manner, is infeasible with conventional large randomized trials \citep{ersek2018implementing,ersek2019critical}. To match the right therapy with the right patients, the number of scientific questions that need to be answered during clinical development has increased substantially. 
Traditionally, oncology drug development comprises a series of clinical trials where each study’s objective is to establish the safety and efficacy of a single investigational therapy over the current standard of care (SOC) in a broad study population \citep{redman2015master, berry2015brave}. A targeted therapy’s safety and benefit over the SOC needs to be established for a long list of considerations specific to the biomarker-defined subpopulation and pathology, including safety, therapy sequence, drug combinations, combination dosing, and the contribution of individual drug components. 
The reality of developing “precision medicines” is that there are fewer subjects, who are harder to find, which may jeopardize study completion and extend timelines. The costs of trials have increased with more extensive tissue sample collection, biomarker assessment and tumor imaging, more expensive comparator drugs, and the generally rising cost of medical care. 
Recent advances in tumor sequencing and genomics affords a more detailed understanding of the underlying biology and pathology \citep{lima2019recent}. Although focusing in on molecularly defined subpopulations, this actually expands the reach of targeted therapies across tumors and lines of therapy which can be matched by specific gene signatures or biomarkers such as high expression of microsatellite instability (MSI-hi), or PD-1/PD-L1.


In 2019, 3,876 immuno-therapy compounds were in clinical development, 87\% of which were oncology agents. This marks a 91\% increase over the 2,030 compounds in development in 2017 \citep{xin2019immuno}. With additional information emerging at an increasing pace, it is expected that today’s clinical protocols will require revisions tomorrow, and may need to accommodate a potential change in SOC, emerging information on safety and efficacy of similar compounds, and a better understanding of the fundamental tumor biology \citep{hirsch2013characteristics, xin2019immuno}.
Therefore, clinical study protocols are required that learn faster from fewer study subjects, expedite the evaluation of novel therapies, use resources judiciously, enable robust hypotheses evaluation, are operationalizable across most clinics, and afford sufficient flexibility to answer multiple research questions and respond to emerging information.
Master protocols have emerged to address this challenge. The term ``master protocol” refers to a single overarching design that evaluates multiple hypotheses, with the objective to improve efficiency and uniformity through standardization of procedures in the development and evaluation of different interventions \citep{renfro2018definitions}. In 2001, the first clinical trial to use a “master protocol” was the study B2225, an Imatinib Targeted Exploration.  \citep{mcarthur2005molecular,park2019systematic}. However, the uptake of master protocols was slow. In 2005, STAMPEDE became only the second study to employ a master protocol design, and by 2010, there were still fewer than 10 master protocol-guided studies in the public domain. The subsequent decade from 2010 to 2019 saw a rapid growth resulting in a 10-fold increased use of master protocols in clinical studies \citep{park2019systematic}. A recent catalyst was the validation of the “master protocol” approach by the regulators. In 2018, the FDA issued a draft guidance for industry that advises on the use of master protocols in support of clinical development, titled “Efficient Clinical Trial Design Strategies to Expedite Development of Oncology Drugs and Biologics” \citep{lal}.
In 2019, 83 clinical trials were in the public domain that utilized master protocols.

\subsection{Umbrella, Platform, and Basket Trials}

The aforementioned rapid growth was also catalyzed by the successful immuno-oncology therapies targeting CTLA-4 in 2011 (ipilimumab), and PD-1 in 2014 (pembrolizumab and nivolumab); see \citet{oiseth2017cancer}. Moreover, clinical development teams were faced with the need to explore quickly and efficiently a broader set of malignancies. Basket trials (59\%) accounted for the largest portion of master protocols in 2019, followed by umbrella trials (22\%), and platform trials (19\%). The growth rate of platform trials outpaced umbrella and basket trials in the late 2010s. The majority of master protocol studies (92\%) focus on oncology, and 83\% enrolled adult populations \citep{park2019systematic}. 
Recently, CDER communicated that the “FDA modernizes clinical trials with master protocols” citing good practice considerations, which is expected to further encourage industry to utilize master protocols to rapidly deliver their drug pipelines; see \citet{food2018master}. Basket trials, umbrella trials, and platform trials are implementation structures of clinical studies, and their designs are defined within a master protocol. Each trial variant provides specific flexibility in the clinical development process, and has its advantages and disadvantages \citep{renfro2018definitions, cecchini2019challenges}. The study design and statistical consideration will need to be weighed on a case-by-case basis so that the clinical hypotheses can be answered as directly as possible.
Key design choices required of clinical development teams are whether to study multiple investigational drugs in one protocol, include a control arm, open multiple cohorts to test for multiple biomarkers, and whether to add or stop treatment arms during the course of the trial. Statistical analysis choices include whether to use Bayesian or frequentist methods to evaluate efficacy,how best to randomize subjects, the selection of appropriate futility and early success criteria, and what covariates to control \citep{renfro2018definitions,food2018master,renfro2017statistical,mandrekar2015improving}.

Basket trials investigate a single drug, or a single combination therapy, across multiple populations based on the presence of specific histology, genetic markers, prior therapies, or other demographic characteristics \citep{food2018master}. They may include expansion cohorts and are especially well-suited for “signal-finding.” They frequently comprise single-arm, open-label Phase I or II studies, enroll 20-50 subjects per sub-study, and use two- or multi-stage decision gates to rapidly screen multiple populations for detecting large efficacy signals (by combining multiple tumor types in one protocol) with acceptable safety profiles \citep{park2019systematic, renfro2018definitions}.
Unlike umbrella and platform trials in which the \textbf{R}ecommended \textbf{P}hase \textbf{II} \textbf{D}ose (RP2D) has been pre-established, a basket trial may enroll first-in-human cohorts for whom the RP2D may be established alongside any safety and efficacy signals \citep{cecchini2019challenges}. While often exploratory, basket trials can have registrational intent. An example is Keynote158 which studied pembrolizumab in solid tumors with high microsatellite instability (MSI-H). 
The simplicity of the basket protocol and its relatively small size needs to be weighed against the design’s lack of control groups and limited information for sub-populations based on pooled sample analyses. Basket trial protocols may be amended to include additional tumor types and study populations \citep{renfro2017statistical}, ineffective cohorts can be excluded in a response-adaptation approach, and new cohorts can be added, but such changes often require a protocol amendment and subsequent patient reconsenting plus retraining of study personnel. \citet{cecchini2019challenges} give a comprehensive review of the challenges from the perspectives of the study sponsor, regulator, investigator, and institutional review boards, and discuss the increased operational complexity and increased cost that accompany the reduction in development time. Despite these limitations, basket protocols have become the most widely used master protocols, as they offer the smallest and fastest option, with a median study size of 205 subjects and a 22.3-month study duration \citep{park2019systematic}. 
Statistical methods which are often used to analyze these trials include frequentist sequential \citep{leblanc2009multiple, park2019systematic} and hierarchical Bayesian \citep{berry2006bayesian, thall2003hierarchical} methods, and the recent approaches that control the family-wise error rates for multi-arm studies \citep{chen2016statistical}, response adaptive randomization \citep{ventz2017bayesian, lin2017comparison}, calibrated Bayesian hierarchical testing \nocite{chu2018bayesian} 
and subgroup design \nocite{chu2018blast} 
(Chu and Yuan, 2018a,b), robust exchangeability \citep{neuenschwander2016robust}, modification of Simon's two stage design to improve efficiency \citep{cunanan2017efficient}, and combination of frequentist and Bayesian approaches \citep{lin2017comparison}.  

Umbrella and platform trials are master protocols with exploratory or registrational intent that match biomarker-selected subgroups with subgroup-specific investigational treatments, and may include the current standard of care for the disease setting as a shared control group. They aim at identifying population subgroups that derive the most clinically meaningful benefit from an investigational therapy, and may enable a smaller, faster, and more cost-effective confirmatory phase III study. 
Umbrella trials are often phase II or phase II/III, have an established RP2D for each investigational therapy, and frequently include biomarker enriched cohorts \citep{renfro2018definitions}. The totality of umbrella trial data enables inference on the predictive and prognostic potential of the studied biomarkers within the given disease setting \citep{renfro2018definitions}. While the study of specific biomarker subsets is a key focus, the inclusion of rare populations can lead to accelerated regulatory approval to fill an unmet need but may result in long accrual and trial durations. It is possible to add or remove investigational treatments and subgroups, but the required protocol amendments can cause considerable logistic challenges for sponsors and investigators \citep{cecchini2019challenges}.
The pre-planned and algorithmic addition or exclusion of treatments during trial conduct is what distinguishes platform trials from umbrella trials \citep{angus2019adaptive}. Platform trials frequently include futility criteria and interim analyses, which provide guidance on whether to expand or discontinue a given investigational therapy. Platform trial cohorts often have an established RP2D for each investigational therapy, and may be expanded directly to a registrational Phase III trial while retaining the flexibility to keep other populations in the study \citep{renfro2018definitions}. Recommendations to continue or discontinue treatments are often derived by using Bayesian and Bayesian hierarchical methods \citep{saville2016efficiencies,hobbs2018controlled}. Some protocols leverage response-adaptive randomization to increase the probability that subjects are assigned to the likely superior treatment for their biomarker type, which may provide ethical and cost advantages over conventional randomization \citep{berry2006bayesian,wen2017response}.
Umbrella and platform trials are 2-5 fold larger and longer than the average basket trial \citep{park2019systematic}, and it is important to weigh the benefits of a smaller Phase I basket trial, which may be amended to provide sufficient data for accelerated approval of a novel therapy as demonstrated by Keynote-001 \citep{kang2017pembrolizumab}, versus the longer and more comprehensive evaluation of multiple investigational agents and subgroups. Another disadvantage co-travelling with the larger size, duration and cost of umbrella and platform trials is the potential change in the treatment landscape and SOC, which may necessitate subsequent modifications to bridge between the control and therapy arms \citep{lai2015adaptive,cecchini2019challenges,renfro2018definitions}.

\section{Group Sequential and Adaptive Designs of Confirmatory Trials of New Treatments}

As pointed out by \citet[ p.77]{bartroff2013sequential}, in standard designs of clinical trials comparing a new treatment with a control (which is a standard treatment or placebo), the sample size is determined by the power at a given alternative, but it is often difficult to specify a realistic alternative in practice because of lack of information on the magnitude of the treatment effect difference before actual clinical trial data are collected. On the other hand, many trials have Data and Safety Monitoring Committees (DSMCs) who conduct periodic reviews of the trial, particularly with respect to incidence of treatment-related adverse events, hence one can use the trial data at interim analyses to estimate the effect size. This is the idea underlying group sequential trials in the late 1970s, and one such trial was the \textbf{B}eta-blocker \textbf{H}eart \textbf{A}ttack \textbf{T}rial (BHAT) that was terminated in October 1981, prior to its prescheduled end in June 1982; see \citet{bartroff2013sequential}. BHAT, which was a multicenter, double-blind, randomized placebo-controlled trial to test the efficacy of long-term therapy with propranolol given to survivors of an acute myocardial infarction (MI), drew immediate attention to the benefits of sequential methods not because it reduced the number of patients but because it shortened a 4-year study by 8 months, with positive results for a long-awaited treatment for MI patients.
The success story of BHAT paved the way for major advances in the development of group sequential methods in clinical trials and for the widespread
adoption of group sequential design. Sections 3.5 and 4.2 of \citet{bartroff2013sequential} describe the theory developed by Lai and Shih (2004) 
\nocite{lai2004power}
for nearly optimal group sequential tests in exponential families to provide a definitive method amidst the plethora of group sequential stopping boundaries that were proposed in the two decades after BHAT, as reviewed in \citet{bartroff2013sequential}.

Lai and Shih's theory is based on (a) asymptotic lower bounds for the sample sizes of group sequential tests that satisfy prescribed type I and type II error probability bounds, and (b) group sequential generalized likelihood ratio (GLR) tests with modified Haybittle-Peto boundaries that can be shown to attain these bounds. Noting that the efficiency of a group sequential test depends not only on the choice of the stopping rule but also on the test statistics, Lai and Shih use GLR statistics that have been shown to have asymptotically optimal properties for sequential testing in one-parameter exponential families and can be readily extended to multiparameter exponential families for which the type I and type II errors are evaluated at $u(\theta)=u_0$ and $u(\theta)=u_1$, respectively, where $u: \Theta \rightarrow \mathbb{R}$ is a continuously differentiable function on the natural parameter space $\Theta$ such that Kullback-Leibler information number $I(\gamma, \theta)$ is increasing in $|u(\theta)-u(\gamma)|$ for every $\gamma$; see Bartroff et al. (2013, Sections 3.7 and 4.2.4). An important consideration in this approach is the choice of the alternative $\theta_1$ (in the one-parameter case, or $u_1$ in the multiparameter exponential families). To test $H_0:\theta\leq\theta_0$, suppose the significance level is $\alpha$ and no more than $M$ observations are to be taken because of funding and administrative constraints on the trial. The FSS (fixed sample size) test that rejects $H_0$ if $S_M\geq c_{\alpha}$ has maximal power at any alternative $\theta>\theta_0$. Although funding and administrative considerations often play an important role in the choice of $M$, justification of this choice in clinical trial protocols is typically based on some prescribed power $1-\beta$ at an alternative $\theta(M)$ ``implied" by $M$. The implied alternative is defined by that $M$ and can be derived from the prescribed power $1-\beta$ at $\theta(M)$. It is used to construct the futility boundary in the modified Haybittle-Peto group sequential test (Bartroff et al., 2013, pp.81-85). 

\subsection{Efficient Adaptive Designs}

Using Lai and Shih's theory of modified Haybittle-Peto group sequential tests, Bartroff and Lai (2008a,b) 
\nocite{bartroff2008efficient} \nocite{bartroff2008generalized} 
developed a new approach to adaptive design of clinical trials. In standard clinical trial designs, the sample size is determined by the power at a given alternative, but in practice, it is often difficult for investigators to specify a realistic alternative at which sample size determination can be based. Although a standard method to address this difficulty is to carry out a preliminary pilot study, the results from a small pilot study may be difficult to interpret and apply, as pointed out by \citet{wittes1990role}, who proposed to treat the first stage of a two-stage clinical trial as an internal pilot from which the overall sample size can be re-estimated. The specific problem they considered actually dated back to Stein's
(1945)
\nocite{stein1945two}
two-stage procedure for testing hypothesis $H_0:\mu_X=\mu_Y$ versus the two-sided alternative $\mu_X\ne \mu_Y$ for the means of two independent normal distributions with common, unknown variance $\sigma^2$. In its first stage, Stein's procedure samples $n_0$ observations from each of the two normal distributions and computes the usual unbiased estimate $s_0^2$ of $\sigma^2$. The second stage samples
$n_1=n_0\vee \lfloor \left(t_{2n_0-2,\alpha/2}
  +t_{2n_0-2,\beta} \right)^{\!\!2} \,2s_0^2/\delta^2 \rfloor$
observations from each population, where $\lfloor \cdot \rfloor$ denotes the greatest integer function, $\alpha$ is the prescribed type I error probability, $t_{\nu,\alpha}$ is the upper $\alpha$-quantile of the $t$-distribution with $\nu$ degrees of freedom, and $1-\beta$ is the prescribed power at the alternatives satisfying $|\mu_X-\mu_Y|=\delta$. The null hypothesis $H_0:\mu_X=\mu_Y$ is then rejected if $$|\bar{X}_{n_1}- \bar{Y}_{n_1}|>t_{2 n_0-2,\alpha/2}\sqrt{2s_0^2/n_1}.$$ Modifications of the two-stage procedure were provided by
\citet{wittes1990role}, \citet{lawrence1992sample}, and \citet{herson1993use}, which represent the ``first generation" of adaptive designs. The second generation of adaptive designs adopts a more aggressive method to re-estimate the sample size from the estimate of $\delta$ (instead of the nuisance parameter $\sigma$) based on the first-stage data. In particular,  \citet{fisher1998self} considers the case of normally distributed outcome variables with known common variance $\sigma^2$. 
Letting $n$ be the sample size for each treatment and $0 < r< n$, he notes that after $r n$ pairs of observations $(X_i, Y_i)$, $S_1 = \sum_{1}^{r n}(X_i - Y_i) \sim N(r n \delta , 2 \sigma^2 r n),$ where $\delta = \mathbb{E}(X_i - Y_i)$. Let $\gamma > 0$ and $n^* = rn + \gamma (1- r) n$ be the new total sample size for each treatment. Under $H_0: \delta = 0$, $$S_2 = \sum \limits_{i = r n^* + 1}^{n^*} (X_i - Y_i) \sim N(0, 2\sigma^2(1-r) \gamma n),$$ hence the test statistic $(2 \sigma^2 n)^{-1/2}(S_1 + \gamma^{-1/2} S_2)$ is standard normal. Whereas Fisher uses a ``variance spending" approach as $1 - r$ is the remaining part of the total variance that has not been spent in the first stage, \citet{proschan1995designed} use a conditional Type I error function $C(z)$ with range $[0,1]$ to define a two-stage procedure that rejects $H_0: \delta = 0$ in favor of $\delta > 0$ if the second-stage $z$-value $Z_2$ exceeds $\Phi^{-1}(1 - C(Z_1))$, where $Z_1$ is the first-stage $z$-value. The type I error of the two-stage test can be kept at $\alpha$ if $\int \limits_{-\infty}^{\infty} C(z) \phi(z) dz = \alpha,$ where $\phi$ and $\Phi$ are the density function and distribution function, respectively, of the standard normal distribution. 

Assuming normally distributed outcomes with known variances, Jennison and Turnbull (2006 a,b) \nocite{jennison2006adaptive} \nocite{jennison2006efficient}
introduced adaptive group sequential tests that choose the $j$th group size and stopping boundary on the basis of the cumulative sample size $n_{j-1}$ and the sample sum $S_{n_{j-1}}$ over the first $j-1$ groups, and that are optimal in the sense of minimizing a weighted average of the expected sample sizes over a collection of parameter values, subject to prescribed error probabilities at the null and a given alternative hypothesis. They showed how the corresponding optimization problem can be solved numerically by using backward induction algorithms, and that standard (non-adaptive) group sequential tests with the first stage chosen appropriately are nearly as efficient as their optimal adaptive tests. They also showed that the adaptive tests proposed in the preceding paragraph performed poorly in terms of expected sample size and power in comparison with the group sequential tests. \citet{tsiatis2003inefficiency} attributed this inefficiency to the use of the non-sufficient ``weighted'' statistic. Bartroff and Lai's (2008a,b) approach to adaptive designs, developed in the general framework of multiparameter exponential families, uses efficient generalized likelihood ratio statistics in this framework and adds a third stage to adjust for the sampling variability of the first-stage parameter estimates that determine the second-stage sample size. The possibility of adding a third stage to improve two-stage designs dated back to \citet{lorden1983asymptotic}, who used crude upper bounds for the type I error probability that are too conservative for practical applications. Bartroff and Lai overcame this difficulty by using new methods to compute the type I error probability, and also extended the three-stage test to multiparameter and multi-arm settings, thus greatly broadening the scope of these efficient adaptive designs. Details are summarized in Chapter 8, in particular Sections 8.2 and 8.3, of Bartroff et al. (2013), where Section 8.4 gives another modification of group sequential GLR tests for adaptive choice between the superiority and non-inferiority objectives of a new treatment during interim analyses of a clinical trial to test the treatment's efficacy, as in an antimicrobial drug developed by the company of one of the coauthors of \citet{lai2006modified}.

\subsection{Adaptive Subgroup Selection in Confirmatory Trials}
\label{subsec:adaptive}

Choice of the patient subgroup to compare the new and control treatments is a natural compromise between ignoring patient heterogeneity and using stringent inclusion-exclusion criteria in the trial design and analysis. \citet{lai2014adaptive} introduce a new adaptive design to address this problem. They first consider trials with fixed sample size, in which $n$ patients are randomized to the new and control treatments and the responses are normally distributed, with mean $\mu_j$ for the new treatment and $\mu_{0j}$ for the control treatment if the patient falls in a pre-defined subgroup $\Pi_j$ for $j = 1,\ldots, J$, and with common known variance $\sigma^2$. Let $\Pi_J$ denote the entire patient population for a traditional randomized controlled trial (RCT) comparing the two treatments, and let $\Pi_1\subset\Pi_2\subset\cdots\subset\Pi_J$ be the $J$ prespecified subgroups. Since there is typically little information from previous studies about the subgroup effect size $\mu_j-\mu_{0j}$ for $j \neq J$, Lai et al. (2014) begins with a standard RCT to compare the new treatment with the control over the entire population, but allows adaptive choice of the patient subgroup $\hat{I}$, in the event $H_J$ is not rejected, to continue testing $H_i:\mu_i\leq \mu_{0i}$ with $i = \hat{I}$ so that the new treatment can be claimed to be better than control for the patient subgroup $\hat{I}$ if $H_{\hat{I}}$ is rejected. Letting $\theta_j=\mu_j-\mu_{0j}$ and $\boldsymbol\theta=(\theta_1,\ldots,\theta_J)$, the probability of a false claim is the type I error
\begin{align}
\alpha(\boldsymbol\theta) =
  \begin{cases}
   P_{\boldsymbol\theta}(\text{reject } H_J) + P_{\boldsymbol\theta}(\theta_{\hat{I}}\leq 0, \text{ accept } H_J \text{ and } \text{reject } H_{\hat{I}})     &    \text{if } \theta_J \leq 0 \\
   P_{\boldsymbol\theta}(\theta_{\hat{I}}\leq 0, \text{ accept } H_J \text{ and } \text{reject } H_{\hat{I}})      &   \text{if }  \theta_J>0,
  \end{cases}\label{alpha_theta}
\end{align}
for $\boldsymbol\theta\in\Theta_0$. Subject to the constraint $\alpha(\boldsymbol\theta)\leq\alpha$, they prove the asymptotic efficiency of the procedure that randomly assigns $n$ patients to the experimental
treatment and the control, rejects $H_J$ if $\mathrm{GLR}_i\geq c_{\alpha}$ for $i=J$, and otherwise chooses the patient subgroup $\hat{I}\neq J$ with the largest value of the generalized likelihood
ratio statistic $\mathrm{GLR}_i=\{n_in_{0i}/(n_i+n_{0i})\}(\hat{\mu}_i-\hat{\mu}_{0i})^2_{+}/\sigma^2$ among all subgroups $i\neq J$ and rejects $H_{\hat{I}}$ if $\mathrm{GLR}_{\hat{I}}\geq c_{\alpha}$, where $\hat{\mu}_i(\hat{\mu}_{0i})$ is the mean response of patients in $\Pi_i$ from the treatment (control) arm and $n_i(n_{0i})$ is the corresponding sample size. After establishing the asymptotic efficiency of the procedure in the fixed sample size case, they proceed to extend it to a 3-stage sequential design by making use of the theory of Bartroff and Lai reviewed in the preceding paragraph. They then extend the theory from the normal setting to asymptotically normal test statistics, such as the Wilcoxon rank sum statistics. These designs which allow
 mid-course enrichment using data collected, were motivated by the design of the DEFUSE 3 clinical trial at the Stanford Stroke Center to evaluate a new method for augmenting usual medical care with endovascular removal of the clot after a stroke, resulting in reperfusion of the area of the brain under threat, in order to salvage the damaged tissue and improve outcomes over standard medical care with intravenous tissue plasminogen activator (tPA) alone. The clinical endpoints of stroke patients are the Rankin scores, and Wilcoxon rank sum statistics are used to test for differences in Rankin scores between the new and control treatments. The DEFUSE 3 (\textbf{D}iffusion and Perfusion Imaging \textbf{E}valuation \textbf{f}or \textbf{U}nderstanding \textbf{S}troke \textbf{E}volution) trial design involves a nested sequence of $J=6$ subsets of patients, defined by a combination of elapsed time from stroke to start of tPA and an imaging-based estimate of the size of the unsalvageable core region of the lesion. The sequence was defined by cumulating the cells in a two-way (3 volumes $\times$ 2 times) cross-tabulation as described by Lai et al. (2014, p. 195). In the upper left cell, $c_{11}$, which consisted of the patients with a shorter time to treatment and smallest core volume, the investigators were most confident of a positive effect, while in the lower right cell $c_{23}$ with the longer time and largest core area, there was less confidence in the effect. The six cumulated groups, $\Pi_1,\ldots,\Pi_6$ give rise to corresponding one-sided null hypotheses, $H_1,\ldots,H_6$ for the treatment effects in the cumulated groups. 

Shortly before the final reviews of the protocol for funding were completed, four RCTs of endovascular reperfusion therapy administered to stroke patients within 6 hours after symptom onset demonstrated decisive clinical benefits. Consequently, the equipoise of the investigators shifted, making it necessary to adjust the intake criteria to exclude patients for whom the new therapy had been proven to work better than the standard treatment. The subset selection strategy became even more central to the design, since the primary question was no longer whether the treatment was effective at all, but for which patients should it be adopted as the new standard of care. Besides adapting the intake criteria to the new findings, another constraint was imposed by the NIH sponsor, which effectively limited the total randomization to 476 patients. The first interim analysis was scheduled after the 200 patients, and the second interim analysis after an additional 140 patients. DEFUSE 3 has a Data Coordinating Unit and an independent Data and Safety Monitoring Board (DSMB). Besides examining the unblinded efficacy results prepared by a designated statistician at the data coordination unit, which also provided periodic summaries on enrollment, baseline characteristics of enrolled patients, protocol violations, timeliness and completeness of data entry by clinical centers, and safety data. During interim analyses, the DSMB would also consider the unblinded safety data, comparing the safety of endovascular plus IV-tPA to that of IV-tPA alone, in terms of deaths, serious adverse events, and incidence of symptomatic intracranial hemorrhage. 

In June 2017 positive results of another trial, \textbf{D}WI or CTP \textbf{A}ssessment with Clinical Mismatch in the Triage of \textbf{W}ake-Up and Late Presenting Stokes undergoing \textbf{N}euro-intervention with Trevo (DAWN), 
which involved patients and treatments similar to those of DEFUSE 3, were announced. Enrollment in the DEFUSE 3 trial was placed on hold; an early interim analysis of the 182 patients enrolled to date was requested by the sponsor (NIH); see \citet{albers2018thrombectomy} who say: ``As a result of that interim analysis, the trial was halted because the prespecified efficacy boundary $(P<0.0025)$ had been exceeded." 
As reported by the aforementioned authors, DEFUSE 3 ``was conducted at 38 US centers and terminated early for efficacy after 182 patients had undergone randomization (92 to the endovascular therapy group and 90 to the medical-therapy group)." For the primary and secondary efficacy endpoints, the results show significant superiority of endovascular plus medical therapies. The DAWN trial ``was a multicenter randomized trial with a Bayesian adaptive-enrichment design" and was ``conducted by a steering committee, which was composed of independent academic investigators and statisticians, in collaboration with the sponsor, Stryker Neurovascular" \citep{nogueira2018thrombectomy}. Early termination of DEFUSE 3 provides a concrete example of importance of a flexible group sequential design that can adapt not only to endogenous information from the trial but also to exogenous information from advances in precision medicine and related concurrent trials. 

We conclude this section with recent regulatory developments in enrichment strategies for clinical trials and in adaptive designs of confirmatory trials of new treatments. In March 2019, the FDA released its Guidance for Industry on Enrichment Strategies for Clinical Trials to Support Determination of Effectiveness of Human Drugs and Biological Products. In November 2019, CDER and CBER of FDA released its Guidance for Industry on Adaptive Designs for Clinical Trials of Drugs and Biologics, which was an update of the 2010 CDER's Guidance for Industry on Adaptive Designs.


\section{Analysis of Novel Confirmatory Trials}

This section describes some advances in statistical methods for the analysis of the novel clinical trial designs of confirmatory trials in Section 2.2. It begins with hybrid resampling for inference on primary and secondary endpoints in Section 2.3.1. Section 2.3.2 considers statistical inference from multi-arm trials for developing and testing biomarker-guided personalized therapies.

\subsubsection{Hybrid Resampling for Primary and Secondary Endpoints}

\citet{tsiatis1984exact} developed exact confidence intervals for the mean of a normal distribution with known variance following a group sequential test. Subsequently,
\cite{chuang1998resampling,chuang2000hybrid} 
noted that even though $\sqrt{n}(\bar{X}_n-\mu)$ is a pivot in the case of $X_i\sim N(\mu,1)$, $\sqrt{T}(\bar{X}_T-\mu)$ is highly non-pivotal for a group sequential stopping time, hence the need for the {\it exact method} of \citet{tsiatis1984exact}, which they generalized as follows. If $\mathcal{F}=\{F_{\theta}:\theta\in\Theta\}$ is indexed by a real-valued parameter $\theta$, an exact equal-tailed confidence region can always be found by using the well-known duality between hypothesis tests and confidence regions. Suppose one would like to test the
null hypothesis that $\theta$ is equal to $\theta_0$. Let $R(\mathbf{X},\theta_0)$ be some real-valued test statistic. Let $u_{\alpha}(\theta_0)$ be the $\alpha$-quantile of the distribution of $R(\mathbf{X},\theta_0)$ under the distribution $F_{\theta_0}$. The null hypothesis is accepted if $u_{\alpha}(\theta_0)<R(\mathbf{X},\theta_0)<u_{1-\alpha}(\theta_0)$. An exact equal-tailed confidence region with coverage probability $1-2\alpha$ consists of all $\theta_0$ not rejected by the test and is therefore given by $\{\theta: u_{\alpha}(\theta)<R(\mathbf{X},\theta)<u_{1-\alpha}(\theta)\}$. The exact method, however, applies only when there are no nuisance
parameters and this assumption is rarely satisfied in practice. To address this difficulty, Chuang and Lai (1998, 2000) introduced a {\it hybrid resampling method} that ``hybridizes" the exact method with Efron's (1987) \nocite{efron1987better} 
bootstrap method to construct confidence intervals. The bootstrap method replaces the quantiles $u_{\alpha}(\theta)$ and $u_{1-\alpha}(\theta)$ by by the approximate quantiles $u^*_{\alpha}$ and $u^*_{1-\alpha}$ obtained in the following manner. Based on $\mathbf{X}$, construct an estimate $\hat{F}$ of $F\in\mathcal{F}$. The quantile $u^*_{\alpha}$ is defined to be $\alpha$-quantile of the distribution of $R(\mathbf{X}^*,\hat{\theta})$ with $\mathbf{X}^*$ generated from $\hat{F}$ and $\hat{\theta}=\theta(\hat{F})$, yielding the confidence region 
$\{\theta: u^*_{\alpha}<R(\mathbf{X},\theta)<u^*_{1-\alpha}\}$ with approximate coverage probability $1-2\alpha$. For group sequential designs, the bootstrap method breaks down because of the absence of an approximate pivot, as shown by \citet{chuang1998resampling}. The hybrid confidence region is based on reducing the
family of distributions $\mathcal{F}$ to another family of distributions $\{\hat{F}_{\theta}:\theta\in\Theta\}$, which is used as the ``resampling family" and in which $\theta$ is the unknown parameter of interest. Let $\hat{u}_{\alpha}(\theta)$ be the $\alpha$-quantile of the sampling distribution of $R(\mathbf{X},\theta)$ under the assumption that $\mathbf{X}$ has distribution $\hat{F}_{\theta}$. The hybrid confidence region results from applying the exact method to $\{\hat{F}_{\theta}:\theta\in\Theta\}$ and is given by 
\begin{align}
\left\lbrace \theta:\hat{u}_{\alpha}(\theta)<R(\mathbf{X},\theta)<\hat{u}_{1-\alpha}(\theta) \right\rbrace.\label{hybrid_CI}
\end{align} 
The construction of \eqref{hybrid_CI} typically involves simulations to compute the quantiles as in the bootstrap method. 

Since an exact method for constructing confidence regions is based on
inverting a test, such a method is implicitly or explicitly linked to an ordering of the sample space of the test statistic used. The ordering defines the $p$-value of the test as the probability (under the null hypothesis) of more extreme values (under the ordering) of the test statistic than that observed in the sample. Under a total ordering $\leq$ of the sample space of $(T,S_T)$, Lai and Li (2006)
\nocite{lai2006confidence}
call $(t,s)$ a $q$th quantile if $P\{(T,S_T)\leq (t,s)\}=q$, which generalizes Rosner and Tsiatis' exact method for randomly stopped sums $S_T$ of of independent normal random variables with unknown mean $\mu$. For the general setting where a stochastic process $\mathbf{X}_u$, in which $u$ denotes either
discrete or continuous time, is observed up to a stopping time $T$, Lai and Li (2006) define $\mathbf{x}=\{ \mathbf{x}_u:u\leq t \}$ to be a $q$th quantile if
\begin{align}
P\{\mathbf{X}\leq \mathbf{x}\}\geq q, \quad P\{\mathbf{X}\geq \mathbf{x}\}\geq 1-q,\label{def:quantile_ordering}
\end{align}
under a total ordering $\leq$ for the sample space of $\mathbf{X}=\{\mathbf{X}_u:u\leq T\}$.
For applications to confidence intervals of a real parameter $\theta$, the choice of the total ordering should be targeted toward the objective of interval estimation. Let $\{U_r:r\leq T\}$ be real-valued statistics based on the observed process $\{\mathbf{X}_s:s\leq T\}$. For example, let $U_r$ be an estimate of $\theta$ based on $\{\mathbf{X}_s:s\leq r\}$. A total ordering on
the sample space of $\mathbf{X}$ can be defined via $\{U_r:r\leq T\}$  as follows:
\begin{align}
\mathbf{X}\geq \mathbf{x} \text{ if and only if } U_{T\wedge t}\geq u_{T\wedge t},\label{Lai_ordering} 
\end{align}
in which $\{u_r:r\leq t\}$ is defined from $\mathbf{x}=\{\mathbf{x}_r:r\leq t\}$ in the same way as $\{U_r:r\leq T\}$ is defined from $\mathbf{X}$ and which has the attractive feature that the probability mechanism generating $\mathbf{X}_t$ needs only to be specified up to the 
stopping time $T$ in order to define the quantile.
Bartroff et al. (2013, p.164) remark that if $U_r = \sqrt{r}(\bar{X_r} - \mu_{0})$ then the Lai-Li ordering is equivalent to Siegmund's ordering and also to the Rosner-Tsiatis ordering, but ``the original Rosner-Tsiatis ordering requires $n_1,\ldots,n_k$ (or the stochastic mechanism generating them to be completely specified" and has difficulties ``described in the last paragraph of Sect. 7.1.3 if this is not the case."

Bartroff et al. (2013, Sections 7.4 and 7.5) 
\nocite{bartroff2013sequential} 
describe how this ordering can be applied to implement resampling for secondary endpoints together with applications to time-sequential trials which involve interim analyses at calendar time $t_j$ $(1\leq j\leq k)$, with $0<t_1<\cdots<t_k=t^*$ (the prescribed duration of the trial), and which have time to failure as the primary endpoint;
Lai et al. (2009) \nocite{lai2009tests} 
have also extended this approach to inference on secondary endpoints in adaptive or time-sequential trials.

\subsection{Statistical Inference from Multi-Arm Trials for Developing and Testing Biomarker-Guided Personalized Therapies }

\citet{lai2013group} first elucidate the objectives underlying the design and analysis of these multi-arm trials that attempt to select the best of $k$ treatments for each biomarker-classified subgroup of cancer patients in Phase II studies, with objectives that include (a) treating accrued patients with the best (yet unknown) available treatment, (b) developing a biomarker-guided treatment strategy for future patients, and (c) demonstrating that the strategy developed indeed has statistically significantly better treatment effect than some predetermined threshold. The group sequential design therefore uses an outcome-adaptive randomization rule, which updates the randomization probabilities at interim analyses and uses GLR statistics and modified Haybittle-Peto rules to include early elimination of inferior treatments from a biomarker class. It is shown by \citet{lai2013group}
to provide substantial improvements, besides being much easier to implement, over the Bayesian outcome-adaptive randomization design used in the BATTLE (\textbf{B}iomarker-integrated \textbf{A}pproaches of \textbf{T}argeted \textbf{T}herapy for \textbf{L}ung Cancer \textbf{E}limination) trial of personalized therapies for non-small cell lung cancer. An April 2010 editorial in \textit{Nature Reviews in Medicine} points out that BATTLE design, which ``allows researchers to avoid being locked into a single, static protocol of the trial" that requires large sample sizes for multiple comparisons of several treatments across different biomarker classes, can ``yield breakthroughs, but must be handled with care" to ensure that ``the risk of reaching a false positive conclusion" is not inflated. As pointed out by Lai et al. (2013, pp.651-653, 662), targeted therapies that target the cancer cells (while leaving healthy cells unharmed) and the ``right" patient population (that has the genetic or other markets for the sensitivity to the treatment) have great promise in cancer treatments but also challenges in designing clinical trials for drug development and regulatory approval. One challenge is to identify the biomarkers that are predictive of response and another is to develop a biomarker classifier that can identify patients who are sensitive to the treatments. We can address these challenges by using recent advances in contextual multi-arm bandit theory, which we summarize below.

The $K$-arm bandit problem,  introduced by \cite{robbins1952some} for the case $K=2$, is prototypical in the area of stochastic adaptive control that addresses the dilemma between ``exploration" (to generate information about the unknown system parameters needed for efficient system control) and ``exploitation" ( to set the system inputs that attempt to maximize the expected rewards from the outputs.) 
Robbins considered the problem of which of $K$ populations to sample from sequentially in order to maximize the expected sum $E \left(\sum_{i=1}^N Y_i\right)$. Let $\mathcal{F}_t$ be the history (or more formally, the $\sigma$-algebra of events) up the time $t$. An allocation rule $\phi=(\phi_1,\ldots,\phi_N)$ is said to be ``adaptive" if $\{\phi_t=k\}\in \mathcal{F}_{t-1}$ for $k=1,\ldots,K$.
Suppose $Y_t$ has density function $f_{\theta_k}$ when ${\phi_t=k}$, and let
$\bm{\theta}=(\theta_1,\ldots,\theta_K)$. Let $\mu_k$ be the mean of the $k$th population, which is assumed to be finite. Then
\begin{align}
E_{\bm{\theta}}\left(\sum_{t=1}^N Y_t\right)=\sum_{t=1}^N\sum_{k=1}^KE_{\bm{\theta}}\{E_{\bm{\theta}}(Y_tI_{\{\phi_t=k\}}|\mathcal{F}_{t-1})\}=\sum_{k=1}^K\mu(\theta_k)E_{\bm{\theta}}T_N(k),\label{1.1}
\end{align}
where $T_N(k)=\sum_{t=1}^NI_{\{\phi_t=k\}}$ is the total sample size from population $k$. If the population $k^*$ with the largest mean were known, then obviously one should sample from it to receive expected reward $N\mu_{k^*}$, where $\mu_{k^*}=\max_{1\leq k \leq K}\mu_k$. Hence maximizing the expected sum $E_{\bm{\theta}}(\sum_{t=1}^N Y_t)$ is equivalent to minimizing the regret, or shortfall from $N\mu_{k^*}$:
\begin{align}
R_N(\bm{\theta})=N\mu_{k^*}-E_{\bm{\theta}}\left(\sum_{t=1}^N Y_t \right)=\sum_{k:\mu(\theta_k)<\mu_{k^*}}\{\mu_{k^*}-\mu(\theta_k)\}E_{\bm{\theta}}T_N(k),
\end{align}
in which the second equality follows from \eqref{1.1} and shows that the regret is a weighted sum of expected sample sizes from inferior populations. Making use of this representation in terms of expected sample sizes, \cite{lai1985asymptotically} derive an the asymptotic lower bound, as $N\rightarrow\infty$, for the regret $R_N(\bm{\theta})$ of uniformly good adaptive allocation rules:
\begin{align}
R_N(\bm{\theta})\geq (1+o(1))\sum_{k:\mu(\theta_k)<\mu(\theta^{*})}\frac{\mu(\theta^*)-\mu(\theta_k)}{I(\theta_k,\theta^*)}\log N,\label{R_Ntheta}
\end{align}
where $\theta^*= \theta_{k^*}$ and  $I(\bm{\theta},\lambda)=E_{\theta}\{\log(f_\theta(Y)/f_\lambda(Y))\}$ is the Kullback-Leibler information number; an adaptive allocation rule is called ``uniformly good" if $R_N(\bm{\theta})=o(N^a)$ for all $a>0$ and $\bm{\theta}$. They show that the asymptotic lower bound (\ref{R_Ntheta}) can be attained by the ``upper confidence bound" (UCB) rule that samples from the population (arm) with the largest upper confidence bound, which incorporates uncertainty in the sample mean by the numbers of observations sampled from the arm (i.e., width of a one-sided confidence interval.)

New applications and advances in information technology and biomedicine in the new millenium have led to the development of {\it contextual multi-arm bandits}, also called bandits with side information or covariates, while the classical multi-arm bandits reviewed above are often referred to as ``context-free" bandits. Personalized marketing (e.g., Amazon) uses web sites to track a customer's purchasing records and thereby to maket products that are individualized for the customer. Recommender systems select items such as movies (e.g., Netflix) and news (e.g., Yahoo) for users based on the users' and items' features(covariates). 
Whereas classical $K$-arm bandits reviewed above aim at choosing $\phi_i$ sequentially so that $ E_{\bm{\theta}}(\sum_{i=1}^N Y_i)$ is as close as possible to $N \max_{1\leq k\leq K} \mu_k$, contextual bandits basically replace $N\mu_k $ by $\sum_{i=1}^N \mu_k(\bm{x_i})$, where $\bm{x_i}$ is the covariate of the $i$th subject, noting that analogous to \eqref{1.1},
\begin{align}
E_{\bm{\theta}} (Y_i)=\sum_{k=1}^KE_{\bm{\theta}}\{E_{\bm{\theta}}(Y_iI_{\{\phi_i=k\}}|\bm{x_i},\mathcal{F}_{t-1})\}=\sum_{k=1}^KE_{\bm{\theta}}(\mu_k(\bm{x_i})I_{\{\phi_i=k\}}).
\end{align}
Assuming $\bm{x_i}$ to be i.i.d. with distribution $G$, we can define $g^*(x)=\arg \max_{1\leq k \leq K} \mu(\theta_k,x)$, $\theta^*(x)=\theta_{k^*(x)}$ and the regret
\begin{equation}
\begin{aligned}
R_N(\bm{\theta},B)&=N \int_B \mu(\theta^*(\bm{x}),\bm{x})dG(\bm{x})-\sum_{i=1}^N \sum_{k=1}^K\int_B \mu(\theta_k,\bm{x})E_{\bm{\theta}}(I_{\{\phi_i=k\}})dG(\bm{x})\\
&=\sum_{k=1}^K\int_B\{\mu(\theta^*(\bm{x}),\bm{x})-\mu(\theta_k,\bm{x})\}E_{\bm{\theta}}T_N(k,\bm{x})dG(\bm{x})\label{1.8}
\end{aligned}
\end{equation}
for Borel subsets $B$ of the support of $G$, where $T_N(k,B)=\sum_{i=1}^NI_{\{\phi_i=k,\bm{x_i}\in B\}}$, noting that the measure $E_{\bm{\theta}}T_N(k,\cdot)$ is absolutely continuous with respect to $G$, hence $E_{\bm{\theta}}T_N(k,\bm{x})$ in \eqref{1.8} is its Radom-Nikodym derivative with respect to $G$. For contextual bandits, an arm that is inferior at $\bm{x}$ may be the best at $\bm{x}'$. Therefore the uncertainty in the sample mean reward at $\bm{x}_t$ does not need to be immediately accounted for, and adaptive randomization (rather than UCB rule) can yield an asymptotically optimal policy.

To achieve the objectives (a), (b) and (c) in the first paragraph of this subsection, Lai et al.\cite[pp.654-655]{lai2013group} use contextual bandit theory which we illustrate below with $J=3$ groups of patients and $K=3$ treatments, assuming normally distributed responses with  mean $\mu_{jk}$ and known variance 1 for patients in group $j$ receiving treatment $k$.  Using Bartroff and Lai's adaptive design (2008a,b) reviewed in Section 2.1,  let  $n_i$ denote the total sample size up to the time of the $i$th interim analysis, $n_{ij}$ denote the total sample size from group $j$ in those $n_i$ patients, and let $n_{ijk}$ be the total sample size from biomarker class $j$ receiving treatment $k$ up to the $i$th interim analysis. Because it is unlikely for patients to consent to being assigned to a seemingly inferior treatment, randomization in a double blind setting (in which the patient and the physician both do not know whether treatment or control is assigned) is needed for informed consent.
Contextual bandit theory suggests assigning the highest randomization probability between interim analyses $i$ and $i+1$ to
$\hat{k}_j^{(i)} = \arg \max_k \hat{\mu}_{jk}$ (which
is the MLE of $k^*_j = \arg \max_k \mu_{jk}$) and eliminating treatment $k$ from the set of $\mathcal{K}_{ij}$ of surviving treatments at the $i$th interim analysis if the GLR statistic $l^i_j(k, \hat{k}^{(i)}_j)$ exceeds 5$\delta_{ij}$, where $\delta_{ij}\to 0$ but
$\sqrt{n_{ij}}\delta_{ij}\rightarrow\infty$, with a randomization scheme 
in which
\begin{align}
\pi_{jk}^{(i)}=\left( 1-\varepsilon|\mathcal{K}_{ij}\setminus \mathcal{H}_{ij}| \right)/|\mathcal{H}_{ij}|, 
\end{align} 
in which $|A|$ denotes the cardinality of a finite set $A$ and  $\mathcal{H}_{ij} = \{k 
 \in \mathbb{X}_{ij}: |\hat{\mu}_{jk}^{(i)} - \hat{\mu}_{j,\hat{k}^{(i)}_j}| \leq \delta_{ij} \}$. Equal randomization (with randomization probability $1/K$) for the $K$ treatments is used up to the first interim analysis. In context-free multi-arm bandit theory, this corresponds to the $\varepsilon$-greedy algorithm which has been shown by \cite{auer2002finite} to provide an alternative to the UCB rule for attaining the asymptotic lower bound for the regret. 
\citet{lai2013group} introduce a subset selection method for selecting a subset of treatments at the end of the trial to be used for future patients, with an overall probability guarantee of $1-\alpha$ to contain the best treatment for each biomarker class, and such that the expected size of the selected subset is as small as possible in some sense. They also develop a group sequential GLR test with prescribed type I error to demonstrate that the developed treatment strategy improves the mean treatment effect of SOC by a given margin.

\subsection{Precision-Guided Drug Development and Basket Protocols}

Janet Woodcock, director of FDA's Center for Drug Evaluation and Research (CDER) and current Acting Commissioner of the FDA, published in 2017 a seminal paper on master protocols of ``mechanism-based precision medicine trials," affordable in cost, time, and sample size, to study multiple therapies, multiple diseases, or both; see \citet{woodcock2017master}.
Table 2 of the paper lists six such trials to illustrate the concept: (i) B2225, a Phase II basket trial, (ii) BRAF V600, an early Phase II basket trial, (iii) NCI-Match, a Phase I followed by Phase II umbrella trial, (iv) BATTLE-1, a Phase II umbrella trial, (v) I-SPY 2, a Phase II platform trial, and (vi) Lung-MAP, a Phase II-III trial with a master protocol to study 4 molecular targets for NSCLC initially, to be trimmed to 3 targets for the PHASE III confirmatory trial. We have discussed the BATTLE (respectively, I-SPY) trials for therapies to treat NSCLC (respectively, breast cancer) in Section 3.2. For NCI-Match, a treatment is given across multiple tumors sharing a common biomarker; see \citet{conley2014molecular} and \citet{do2015overview}.
\citet{hyman2015vemurafenib}
describe the BRAF V600 basket trial, after noting that (a) BRAF V600 mutations occur in almost 50\% of cutaneous melanomas and result in constitutive activation of downstream signaling through the MAPK (mitogen-activated protein kinase) pathway, based on previous studies reported by \citet{davies2002mutations} and \citet{curtin2005distinct}; 
(b) Vemurafenib, a selective oral inhibitor of BRAF v600 kinase produced by Roche-Genentech, has been shown to improve survival of patients with BRAF V600E mutation-positive metastic melanoma according to \citet{chapman2011improved};
and (c) efforts by the Cancer Genome Atlas and other initiatives have identified BRAF V600 mutations in non-melanoma cancers \citep{de2010effects, van2011cetuximab,weinstein2013cancer, kris2014using}. 
They point out that ``the large number of tumor types, low frequency of BRAF V600 mutations, and the variety of some of the (non-melanoma) cancers make disease specific studies difficult (unaffordable) to conduct." \citet{hyman2015vemurafenib} therefore use six ``baskets" (NSCLC, ovarian, colorectal, and breast cancers, multiple myeloma, cholangiocarncinoma) plus a seventh (``all-others") basket which ``permitted enrollment of patients with any other BRAF V600 mutation-positive cancer" 
in their Phase II basket trial of Vemurafenib. The Phase II trial uses Simon's two-stage design ``for all tumor-specific cohorts in order to minimize the number of patients treated if vemurafenib was deemed ineffective for a specific tumor type." The primary efficacy endpoint was response rate at week 8. ``Kaplan-Meier methods were used to estimate progression-free and overall survival. No adjustments were made for multiple hypothesis testing that could result in positive findings."

In the BRAF V600 trial, 122 adults received at least one dose of Vemurafenib (20 for NSCLC, 37 for colorectal cancer, 5 for multiple myeloma, 8 for cholangocarcinoma, 18 for ECD or LCH, 34 for breast, ovarian, and ``other" cancers), and 89\% of these patients had at least one previous line of therapy. Vemurafenib showed (a) ``efficacy in BRAF V600
mutation-positive NSCLC” compared to standard
second-line docetexal in molecularly unselected patients,
and (b) for ECD or LCH ``which are closely related orphan
diseases with no approved therapies,” the response rate
was 43\% and none of the patients had disease
progression while receiving therapy, despite a median
treatment duration of 5.9 months.
\citet{hyman2015vemurafenib} point out that ``one challenge in interpreting the results of basket studies
is drawing inferences from small numbers of patients.”
Following up on this point,  \citet{berry2015brave} discusses other challenges for inference from basket trials. In particular, he points out that even though patients have the same biomarker, different tumor sites and
tumor types may have different response rates and simply pooling trial
results across tumor types may mislead interpretation. On the other hand, different tumors may have similar response rates
and hierarchical Bayesian modeling can help borrow information across
these types to compensate for the small sample sizes.
 
 We include here another basket trial led by our former Stanford colleague, Dr. Shivaani Kummar. She collaborated with investigators at Loxo Oncology in South San Francisco, and other investigators at UCLA, USC, Harvard, Cornell, Vanderbilt, MD
Anderson, and Sloan Kettering, to design and conduct a basket
trial involving seven specified cancer types and an eighth basket (``other
cancers”) to evaluate the efficacy and safety of larotretinib, a highly
selective TRK inhibitor produced by Loxo Oncology in South San
Francisco, for adults and children who had TRK fusion-positive cancers.
A total of 55 patients were enrolled into one of three protocols and
treated with larotretinib: a Phase I study involving adults, a Phase I-II
study involving adults and children, and a Phase II study involving
adolescents and adults with TRK fusion-positive tumors. The Phase II
study uses the recommended dose of the drug twice daily. The
dose-escalation Phase I study and the Phase I portion of the Phase I-II
study do not require the subjects to have TRK fusions although the
combined analysis only includes ``patients with prospectively identified
TRK fusions.” 
The primary endpoint for the combined analysis was the overall
response assessed by an independent radiology committee. Secondary
endpoints include duration of response, progression-free survival, and
safety.
At the data-cutoff date 7/17/2017, the overall response rate was 75\%,
and 7 of the patients had complete response while 34 had partial
response; see \citet{drilon2018efficacy}. In the accompanying editorial of
that issue in \textit{NEJM}, \citet{andre2018developing} says that ``this study is an illustration
of what is likely to be the future of drug development in rare genomic
entities” 
and that according to the Magnitude of Clinical Benefit Scale
for single-arm trials recently developed by the European Society of
Medical Oncology, ``studies that show rates of objective response of
more than 60\% and a median progression-free survival of more than 6
months, as the study conducted by Drilon et al. does, are considered to
have the highest magnitude of clinical benefit” 
in line with the pathway
for single-arm trials of treatments of rare diseases with well-established
natural histories to receive approval from regulatory agencies. 
\citet{andre2018developing} also mentions that the study by Drilon et al. 
``did not find any difference in efficacy among the 12 tumor
histotypes (including those in the all-other basket),” 
proving a successful ``trans-tumor approach” in the case of TRK
fusions with larotrectinib, but that ``some basket trials have
not shown evidence of trans-tumor efficacy of targeted
therapies, notably BRAF inhibitors.” 
He points out the
importance of developing ``statistical tools to support a
claim that a drug works across tumor types” and to provide
``a more in-depth understanding of the failure of some
targets in a trans-tumor approach.” 

BioPharma Dive, a company in Washington, D.C. that provides news and analysis of clinical trials, drug discovery, and development, FDA regulations and approvals, for biotech and biopharmaceutical corporations, has a 2019 article sponsored by Paraxel, a global provider of biopharmaceutical services headquarted in Waltham, MA, highlighting that ``in the past five years, we've seen a sharp increase in the number of trials designed with a precision medicine approach," and that ``in 2018 about one of every four trials approved by the FDA was a precision medicine therapy;" see \citep{BPDthatworks}.
Moreover, ``developing these medicines requires changes to traditional clinical trial designs, as well as the use of innovative testing procedures that result in new types of data," and ``the FDA has taken proactive steps to modernize the regulatory framework" that "prioritizes novel clinical trials and real-world data solutions to provide robust evidence of safety and efficacy at early stages." The February 12, 2020, news item of BioPharma Dive is about Merck's positive results for its cancer drug Keytruda, when combined with chemotherapy, in breast cancer patients on whom a certain amount of tumor and immune cells express a protein that make Keytruda truly effective for this difficult-to-treat form of breast cancer called ``triple negative." The news item the following day is that the FDA granted BMS's CAR-T treatment (called liso-cel) of a type of lymphoma priority reviews, setting up a decision by August 17, 2020; see \cite{BPDBMS,BPDMerck}. Liso-cel was originally developed by the biotech company Juno Therapeutics before its acquisition by Celgene in 2018. In Jan 2019, BMS announced its \$74 billion acquisition of Celgene and completed the acquisition in November that year after regulatory approval by all the government agencies required by the merger agreement.

Since 2016, Stanford University has held an annual drug discovery symposium, focusing on precision-guided drug discovery and development. We briefly describe here the work of Brian Kobilka, one of the founding conference organizers and the director of Kobilka Institue of Innovative Drug Discovery (KIDD) at The Chinese University of Hong Kong, Shenzhen, and his former mentor and Nobel Prize co-winner Robert Lefkowitz. In a series of seminal papers from 1981 to 1984 published by Lefkowitz and his postdoctoral fellows at the Howard Hughes Medical Institute and Departments of Medicine and Biochemistry at Duke University, the $\beta_2$-subtypes of the pharmacologically important $\beta$-andrenergic receptor ($\beta$AR) were purified to homogeneity and demonstrated to retain binding activity. Dixon, Sigal, and Strader of Merck Research Laboratories subsequently collaborated with Lefkowitz, Kobilka and others on their team at Duke to derive an amino-acid sequence of peptides which indicated significant amino-acid homology with bovine rhodopsin and were able to find a genomic intronless clone in 1986. In his Dec. 2012 Nobel Lecture
\citep{lefkowitz2013brief},
Lefkowitz highlights the importance of the discovery, saying: "Today we know that GPCRs (G protein coupled receptors), also known as seven transmembrane receptors, represent by far the largest, most versatile and most ubiquitous of the several families of plasma membrane receptors $\ldots$ Moreover, these receptors are the targets for drugs accounting for more than half of all prescription drug sales in the world \citep{pierce2002seven}." Kobilka highlights in his Nobel lecture \citep{kobilka2013structural} his efforts to understand the structural basis of $\beta_2$AR using advances in X-ray crystallography and later in electron microscopy to study the crystal structure of $\beta_2$AR. He concludes his Nobel lecture by saying: ``While the stories outlined in this lecture have advanced the field, much work remains to be done before we can fully understand and pharmacologically control signaling by these fascinating membrane proteins." This work is continued at the Kobilka Institute of Innovative Drug Discovery and by his and other groups at Stanford, Lefkowitz's group at Duke, and other groups in other centers in academia and industry, in North America, Asia, and Europe.

\subsection{Discussion and New Opportunities for Statistical Science}

\citet{woodcock2017master} point out new opportunities for statistical science in the design and analysis of master protocols; ``With multiple
questions to address under a single protocol, usually in an
area of unmet need, and an extensive infrastructure in
place to handle data flow, master protocols are a natural
environment for considering innovative trial designs. The
flexibility to allow promising new therapies to enter and
poor-performing therapies to discontinue usually requires
some form of adaptive design, but the level of complexity of
those adaptations can vary according to the objectives of
the master protocol.”
 They also point out that 
``two
types of innovation are hallmarks of master protocols: the
use of a trial network with infrastructure in place to
streamline trial logistics, improve data quality, and facilitate
data collection and sharing; and the use of a common
protocol that incorporates innovative statistical approaches
to study design and data analysis, enabling a broader set
of objectives to be met more effectively than would be
possible in independent trials”. Recent advances in hidden Markov models and MCMC schemes that we are developing for cryo-EM analysis at Stanford is another example of new opportunities for statistical science in drug discovery. This will be coupled with innovative designs for regulatory submission. It is an exciting interdisciplinary team effort, merging statistical science with other sciences and engineering.

\chapter[Bandit Theory for Learning Heathcare Systems]{Bandit Theory: Applications to Learning Healthcare Systems and Clinical Trials}

\section{Introduction and Background}

In this section we review multi-arm bandit theory with covariate information, also called ``contextual multi-arm bandits," to pave the way for it to have major impact on the future of clinical research, as the medical community grapples with the challenges of generating and applying knowledge at point of care in fulfillment of the concept of the ``learning healthcare system" (LHS) \citep*{chamberlayne1998creating}.
“A learning healthcare system is one that is designed to generate and apply the best evidence for the collaborative healthcare choices of each patient and provider; to drive the process of discovery as a natural outgrowth of patient care; and to ensure innovation, quality, safety, and value in health care” \citep*{olsen2007institute}.
The first branch of Tze Lai’s work discussed below deals with methods for incorporating true experimental strength into efforts to explore the comparative effects of different treatments, while exploiting what is learned to improve outcomes in patients.

\subsection{The Multi-Armed Bandit Problem}
\label{subsec:MAB}

The name ``multi-arm bandit" suggests a row of slot machines, which in the 1930s were nicknamed ``one-armed bandits." (Presumably the name is inspired by their pull-to-play levers and the often large house edge.) For a gambler in an unfamiliar casino, the ``multi-arm bandit problem" would refer to a particular challenge: to maximize the expected winnings over a total of $T$ plays, moving between machines as desired. The distribution of payouts from pulling each arm may be unknown and different for each machine. How should the gambler play? Research into the MAB problem and its variants has led to foundational insights for problems in sequential sampling, sequential decision-making, and reinforcement learning.

Mathematical analysis of the MAB problem has been motivated by medical applications since \cite{thompson1933likelihood}, with different medical treatments playing the role of bandit machines. Subsequent theory has found wide application across disciplines including finance, recommender systems, and telecommunications \citep{bouneffouf2019survey}. According to \cite{whittle1979discussion}, the bandit problem was considered by Allied scientists in World War II, but it ``so sapped [their minds] that the suggestion was made that the problem be dropped over Germany, as the ultimate instrument of intellectual sabotage." It was \cite{lai1985asymptotically} who gave the first tractable asymptotically efficient solution.

Given a set of arms $k \in 1,...K$, Lai and Robbins frame the question: How should we sample $y_1$, $y_2$,... sequentially from the $K$ arms in order to achieve the greatest possible expected value of the sum $S_T = y_1 + ... + y_T$ as $T \to \infty$?
They model each sample from arm $k$ as an independent draw from a population $\Pi_k$ from a family of densities $f_{\theta_k}$ indexed by parameter $\theta_k$. Then, they formalize the space of (possibly random) strategies $\phi \in \Phi$, defining $\phi$ to be an \textit{adaptive allocation rule} if it is a collection of random variables that makes the arm selection at each timestep, $\boldsymbol{\phi} := (\phi_1,\phi_2, ...\phi_T)$. Thus,
each $\phi_t$ is a random variable on $\{ 1,...,K\}$, where the event $\{\phi_t = k\}$ (``arm $k$ is chosen at time $t$") belongs to the $\sigma$-field $F_{n-1}$ generated by prior decisions and observations $(\phi_1, x_1, \phi_2, x_2,...\phi_{t-1},x_{t-1})$. In this framework, \cite{lai1985asymptotically} define the \textit{cumulative regret} of an adaptive allocation, rule which measures the strategy's expected performance against the best arm, equivalent to $$R_T(\boldsymbol{\phi},\boldsymbol{\theta}) := \sum \limits_{t = 1}^{T} \mu^*(\boldsymbol{\theta}) - \mathbb{E}\left[ \mu(\theta_{\phi_t})\right],$$
where 
$\mu(\theta_{k})$ is the expected value of arm $k$, and $\mu^*(\boldsymbol{\theta}) := \max_{k}\{\mu(\theta_k)\}$. \cite{lai1985asymptotically} give a strategy that achieves an expected cumulative regret of order $O(\log T)$, and provide a matching lower bound to show it is nearly optimal. This strategy creates an \textit{upper confidence bound} (UCB) for each arm, where the estimated return is given a bonus for uncertainty. A simple example of a UCB is the UCB1 o \cite{auer2002finite}, which at round $t$, picks the arm maximizing
$$\bar{y}_{k,t} + \sqrt{2 \ln(t) / n_{k,t}},$$
where the rewards $y_t$ are in $[0,1]$, $\bar{y}_{k,t}$ is the average of the observed rewards from arm $k$, and $n_{k,t}$ is the number of samples observed from arm $k$. Typically, UCBs are designed so that inferior arm(s) are discarded with minimal investment, and the best arm(s) are guaranteed to remain in play; a key contribution of \cite{lai1985asymptotically} was to show how such statements can be quantified using Chernoff bounds (or other concentration inequality arguments), and then converted into an upper bound on the cumulative regret. Their approach has been generalized and extended to yield algorithms and regret guarantees across a variety of applications, with UCBs acting as a guiding design principle.

The richness of the bandit problem has generated a multitude of other approaches. By adding to the above model a prior distribution for the arm parameters $\boldsymbol{\theta}$, the bandit problem can be framed as a Bayesian optimization over $\boldsymbol{\phi}$ to find the allocation strategy that minimizes the expected regret $\int R_T(\boldsymbol{\theta},\boldsymbol{\phi}) dp\boldsymbol{\theta}$. This optimization can, in principle, be solved with dynamic programming (as in \cite{cheng2007optimal}); however, dynamic programming does not scale well to large or complicated experiments, because the number of possible states explodes. Using results from \cite{whittle1980multi}, \cite*{villar2015multi} show how the computation can be reduced considerably by framing the optimal solution as an index policy.

When solving for the optimal strategy is not feasible, the heuristic solution of Thompson sampling is a popular choice, with good practical and theoretical performance  \citep*{chapelle2011empirical,kaufmann2012thompson,russo2016information}. 
The decision rule proposed by \cite{thompson1933likelihood} is an adaptive allocation rule, where $\phi_{t}$, given all data observed prior to time $t$, is nondeterministic and chooses arm $k$ with probability equal to its posterior chance of being the best arm. That is, $\phi_t = k$ with probability 
$p_{k,t} := P_{\mathbb{F}_t}\left\{k^* = k \right\}$, where 
$P_{\mathbb{F}_t}$ is the posterior probability distribution given $(\phi_1,x_1,...\phi_{t-1},x_{t-1})$, and 
$k^*:= \argmax_k (\mu(\theta_k))$ is the index of the best arm (which is a random variable). If the best arm is not unique, the tie should be broken to ensure the uniqueness of $k^*$. In fact, a Thompson allocation can be performed with just one sample from the posterior $\mathbb{F}_t$, as shown in the following workflow:

\medskip

\SetAlgoLined
\LinesNumbered
\begin{algorithm}[H]
 \caption{Bayesian Workflow with Thompson Sampling}
 
Assume a likelihood model parametrized by $\boldsymbol{\theta}$, such that $\boldsymbol{\theta}$ determines the arm means by $\boldsymbol{\mu}(\boldsymbol{\theta}) = (\mu_1(\boldsymbol{\theta}),\ldots,\mu_k(\boldsymbol{\theta}))$\;

Assume a prior $\mathbb{F}_{1}$\;

\For{each sample $t \in \{1, \ldots,T\}$}{
   Draw from the posterior a sample of the vector of arm means $sample$ $\boldsymbol{\theta'}  \sim \mathbb{F}_{t}$\ ; $set$ $\boldsymbol{\mu '} := (\mu_1(\boldsymbol{\theta'}),\ldots,\mu_k(\boldsymbol{\theta'}))$;
   
   Allocate to the arm corresponding to the largest entry of $\boldsymbol{\mu'}$: \newline
   $set$ $\phi_t := \arg\!\max _k\{\mu_k'\}$ (breaking ties at random)\;
   Receive from arm $\phi_t$ the next payoff $x_t$\;
   Given the new observation, update posterior to $\mathbb{F}_{t+1}$
 }
\end{algorithm}

Exact sampling from the posterior is not always tractable. A popular technique for sampling the posterior approximately is the Markov Chain Monte Carlo (MCMC) method. The convergence properties of MCMC to the posterior distribution, and in particular the number of steps that must be run to achieve accurate sampling, are well understood only in special cases \citep*{diaconis2009markov,dwivedi2018log}. Where theory falls short, practitioners may appeal to a variety of diagnostics tools to provide evidence of convergence to the posterior \citep{roy2020convergence}.

There are many other approaches to the bandit problem, including epsilon-greedy \citep*{sutton1998introduction}, knowledge gradient \citep*{ryzhov2012knowledge}, and information-directed sampling (\cite{russo2014learning}).

\subsection{Contextual MABs and Personalized Medicine}

For an LHS that continuously seeks to improve and personalize treatment, the important question is not \textit{which} treatment is best, but \textit{for whom} each treatment is best. To address this question, one must augment the bandit model with information about each patient. Calling this side information ``covariates" or ``\textit{contexts}," one arrives at the CMAB problem.

CMABs have found great success in the internet domain for problems such as serving ads, presenting search results, and testing website features. In contrast, applications in medicine have lagged (with the prominent exception of mobile health \citep*{greenewald2017action,xia2018price}).
The design of trials in an LHS brings new challenges to the CMAB framework, such as ethical requirements, small sample sizes (roughly $10^2$ \textendash $10^4$ patients, in comparison to $10^4$ \textendash $10^9$ clicks for internet applications), requirements for medical professionals to inspect and understand processes, feedback times, and demand for generalizable conclusions.
In section 3.4 we return to this topic. Section 3.2 considers adaptive randomization in an LHS. Section 3.3 discusses inference for MABs in an LHS.

\section{Adaptive Randomization in an LHS}

In an LHS, the arms of an MAB are treatments and the rewards are patient outcomes. Thus, minimizing the cumulative regret corresponds to maximizing patients' measured quality of care, a primary function of the LHS. However, typically, there is a secondary goal of learning from a trial: useful takeaways may include confidence intervals for the treatment effects, developing a treatment guide, or making recommendations for non-participating patients in parallel with the trial.

The goals of regret minimization and knowledge generation, often framed as ``exploitation vs. exploration," are indeed in fundamental conflict:
\cite*{bubeck2011pure} formalized a notion of exploration-based experiments, where recommendations are made outside the trial. They define the \textit{simple regret} to be $$r_T = \mu^* - \mu_{c,T},$$ where $\mu_{c,T}$ is the expectation of the recommended arm after round $T$, and $\mu^*$ is the expectation of the best arm. Bubeck, Munos, and Stoltz show that upper bounds on the cumulative regret $R_T$ lead to lower bounds on $r_T$, and vice versa. In this sense, algorithms that minimize the cumulative regret occupy an extreme point of a design space: they maximize the welfare of trial patients, but sacrifice knowledge about inferior treatments. At the other extreme point of the design space, an ideal trial for knowledge generation, with two arms of equal variance, will split the sample sizes equally, consigning half of the patients to the inferior treatment.

Most practical implementations of adaptive randomization in clinical trials use modified bandit algorithms. A common prescription is to lead with a first phase of equal randomization. Or, allocation probabilities may be shrunk toward $1/K$ in some fashion. \cite{wathen2017simulation} discuss the design options of restricting allocations to [.1, .9], leading with a period of equal randomization to prevent the algorithm from ``getting stuck" on a worse arm, and altering the Thompson sampling to allocate with probability proportional to $p_{k,t}^{c}$, for $c \in (0,1]$. \cite*{villar2015multi} consider forced sampling of the control arm every $1/K$ patients. \cite{kasy2019adaptive} modify the Thompson sampling to tamp down selection of the best arm(s), asymptotically leading to equal randomization between the best candidates. 
\cite*{lai2013group} give a design that maintains a preferred set of arms, randomizing equally between them, and adaptively drops arms from this set at interim analyses. These various design choices and algorithmic tweaks are typically investigated and tuned by simulation. Even without explicit modification to the standard bandit approach, most medical applications will have a \textit{delay} between the treatment assignment and the observation of an outcome; the resulting reduction in available information leads to more exploration for most algorithms.

There are many benefits to using nearer-to-equal randomization probabilities. First, balancing sample sizes between a pair of arms serves inference goals such as increased power of hypothesis tests, shorter confidence intervals, and more accurate future recommendations. Second, closer-to-equal randomization may improve the information for interim decisions such as early stopping and sample size re-estimation. Third, without tuning, there may be an unacceptably high chance of sending a majority of patients to the wrong arm \citep*{thall2015statistical}. Fourth, more equal randomization can help detect violated assumptions, such as time trends or a model misspecification. Fifth, the possibility of violated assumptions suggests treating data as slightly less informative. Finally, probabilities nearer $1/2$ are helpful for inverse-probability weighting and randomization tests. 

On the other hand, when a treatment is strongly disfavored for a patient, ethical health care requires setting its randomization chance to zero. This may be achieved by thresholding allocation probabilities according to some rule, or suspending or dropping treatment arms at interim analyses. Furthermore, more equal randomization comes at an opportunity cost to the welfare of trial participants. Practical trial design in an LHS must seek a balance between these competing objectives of knowledge generation and participant welfare.

\subsection{Inference for MABs in an LHS}
\label{sec:MABinference}

The LHS may desire several forms of knowledge from an adaptive randomization trial, including confidence intervals for the outcomes of arms (and their differences), guarantees about selecting arms correctly, and recommendations for treatments in non-participating patients.

Frequentist inference under adaptive randomization designs can be challenging. Owing to adaptive sampling, the distribution of standard estimates for the mean of an arm is typically nonGaussian, and not pivotal with respect to the treatment effect. Concentration techniques for UCBs, such as Chernoff bounds, can be applied for confidence bounds that may hold uniformly over possible stopping times \citep*{jamieson2014best,zhao2016adaptive, karnin2013almost}. The concentration approach has been extended to FDR control with the always-valid p-values framework \citep*{johari2015always,yang2017framework}. Furthermore, self-normalization techniques from de la \cite*{pena2008self} permit extensions to large classes of distributions. However, confidence intervals from concentration bounds may be conservative, slack by a constant or logarithmic factor of width.

In confirmatory trial design, adaptivity may be managed by dividing the trial into segments, each having constant randomization probabilities so that Gaussian theory can be used (with numerical integration for stopping boundaries to compute the type-I error and power at fixed alternatives). 
\cite*{lai2013group} and \cite{shih2013sequential} show how to do this for their MAB-inspired designs.
Alternatively, \cite{korn2011outcome} suggest block-randomization and block-stratified analysis. Compared to the constantly changing allocation strategies of the standard bandit algorithms, discretization of strategy can come at a moderate or minimal cost, depending on the design and goals.

For analyzing MAB designs with a constantly updating allocation strategy, a key idea for constructing valid frequentist p-values is the randomization test. The randomization test assumes the sharp null hypothesis that the treatment has exactly zero effect, and relies on probabilistic randomization in the allocation algorithm to generate power. In exchange, with other minimal assumptions, it grants valid p-values, even in the presence of time trends and other confounders in the patient population \citep{simon2011using}. To form confidence intervals, a sharp additive model for the treatment effect may be considered. Confidence bounds then follow by inverting the randomization test, as in \cite*{ernst2004permutation}.

Another tool for constructing confidence intervals is hybrid resampling, by \cite{lai2006confidence}. This procedure considers families of different shifts and scales of the observed data, and simulates via resampling to infer which distributions are consistent with the observed treatment effects. Lai and Li show that for group sequential trials, confidence intervals from hybrid resampling can have more accurate coverage than that of standard normal approximations.

\cite*{hadad2019confidence} suggest a double-robust estimation approach. In addition to using an augmented inverse-probability weight (AIPW) model, they propose further adaptively re-weighting the data to force the treatment effect estimate into an asymptotically Gaussian distribution. Double-robust estimation may help to correct for time trends or other confounding. However, data re-weighting comes at a cost to efficiency, as pointed out by \cite{tsiatis2003inefficiency}.

Finally, if one assumes a prior and enters the Bayesian framework, posterior inference is a highly flexible approach to analysis. Because Bayes' rule decouples the experimenter's allocation decisions from the rest of the likelihood, the standard Bayesian workflow can be applied to the data without concern for the adaptivity of the design \citep{berger1988likelihood}. Subject to typical caveats on prior selection and accurate posterior sampling, posterior inference can yield Bayes factors for testing, credible intervals for treatment effects, and decision analysis for treatment recommendations.

\subsection{Linear And More General Models for the Reward in Personalized Treatments in LHS}

We now return to the contextual MABs (CMABs) for the reward in personalized treatments in LHS introduced in Section 3.1.2. First, we focus on a correctly specified linear model in Section 3.4.1. 
This assumption derives some justification from the features of an LHS: assuming that covariates are continuous and low dimensional, the patient population of greatest interest is expected to occupy a small region of the covariate domain, owing to the systematic filtering of equipoise requirements and further shrinking of the population under experimental focus as ``exploiting" increases. Additionally, the conditional expectation of the response is typically a smooth function of the covariates. Therefore, assuming both smoothness of conditional expectation and locality of the studied population, Taylor's theorem implies approximate correctness of the linear model. Similar arguments can be applied to logistic models and other smooth model classes.

\subsubsection{Linear Models for the Reward}
\label{subsec:lm}

If at step $t$ we observe a context vector $x_{t}$ of length $d$,  sample from arm $\phi_t = k$, and receive reward $y_t$, we may consider the following simple linear model for the expected reward:
$$E[y_t | x_t,\phi_t = k] = x_{t} ^T \theta^*_k,$$ \label{linsimple}
where $\theta_k^*$ is an unknown parameter vector of length $d$. The LinUCB algorithm of \cite{li2010contextual} brings the UCB of \cite{lai1985asymptotically} to this linear model. Assuming the linear model parameters are not shared between arms and that contexts do not depend on the arm chosen (see \cite{li2010contextual} for the general case,) they suggest estimating $\theta_k^*$ for each arm using a ridge regression $\hat{\theta}_k$. That is, if $X_{k,t}$ is a design matrix whose rows are the contexts of the individuals previously assigned to arm $k$ before time $t$ and $Y_{k,t}$ is a vector of their rewards, the ridge estimator with tuning parameter $\lambda$ is
$$\hat{\theta}_{k,t} = (X_{k,t}^TX_{k,t} + \lambda I_d)^{-1}X_{k,t}^TY_{k,t}.$$

Next, \cite{li2010contextual} construct a UCB for the expected reward around the ridge regression prediction, suggesting the confidence interval

$$|x_{t}^T \hat{\theta}_{k,t} - x_t ^T \theta^*_{k}| \leq \alpha \sqrt{x_{t}^T(X_{k,t}^TX_{k,t} + \lambda I_d)^{-1}x_{t}}$$
where $\lambda$ is set to one and $\alpha$ is a tuning parameter. This confidence interval implicitly assumes a correctly specified linear model and independence of $Y_{k,t}$ given $X^T_{k,t}$, an assumption which is typically broken by the allocation mechanism unless ($x_t,y_t)$ is independent and identically distributed (i.i.d.) for all $t$. Nevertheless, analogously to the basic UCB algorithm, they propose the LinUCB algorithm, which chooses the arm with the highest UCB,

$$\phi_{t}^{UCB} := \argmax_k \left\{x_{t}^T \hat{\theta}_{k,t} + \alpha \sqrt{x_{t}^T(X_{k,t}^TX_{k,t} + \lambda I_d)^{-1}x_{t}} \right\}$$.

LinUCB is easy to implement and has proven popular in applications, inspiring further improvements and competitors. \cite{chu2011contextual} analyze a theoretical fix to LinUCB and give a regret analysis for a modified algorithm of order $O\left( \sqrt{Td \ln^3(KT \ln(T)/\delta)}\right)$. They also give a  nearly-matching general lower bound for the problem of order $\Omega \left(\sqrt{KT}\right)$.

Alternatively, \cite{Abbasi-Yadkori2011}, working within a more general framework called ``linear bandits" or ``linear stochastic bandits," construct self-normalized confidence sets for the arm parameters. In the linear bandit, rather than choosing among a discrete set of arms, one chooses the context $x_{t}$ from a set $D_t$, and the rewards are modeled as $y_t = x_{t}^T\theta^* + \eta_t$. Note that model \eqref{linsimple} can be embedded within the linear bandit by sufficiently increasing the dimensions of $x_t$ and $\theta^*$ and taking $D_t$ as an appropriate finite set of $K$ vectors. \cite{Abbasi-Yadkori2011} assume that, conditioned on data prior to time $t$, $\eta_t$ is mean-zero and $R$-sub-Gaussian for some $R \geq 0$. Further, it is assumed that $\| \theta^* \|_2 \leq S$, for some $S \geq 0$. Then, defining $X_t$ as a $(t-1) \times d$ matrix whose rows consist of the contexts $x_{s}^T$, for $s = 1,\ldots,t-1$, defining the reward vector $Y_t$ as a vector of length $(t-1)$ of the corresponding rewards $y_s$, for $s = 1,\ldots,t-1$, and denoting $\bar{V}_t :=\lambda I_d + X_t^T X_t$, for all $t \geq 1$, one may write the ridge estimator as $$\hat{\theta}_t := \bar{V}_t^{-1}X_t^T Y_t.$$ \cite{Abbasi-Yadkori2011} then derive the confidence set

$$C_t := \left\{\| \hat{\theta}_t - \theta^* \|_{\bar{V}_t} \leq R \sqrt{2 \log \left( \frac{ \det(\bar{V}_t)^{1/2} \det(\lambda I_d)^{-1/2}}{\delta}\right)} + \lambda^{1/2}S \right\}$$
where $\| \cdot \|_{\bar{V}_t}$ is a matrix weighted 2-norm. The collection of these sets, $\mathbb{C} := \bigcap \limits_{t \geq 1} C_t$, provides $1-\delta$ uniform confidence that $\theta^* \in \mathbb{C}$, regardless of an adaptive mechanism for the context choice. \cite{Abbasi-Yadkori2011} leverage this confidence approach into a strategy that generalizes the UCB. They follow the underlying principle of ``optimism in the face of uncertainty" to select the context $$x_t := \argmax \limits_{x \in D_t} \max \limits_{\theta \in C_t} x^T\theta ,$$
and prove regret guarantees for the linear bandit with this algorithm. For a $K$-arm trial designer, a key takeaway is that uniform confidence sets offer an approach to model inference (noting that practical use requires strong modeling assumptions, a choice of $\lambda$, and bounds for the unknown parameters $R$ and $S$).

A different approach to the CMAB problem is to generalize the $\epsilon$-greedy algorithm: periodic exploration can be used to estimate a model, and to verify that estimates based on adaptive data collection are not far off. Under the simple linear model \eqref{linsimple}, \cite{goldenshluger2013linear} propose maintaining two sets of linear model estimates: $\hat{\theta}^*_k$, estimated on a small amount of equal-randomized data, and $\tilde{\theta}^*_k$, based on all of the (adaptively allocated) data. If the estimated rewards from equal randomization $x^T_t \hat{\theta}^*_k$ are well separated, the arm with the largest estimate is chosen. Else, the arm with the largest value of $x^T_t \tilde{\theta}^*_k$ is chosen. Under strong assumptions including $K=2$ arms, i.i.d samples, and a
\textit{margin condition} that ensures that the decision boundary between the arms is sharp, that is,
$$\mathbb{P}\left\{|(\theta^*_1 - \theta^*_2)^TX_t|\leq \rho \right\} \leq L\rho, \forall \rho \in (0,\rho_0],$$
they derive a cumulative regret bounded by $O(d^3 \log T)$. \cite{bastani2019online} improve these bounds and extend this approach to high-dimensional sparse linear models using $L^1$ penalization. \cite*{bastani2020mostly} also show that under certain conditions, a pure greedy approach can yield rate-optimal regret.

\subsection{More General Models for the Reward}

The Bayesian workflow for the MAB naturally extends to linear models and beyond. \cite{russo2014post} show that for several classes of well-specified Bayesian problems with contexts, Thompson sampling achieves near-optimal performance and behaves like a problem-adaptive UCB. A variety of competitive risk bounds have been proven for Thompson sampling \citep*{agrawal2012analysis,agrawal2013thompson,kaufmann2012thompson,korda2013thompson}. In empirical studies, Thompson sampling often outperforms competitors by a small margin \citep{scott2010modern,chapelle2011empirical,dimakopoulou2017estimation}.

An alternative for the nonBayesian is what we call ``pseudo-Thompson bootstrapping." Given a black box algorithm that models the outcomes under each arm, the idea is to bootstrap-resample data to generate variation in the model's estimates. Pretending that this resampling distribution is a posterior, one can drop the estimated ``probabilities" of arm superiority into the Thompson rule and hope to recover its performance advantages. While this technique approximates Thompson sampling for some known cases \citep{eckles2014thompson}, its general theoretical properties remain unclear. The main appeal of the approach is to offer a wrapper for popular estimation algorithms for large data sets, including regression trees, random forests, and neural networks \citep{elmachtoub2017practical,osband2016deep}). 

\cite{vaswani2019old} propose the RandUCB algorithm, which gives LinUCB nondeterministic allocation probabilities by perturbing the confidence bound randomly in a way that somewhat resembles bootstrapping. For the linear model, RandUCB can be viewed as a generalization of Thompson sampling under a Gaussian model. Vaswani et al. also prove competitive regret guarantees for RandUCB.

Finally, there are nonparametric methods that leverage the smoothness of the expected response. \cite{rigollet2010nonparametric} discretize space into buckets, and run MABs on each of them independently. \cite{lu2010contextual} give a contextual bandit that clusters data adaptively and provides guarantees under Lipschitz assumptions. \cite{lai2020} perform a local linear regression and pair it with $\epsilon$-greedy randomization and arm elimination, meeting minimax lower bounds on regret under certain regularity conditions, which will be discussed further in the next chapter where we provide new advances in CMABs, further discussion and references.

\chapter[Bandit Processes with Non-Denumerable Arms]{Bandit and Covariate Processes, with Finite and Non-Denumerable Set of Arms}

\section{Introduction}

In this chapter we describe the work of \cite{KimLaiXu} in greater detail and discuss several major ideas there which we extend in this chapter to provide for-reaching generalization of the CMAB problem and obtain definitive solutions that are remarkably simple and nonparametric. Our investigation was inspired by a seminal paper of Larry Shepp and his coauthors in 1996, who considered a non-denumerable set of arms for the bandit process; see \cite{berry1997}. Earlier, \cite{yakowitz1991nonparametric}, \cite{yakowtiz1992theory}, and \cite{yakowitz1995nonparametric} also considered nonparametric CMABs in the setting of a non-denumerable set of Markov decision processes. Section 4.3 not only unifies these approaches but also provides a definitive asymptotic theory. Key to this theory is section 4.2 on the ``transformational" insights of the aforementioned work of \cite{KimLaiXu}.

\section{From Index Policies in K-Armed Bandits to arm randomization and elimination rules for CMABs}

Kim, Lai and Xu \cite{KimLaiXu} have recently developed a definitive nonparametric $k$-armed contextual bandit theory for discrete time $T = \{1, 2, \dots, n\}$. We now extend the theory to the general framework of $\{(y_t, \phi_t, \bm{x}_t: t \in T\}$, in which 
$T = \{1, 2, \dots, n\}$ or $(0, n]$, $y_t$ is the bandit process, $\phi_t$ is the indicator of the arm selected to generate $y_t$ and $\bm{x}_t$ is the covariate process such that $\phi_t$ and $\bm{x}_t$ are 
$\F_t$-measurable; $\phi_t$ is assumed to be \textit{c\`adl\`ag} for the case $T = (0, n]$. 
There are three key ingredients in this nonparametric $k$-armed contextual bandit theory, which we consider in the next three subsections.

\subsection{Lower bound of the regret over a covariate set}

As in \cite{KimLaiXu}, the covariate vectors $\bm{x}_t$ are assumed to be stationary with a common distribution $H$ so that letting $\mu_j(\bm{x}) = \E(y_t | \phi_t = j, \bm{x}_t = \bm{x})$
and $\mu^*(\bm{x}) = \max_{1 \leqslant j \leqslant k} \mu_j(\bm{x})$, the regret of an adaptive allocation rule $\phi = (\phi_t : t \in T)$ over $B \subset \text{supp}H$ can be expressed as
\begin{equation}\label{5}\tag{4.1}
    R_{n, \phi}(B) = \sum_{j=1}^k \int_B (\mu^*(\bm{x}) - \mu_j(\bm{x})) E \tau_n (j, \bm{x}) dH(\bm{x})
\end{equation}
where $\E \tau_n(j, \bm{x})$ is the Radon-Nikodym derivative of the measure $\E \tau_n(j, \cdot)$ with respect to $H$. An adaptive allocation rule is called ``uniformly good" over $B \subset \text{supp}H$ if
\begin{equation}\label{6}\tag{4.2}
    R_n(\bm{\theta}, B) = o(n^a) \ \ \ \ \ \ \text{for every } a > 0 \text{ and } \bm{\theta} \in \Theta^k.
\end{equation}
in analogy with classical (context-free) multi-armed bandit theory reviewed above. Under certain regularity conditions on the nonparametric family $\mathcal{P}$ generating the data, it is shown in Supplement S1 of \cite{KimLaiXu} that $\mathcal{P}$ consists of a least favorable parametric subfamily (a cubic spline with even spacing between knots of the order $n^{-1/5}$ for univariate covariates and tensor products of these univariate splines for multivariate covariates) such that the regret over $B$ that contains leading arm transitions has lower bound of the order of $(\log n)^2$.

\subsection{Epsilon-Greedy Randomization in lieu of UCB or Index Policy}

The UCB rule in \cite{lai1987}, based on the upper confidence bound 
\begin{equation}\label{3}\tag{4.3}
        U_{j, t} = \inf \Big\{\theta: \theta \geqslant \hat{\theta}_{j, t} \text{ and } I\Big(\hat{\theta}_{j, t}, \theta \Big) \geqslant t^{-1} g(t/n) \Big\},
\end{equation}
($\inf \emptyset := \infty$,) in which $I(\lambda,\theta)$ is the Kullback-Leibler information number, and $\hat{\theta}_{j,n}$ is the MLE of $\theta_j$ up to stage $n$, to approximate the index policy of \cite{gittins1979bandit} and \cite{whittle1980multi} in classical (context-free) parametric multi-armed bandits, basically samples from an inferior arm until the sample size from it reaches a threshold defined by \eqref{3} involving the Kullback–Leibler information number. For contextual bandits, an arm that is inferior at $\bm{x}$ may be best at another $\bm{x}'$. Hence the index policy that samples at stage $t$ from the arm with the largest upper confidence bound (which modifies the sample mean reward by incorporating its sampling variability at $\bm{x}_t$) can be improved by deferral to future time $t'$ when it becomes the leading arm (based on the sample mean reward up to time $t'$). Instead of the UCB rule, Kim, Lai and Xu \cite{KimLaiXu} use the $\epsilon$-greedy algorithm in reinforcement learning \cite{sutton2018reinforcement} for nonparametric contextual bandits as follows. Let $K_t$ denote the set of arms to be sampled from and
\begin{equation}\label{7}\tag{4.4}
    J_t = \Big\{j \in K_t : \Big| \hat{\mu}_{j, t-1} \big(\bm{x}_t\big) - \hat{\mu}^*_{t-1}\big(\bm{x}_t\big)\Big| \leqslant \delta_t \Big\},
\end{equation}
where $\hat{\mu}_{j, s}(\cdot)$ is the regression estimate of $\mu_j(\cdot)$ based on observations up to time $s$, $\hat{\mu}^*_s (\cdot) = \max_{j \in K_s} \hat{\mu}_{j, s}(\cdot)$, and $\delta_t$ is used to lump treatments with effect sizes close to that of the apparent leader into a single set $J_t$. At time $t$, choose arms randomly with probabilities $\pi_{j,t} = \epsilon / |K_t \setminus J_t|$ for $j \in K_t \setminus J_t$ and $\pi_{j,t} = (1 - \epsilon)/|J_t|$ for $j \in J_t$, where $|A|$ denotes the cardinality of a finite set $A$. The set $K_t$ is related to the arm elimination scheme described in the next subsection. The estimate $\hat{\mu}_{j, s}(\cdot)$ uses local linear regression with bandwidth of the order $n^{-1/5}$ that has been shown by Fan \cite{fan1993} to have minimax risk rate for univariate covariates and by Ruppert and Wand \cite{ruppert1994} for multivariate covariates.

\subsection{Arm Elimination via Welch's Test}

First note that \eqref{7} lumps treatments whose effect sizes are close to that of the apparent leader into a single set $J_t$ of leading arms $j \in J_t$. Such lumping is particularly important when the covariates are near leading arm transitions at which a leading arm can transition to an inferior one due to transitions in the covariate values. The choice $\delta_t^2 = (2\log t) / t$ in \cite{KimLaiXu} is especially effective in the vicinity of leading arm transitions, as will be explained in the next paragraph. Hence the transition does not change its status as a member of the set of leading arms so that the $\epsilon$-greedy randomization algorithm still chooses it with probability $(1 - \epsilon) / |J_t|$.

We next describe the arm elimination criterion of \cite{KimLaiXu}. Choose $n_i \sim a^i$, for some integer $a > 1$. For $n_{i-1} < t \leqslant n_i$, eliminate the surviving arm $j$ if 
\begin{equation}\label{8}\tag{4.5}
    \hat{\mu}_{j, t-1} (\bm{x}_t) < \hat{\mu}^*_{t-1}(\bm{x}_t) \text{ and } \Delta_{j, t-1} > g(n_{j, t-1} / n_i),
\end{equation}
where $n_{j, s} = T_s(j)$, $g$ is given in \eqref{3}, and $\Delta_{j, t-1}$ is the square of the Welch $t$-statistic based on $\{(\bm{x}_\ell, y_\ell) : 1 \leqslant \ell \leqslant t - 1\}$; that is, 
\begin{equation}\label{9}\tag{4.6}
    \Delta_{j, t-1} = \sum_{\ell = 1}^{t-1} I_{\{\phi_\ell = j\}} \frac{\Big(\hat{\mu}_{j, \ell-1} (\bm{x}_\ell) - \tilde{\mu}_{j, \ell-1} (\bm{x}_\ell)\Big)_+^2}{\Big(y_\ell - \hat{\mu}_{j, \ell-1}(\bm{x}_\ell)\Big)^2 + \Big(y_\ell - \tilde{\mu}_{j, \ell-1}(\bm{x}_\ell)\Big)^2}\ ,
\end{equation}
where $a_+ = \max(a, 0)$ and $\tilde{\mu}_{j, s}(\cdot) = \max_{j' \in K_s} \hat{\mu}_{j'}(\cdot)$ if $j \in K_s \setminus J_s$, which corresponds to the local linear regression estimate of $\mu_j(\cdot)$ under the null hypothesis $H_{j, s}$ that $\mu_j(\bm{x}_s)$ is not significantly below $\max_{i \in K_s} \mu_i(\bm{x}_s)$ if $j \in J_s$. The Welch $t$-statistics are self-normalized statistics, for which exponential bounds have been established; see Section 15.1 of \cite{pena2008self}. The transition of a leading arm to an inferior one due to transitions in the covariate value does not change the status as a member of the ``lumped set" of leading arms under the regularity conditions assumed by \cite{KimLaiXu}:

\begin{enumerate}[(C1)]
    \item The common distribution $H$ of the i.i.d. covariate vectors $\bm{x}_t$ has a positive density function $f$ (with respective to Lebesgue measure) which is continuously differentiable on a hyperrectangle in $\mathbb{R}^p$.
    \item The regression function $m(\bm{x}) := \E(Y_t | \bm{x}_t = \bm{x})$ is twice continuously differentiable and $\sigma^2(\bm{x}) := \text{Var}\big(Y_t|\bm{x}_t = \bm{x}\big)$ is positive and continuous on $\text{supp}H$ (i.e., the hyperrectangle in (a)).
    \item The kernel $\Psi$ used to define local linear regression estimate of $m(\bm{x})$ is bounded, continuous and $\int|\bm{u}|^r \Psi(\bm{u}) \text{d}\bm{u} < \infty$ for all $r \geqslant 1$, $\int u_i K(\bm{u}) \text{d}\bm{u} = 0$ for $i = 1, \dots, p$.
\end{enumerate}
For univariate covariates, \cite{fan1993} has shown that
\begin{equation*}
    \hat{m}(x) = \sum_{\ell = 1}^n w_\ell(x) y_\ell \bigg/ \sum_{\ell = 1}^n \bigg(w_\ell(x) + n^{-2} \bigg),
\end{equation*}
with
\begin{equation*}
        w_\ell (x) = \Psi \big((x - x_\ell\big) \big/ b_n\big) \big\{ s_{n, 2} - (x - x_\ell)s_{n, 1}\big\},\ s_{n, j} = \sum_{\ell = 1}^n \Psi \big((x-x_\ell) \big/ b_n\big) \big(x - x_\ell\big)^j
\end{equation*}
for $j = 0, 1, 2$, and $b_n \approx n^{-1/5}$, under conditions somewhat stronger than (C1), (C2), and (C3), and \cite{ruppert1994} have extended his result to multivariate covariates.

The adaptive allocation rule that uses the preceding local linear regression estimates in conjunction with the arm elimination rule defined by \eqref{8} and \eqref{9} and the $\epsilon$-greedy randomization algorithm of the preceding subsection is denoted by $\phi_{opt}$ in \cite{KimLaiXu},
where it is shown under (C1) -- (C3) that $\phi_{opt}$ attains the asymptotic minimax rate (as $n \rightarrow \infty)$ for the regret of uniformly good adaptive allocation rules by the following argument. First the sample size of the local linear regression estimate $(\hat{\mu}_{j, t-1}(\cdot) - \tilde{\mu}_{j, t-1}(\cdot))_+$ for $n_{i-1} < t \leqslant n_i$ and $j \in K_t$ is of the order $n_i^{4/5}$ if the selected bandwidth has order $n_i^{-1/5}$. Next consider the parametric model of a cubic spline with evenly spaced knots (with the bandwidth as the spacing) in the univariate case and tensor product of these univariate splines for multivariate covariates, and with the true density function of $\epsilon_t := \big(y_t - m(\bm{x}_t)\big) \big/ \sigma(\bm{x}_t)$. This parametric subfamily is shown to be least favorable and has minimax risk of order $n_i^{4/5}$. It is also shown in \cite{fan1993}, \cite{ruppert1994}, and \cite{KimLaiXu} that the local linear estimator has minimax risk of the order $n_i^{4/5 + o(1)}$, hence the term ``asymptotic minimax rate". Moreover, \cite{KimLaiXu} considers minimax theory of the statistical decision problem, with the risk function of an adaptive allocation rule $\phi$ over a set $B$ of covariate values defined by the regret \eqref{5}. For the least favorable parametric subfamily, the minimax risk is of the order $(\log n)^2$ and is attained by $B$ that contains leading arm transitions and the parametric contextual bandit rule in Section 1.3 of \cite{KimLaiXu}. For nonparametric contextual bandits that need to estimate $\mu_j(\cdot)$ nonparametrically, the minimax risk of $\phi_{opt}$ is of order $(\log n)^{2 + o(1)}$, hence $\phi_{opt}$ attains the asymptotic rate of the risk of adaptive allocation rules.

For continuous-time processes with index set $T = (0, n]$, \cite{bosq1997nonparametric} has developed a corresponding minimax theory for local linear regression estimators
\[ \hat{m}_s(\bm{x}) = \int_0^s y_t \Psi\Big( \big(\bm{x}-\bm{x}_t\big) \big/ b_t\Big) dt \bigg/ \int_0^s \Psi\Big( \big(\bm{x}-\bm{x}_t\big) \big/ b_t\Big) dt\]
of $m(\bm{x}) = \E\big(y_t|\bm{x}_t = \bm{x}\big)$, under smoothness conditions similar to (C2) and (C3). Hence $\phi_{opt}$, with this modification, which is continuous, for continuous-time processes and with the Welch $t$-statistic for arm $j$ defined by
\[ \Delta_{j, s} = \int_0^s I_{\{\phi_t = j\}} \frac{\Big(\hat{\mu}_{j, t}(\bm{x}_t) - \tilde{\mu}_{j, t}(\bm{x}_t)\Big)^2_+}{\Big(y_t - \hat{\mu}_{j, t}(\bm{x}_t)\Big)^2 + \Big(y_t - \tilde{\mu}_{j, t}(\bm{x}_t)\Big)^2}dt,\]
still attains the asymptotic rate of the risk of adaptive allocation rules. Note that the $\epsilon$-greedy algorithm together with \eqref{7} of the preceding subsection and the arm elimination scheme based on the Welch $t$-statistic above for diffusion processes avoids the technical difficulties of defining index policies in \cite{karatzas1984} and \cite{kaspi1998}. Although these references do not consider accompanying covariate processes, the $\epsilon$-greedy algorithm and Welch $t$-test are applicable to context-free multi-armed bandits that \cite{karatzas1984} and \cite{kaspi1998} consider.

\section{Multi-Arm Contextual Bandits, with non-denumerable set of arms}

Mallows and Robbins \cite{mallows1964some} were the first to extend context-free multi-armed bandits from a finite to countably infinite set of arms. They extended the method in the first paragraph of Section 2, which is often called a ``forcing scheme" as it involves a designated set of sparse times for forced sampling to boost the sample size from each arm, to achieve $\lim_{n \rightarrow \infty} n^{-1} \E s_n = \sup_{k \geqslant 1} \mu_k$, assuming certain regularity conditions and using the same notation as that in the first paragraph of Section 2. Yakowitz and Lowe \cite{yakowitz1991nonparametric} extended the definition of regret and UCB rules in Lai and Robbins \cite{lai1985asymptotically} to nonparametric setting and then to a countably infinite set of arms, for which they improved the order $o(n)$ for the regret in the forcing scheme of \cite{mallows1964some} to order $O(n^{1/r}\log n)$ if $\sup_{k \geqslant 1} \E\big(|y_t|^r | \phi_t = k\big) < \infty$. Subsequently Lai and Yakowitz \cite[Section III]{LaiYakowitz} developed an adaptive allocation rule $\tilde{\phi}$ (which depends on ($\alpha_i : i \geqslant 1$)) such that the regret
\begin{equation}\label{3.1}\tag{4.7}
    R_n = \sum_{j: \mu_j < \mu^*} (\mu^* - \mu_j) \E \tau_n(j), \text{   where } \mu^* = \sup_{k \geqslant 1} \mu_j,
\end{equation}
is of the order $O(\alpha_n \log n)$. The $\tau_n(j)$ in \eqref{3.1} is the total number of times $(\leqslant n)$ that the adaptive allocation rule samples from arm $j$, as in (4.1) for parametric $k$-armed bandits. The $\alpha_i$ are nondecreasing positive numbers such that $\alpha_i \rightarrow \infty$ and $\alpha_{2i} = O(\alpha_i)$. The adaptive allocation rule assumes certain exponential bounds that are valid only for some $\gamma > 0$ and $c > 0$; see A1) and A2) of \cite[p.1200]{LaiYakowitz}; without knowledge of these contents in practice, \cite[Theorem 2]{LaiYakowitz} chooses their upper bounds to define $\tilde{\phi}$ so that its regret is of the order $O(\alpha_n \log n)$. Section 4.3.1 removes this assumption by using exponential inequalities for the self-normalized Welch $t$-statistics. Moreover, $\tilde{\phi}$ is basically a UCB-type rule. Whereas the last paragraph of Section 2 explains the technical difficulties of UCB-type and index policies for continuous-time diffusion processes, here their difficulties arise from the exponential inequalities assumed in A1) and A2) for cumulative (i.e., partial sums of) rewards in \cite{LaiYakowitz}. Again, as in Section 4.2, we next show how the $\epsilon$-greedy randomization algorithm of Section 4.2.2 and the arm elimination scheme involving the self-normalized Welch $t$-statistics of Section 4.2.3 can be used to circumvent these difficulties of UCB-type policies. We then proceed from a countably infinite to a non-denumerable set of arms, therefore improving the results of \cite{LaiYakowitz} and \cite{berry1997}. Section 4.3.2 further generalizes the methods and results to nonparametric contextual bandits with a non-denumerable set of arms, and the proof of the procedure developed therein is given in the Appendix.

\subsection{Context-free Multi-armed Bandits with Infinitely Many Arms}

To continue the discussion on context-free multi-armed bandits in the preceding paragraph, we first review an idea of Mallows and Robbins \cite[Sections III and IV]{mallows1964some}, its subsequent enhancement by \cite{lai1978class} and the extension by \cite{de1999theory}. Lemma 1.3 of \cite{de1999theory}, which provides a survey of decoupling inequalities, states that $\pp\big(\cup_{i=1}^n U_i\big) \leqslant 2\pp\big(\cup_{i=1}^n V_i\big)$  for independent events $V_1, \dots, V_n$ such that $\pp(U_i) \leqslant \pp(V_i)$ for all $1 \leqslant i \leqslant n$. From this it follows that
\begin{equation}\label{3.2}\tag{4.8}
        \pp\bigg( \max_{1 \leqslant i \leqslant n} Y_i > y\bigg) \leqslant 2 \pp\bigg(\max_{1 \leqslant i \leqslant n} \tilde{Y}_i > y\bigg) \text{  for all } y,
\end{equation}
where $\tilde{Y}_i$ are independent random variables such that $\tilde{Y}_i$ has the same distribution as $Y_i$. Theorem 6.1 of \cite{de1999theory} provides decoupling inequalities for randomly stopped sums of independent random variables $Y_i$ taking values in a Banach space with norm $||\cdot||$. Let $\{\tilde{Y}_i\}$ be an independent copy of $\{Y_i\}, S_n = \sum_{i=1}^n Y_i, \tilde{S}_n = \sum_{i=1}^n \tilde{Y}_i$. Then for $\alpha > 0$ and nondecreasing continuous functions $\Phi:[0, \infty) \rightarrow [0, \infty)$ such that $\Phi(0) = 0$ and $\Phi(cx) \leqslant c^\alpha \Phi(x)$ for all $c \geqslant 2$ and $x \geqslant 0$, there exist $0 < b_\alpha < B_\alpha$ depending only on $\alpha$ such that 
\begin{equation}\label{3.3}\tag{4.9}
    b_\alpha \E\max_{1 \leqslant n \leqslant T} \Phi \Big( ||\tilde{S}_n||\Big) \leqslant \E \max_{1 \leqslant n \leqslant T} \Phi\Big(||S_n||\Big) \leqslant B_\alpha \E\max_{1 \leqslant n \leqslant T} \Phi \Big( ||\tilde{S}_n||\Big)
\end{equation}
in which $T$ is a stopping time based on $\{Y_1, Y_2, \dots\}$. Note that $(\tilde{Y}_1, \dots, \tilde{Y}_n)$ is independent of $Y_1, Y_2, \dots, Y_n$ and therefore also independent of $T$. The same argument can be used to prove the existence of $0 < b_\alpha < B_\alpha$ such that
\begin{equation}\label{3.4}\tag{4.10}
    b_\alpha \E \Phi \Big( \Big|\Big|\sum_{t=1}^n\psi_t\tilde{Y}_t\Big|\Big|\Big) \leqslant \E  \Phi\Big(\Big|\Big|\sum_{t=1}^n\psi_t{Y}_t\Big|\Big|\Big) \leqslant B_\alpha \E \Phi \Big( \Big|\Big|\sum_{t=1}^n\psi_t\tilde{Y}_t\Big|\Big|\Big)
\end{equation}
in which $\psi_1, \psi_2, \dots$ are bounded real-valued random variables such that $\psi_t$ is $\F_{t-1}$-measurable, where $\F_{t-1}$ is the $\sigma$-field generated by $Y_1, \dots, Y_{t-1}$. From \eqref{3.2}, it follows that if the $Y_i$ have a common distribution function $F$ then the right-hand side of \eqref{3.2} is equal to $2(1 - F^n(y))$, while the left-hand side is majorized by $\min\{1, n(1 - F(y))\}$. Lai and Robbins \cite{lai1978class} call the $Y_i$ ``maximally dependent" if $M_n := \max(Y_1, \dots, Y_n)$ is the stochastically largest possible under the common marginal distribution function $F$ of $Y_i$, i.e., 
\begin{equation}\label{3.5}\tag{4.11}
    \pp(M_n > y) = \min\{1, n(1 - F(y))\} \text{   for all } y,
\end{equation}
and describe a construction of $Y_1, Y_2, \dots$, such that \eqref{3.5} holds for every $n$. They point out that one motivation that led them to study maximally dependent random variables was that $m_n := \E M_n = a_n + n\int_{a_n}^\infty (1 - F(y))dy$ for \eqref{3.5}, with $a_n = \inf\{y: F(y) \geqslant 1 - n^{-1}\}$, is much more tractable than $\tilde{m}_n := n \int_{-\infty}^\infty y F^{n-1}(y)dF(y)$ for large $n$. In particular, Theorem 5 of \cite{lai1978class} states that under maximal dependence, the following statements are equivalent:
\begin{equation}\label{3.6}\tag{4.12}
    \lim_{n \rightarrow \infty} \E \big| \big(M_n / a_n\big) - 1\big|^p = 0 \text{   for all } p > 0,
\end{equation}
\begin{equation}\label{3.7}\tag{4.13}
    \lim_{y \rightarrow \infty} \Big(1 - F\big(cy\big)\Big) \Big/ \Big( 1 - F\big(y\big)\Big) = 0 \text{   for every } c > 1,
\end{equation}
and that under independence, \eqref{3.6} is equivalent to the stronger condition:
\begin{equation}\label{3.8}\tag{4.14}
    \eqref{3.7}\text{ holds and }\int_{-\infty}^0 |y|^r dF(y) < \infty\text{ for some }r > 0.
\end{equation}

We now describe how the preceding overview of maximal dependence and decoupling inequalities can be applied to implement the $\epsilon$-greedy algorithm and the arm elimination scheme when the set of arms is countably infinite, using the same notation as in Section 2.3 in which $\hat{\mu}_{j, \ell - 1} (\bm{x}_\ell), \tilde{\mu}_{j, \ell - 1} (\bm{x}_\ell)$ and $\hat{\mu}_{j, t-1}^*(\bm{x}_t)$ are replaced by $\hat{\mu}_{j, \ell-1}, \tilde{\mu}_{j, \ell-1}$ and $\hat{\mu}_{j, t-1}^*$ (because covariates are absent in the present context-free setting). As in \cite[p.1202]{LaiYakowitz}, assume that there is an arm in the countably infinite set that has the largest expected reward $\mu^*$, i.e.,
\begin{equation}\label{3.9}\tag{4.15}
    L := \{j : \mu_j = \mu^*\} \neq \emptyset
\end{equation}
Under \eqref{3.9}, there are two cases: (a) $\sup_{j \not\in L} \mu_j < \mu^*$, and (b) $\sup_{j \not\in L} \mu_j = \mu^*$. As we have already pointed out in the introductory paragraph of Section 3, \cite{LaiYakowitz} considers case (a) under \eqref{3.9} and shows that regret to be order $O(\alpha_n \log n)$ for any nondecreasing sequence $\alpha_n \rightarrow \infty$ by choosing the UCB-type adaptive allocation rule $\tilde{\phi}$ to depend on ($\alpha_i : i \leqslant 1$). It will be shown that we can achieve order $O(\log n)$ for the regret by using an alternative adaptive allocation rule that we now describe. Moreover, this adaptive allocation rule will be shown to have regret of order $O((\log n)^2)$ for case (b) under \eqref{3.9}. It uses an increasing set (indexed by time $n$) of $k_n$ arms and defines the leading arm $j_n$ by
\begin{equation}\label{3.10}\tag{4.16}
    \overline{Y}_{T_n(j_n)} (j_n) = \max\Big\{\overline{Y}_{T_n(j)} (j): 1 \leqslant j \leqslant k_n \text{ and } T_n(j) \geqslant n / (2k_n)\Big\}.
\end{equation}
For $n + 1 \equiv j\pmod{k_n}$ with $j \in \{1, 2, \dots, k_n\}$, it samples from arm $j$ or arm $j(n)$ according as
\begin{equation}\label{3.11}\tag{4.17}
    \overline{Y}_{T_n(j)}(j) \geqslant \overline{Y}_{T_n(j_n)} (j_n) \text{ or otherwise}.
\end{equation}
Note that \eqref{3.10} is also used to define the leading arm by \cite{LaiYakowitz} which, however, uses an upper confidence bound $\overline{Y}_{T_n(j)}(j) + a_{n, T_n(j)}$ in lieu of $\overline{Y}_{T_n(j)}(j)$ in \eqref{3.11}. Our adaptive allocation rule uses arm elimination instead of UCB, but still uses \eqref{3.10} and \eqref{3.11} to find an arm that belongs to $L$ quickly before applying the $\epsilon$-greedy and arm elimination schemes. Mallows and Robbins \cite[p.96]{mallows1964some} define $M(0, n)$ as the maximum expected sum of $n$ rewards $Y_1, Y_2, \dots, Y_n$ when $F$ is known and the initial reward is drawn from an arm with reward distribution $F$ so that thereafter one can sample from that arm or choose a new arm. They then show that ``in certain cases", 
\[ \lim_{n \rightarrow \infty} \E \big(Y_1 + \cdots + Y_n\big) \big/ M(0, n) = 1, \pp\big\{\lim_{n \rightarrow \infty} \big(Y_1 + \cdots + Y_n\big) \big/ M(0, n)\big\} = 1,\]
$M(0, n) \sim nv(n)$, where $v(n)$ is the exptectation of the maximum of $n$ i.i.d. random variables with common distribution frunction $F$; see \cite[p.97]{mallows1964some}. These arguments have been made more precise and general by \cite{lai1978class} and extended to dependent arms by \cite{de1999theory}, as has been reviewed in the preceding paragraph. Under assumption \eqref{3.9} \cite{LaiYakowitz} describe the procedure of \cite{mallows1964some} more formally via the machine learning algorithm mentioned earlier in this paragraph. They also consider dependent arms, motivated by the applications of \cite{yakowitz1991nonparametric} and \cite{yakowtiz1992theory}. Hence the extension via decoupling inequalities are important, as will be discussed further in Section 3.2.

The applications in \cite{yakowitz1991nonparametric} and \cite{yakowtiz1992theory} actually involve a non-denumerable set of arms and the study of a countably infinite set of arms in \cite{mallows1964some} and \cite{yakowitz1991nonparametric} is intended as an intermediate step to provide insights in transitioning from classical $k$-armed bandits to a non-denumerable set of bandit arms. In particular, the key insight is assumption \eqref{3.9} and the use of an increasing set of $k_n$ arms. We defer this discussion and the related details of Section 4.3.2 in which we provide a far-reaching generalization of $k$-armed nonparametric contextual bandit theory in Section 4.2 to non-denumerable set of arms.

\subsection{Bandit and covariate processes when the set of arms is non-denumerable}

To extend the asymptotic minimax theory of nonparametric contextual bandits for $k$ arms in Section 4.2 to a non-denumerable set of arms, we assume besides the regularity conditions (C1), (C2), and (C3) in Section 4.2.3 a generalization (C4) of assumption \eqref{3.9} in context-free multi-armed bandits with a countably infinite set of arms to contextual bandits, modifying the following idea of \cite[p.1203]{LaiYakowitz} based on the continuity of the expected rewards over the non-denumerable set of arms. They let $\mathcal{G}$ be the set of adaptive allocation strategies, and assume that $\mathcal{G}$ is a metric space and for some probability measure $\nu$ on the Borel $\sigma$-field,
\begin{equation}\tag{4.18}\label{3.12}
    \nu(C_\delta) > 0 \text{ for every open ball } C_\delta \text{ with radius }\delta.
\end{equation}
Letting $\mu_g(\bm{x})$ denote the expected reward (for a single pull of an arm) for $g \in \mathcal{G}$ when covariate $\bm{x}$ is observed and $\mu^*(\bm{x}) = \sup_{g \in \mathcal{G}} \mu_g(\bm{x})$, which may not be attained (as in \cite[Section IVB]{LaiYakowitz}). It should be noted that \cite[Section IVB]{LaiYakowitz} actually considers a non-denumerable set $\mathcal{G}$ of stationary control laws $g$ for which the measure $\nu$ on the Borel $\sigma$-field can be assumed to be known. Here our objective is to extend the theory of $k$-armed nonparametric contextual bandits in Section 2 to a non-denumerable set of arms. The smooth density condition (C1) over the covariate space can also be extended to neighborhoods of arms that have expected rewards close of $\mu^*$ via the following augmentation, in which $L_{\delta, \bm{x}}$ denotes the set of arms $g$ for which $\mu_g(\bm{x}) \geqslant \mu^*(\bm{x}) - \delta$ and $\bm{x} \in B$:
\begin{enumerate}[(C1)]
    \setcounter{enumi}{3}
    \item $\exists \delta > 0$ and regular conditional probability measure $\nu(\cdot | \bm{x})$ on $L_{\delta, \bm{x}}$ (with the Borel $\sigma$-field) that has a positive and continuously differentiable density function with respect to Lebesgue measure.
\end{enumerate}
As in Section 2.3, let $n_i \sim a^i$ for some integer $a > 2$. Our approach involves enlarging the set of leading arms $J_t$ (defined by \eqref{7} for $n_i \leqslant t \leqslant n_{i+1}$) at time $n_i$ before applying the $\epsilon$-greedy randomization algorithm and the arm elimination scheme for that block of times. Let $k_i$ be the cardinality of $K_{n_i}$, $v_i$ be the volume of a $p$-dimensional ball with radius $\delta_i < \delta \wedge 1$, and $\ell_i \sim \delta_i^{-p}$ be a positive integer that represents the length of a consecutive block of times $t$ to search for $g$ to be added to the set of leading arms with $(\hat{\mu}^*_{t-1} - \hat{\mu}_{g, t-1})(\bm{x}_t) \leqslant \delta_i$, where $\hat{\mu}_{g, t-1}$ and $\hat{\mu}^*_{t-1}$ are local linear regression estimates of $(\mu^* - \mu_g)(\bm{x}_t)$ (as in the second paragraph of Section 2.3) based on observations up to time $t-1$. 

Specifically, under conditions (C1)--(C4) for the non-denumerable set of arms, we proceed as follows at times $t \in \{n_i, \dots, n_i + \ell_i - 1\}$. If
\begin{equation}\tag{4.19}\label{3.13}
    (\hat{\mu}^*_{t-1} - \hat{\mu}_{g, t-1}) (\bm{x}_t) \leqslant \delta_i \text{ for some } g \in K_t \setminus J_t,
\end{equation}
redefine $J_t$ by including $g$ and stop, setting $k_{i+1} = k_i + 1$. If the complement of \eqref{3.13} occurs, i.e., if ($\hat{\mu}^*_{t-1} - \hat{\mu}_{g, t-1}) (\bm{x}_t) > \delta_i$ for all $n_i \leqslant t \leqslant n_i + \ell_i - 1$ and $g \in K_t \setminus J_t$, repeat the search procedure for $n_i + \ell_i \leqslant t < n_i + 2\ell_i$ and then for $n_i + 2\ell_i \leqslant t < n_i + 3\ell_i, \dots$, until $n_i + \lfloor n_i / \ell_i \rfloor \ell_i$ so that \eqref{3.13} holds with $g$ that can be included in $J_t$ and $k_{i+1} = k_i + 1$, or stop with $J_{n_{i+1}} = J_{n_i}$ and $k_{i+1} = k_i$ if no such $g$ has been found up to time $n_i + \lfloor n_i / \ell_i \rfloor \ell_i$. Note that $n_{i+1}-n_i \sim (a-1) n_i \geqslant n_i$. Labeling the arms in $K_{n_i}$ as $\{1, 2, \dots, k_i\}$ in the chronological order they are added to $K_{n_i}$, if $t \equiv j\pmod{k_i}$ with $j \in \{1, 2, \dots, k_i\}$ for $n_i \leqslant t < n_{i+1}$, sample from arm $j$ or arm $j_t$ according to \eqref{3.11}, where $j_t$ has the largest estimated mean reward at the covariate $\bm{x}_t$.  This adaptive allocation rule is denoted by $\phi_{opt}^\infty$ as it is an extension of $\phi_{opt}$ in Section 2.3 to infinitely many (even non-denumerable) arms. In the Appendix we state and prove theorems on the optimality of $\phi_{opt}^\infty$ and relate them to the results of \cite{berry1997}, \cite{LaiYakowitz}, \cite{yakowitz1991nonparametric}, \cite{yakowtiz1992theory}.

\subsection{Theorem 1 and 2}

We first present Theorem 1 showing that $\phi_{opt}^\infty$ attains the asymptotically minimal rate for the regret under (C1)--(C4), generalizing the theory in Section 2 for $k$ arms to a non-denumerable set of arms. We then apply the result to the problem introduced by Shepp and coauthors \cite{berry1997} in Theorem 2, and to Markov decision processes introduced by Yakowitz and coauthors \cite{LaiYakowitz, yakowitz1991nonparametric, yakowtiz1992theory, yakowitz1995nonparametric}.

\begin{theorem}
    Assume \emph{(C1)--(C4)} and that $\mu^*$ is finite. Define the risk of an adaptive allocation rule $\phi$ over a covariate set $B$ by
    \begin{equation}\tag{A.1}\label{A1}
        R_{n, \phi}(B) = \sum_{t=1}^n \int_B \E \tau_n(\phi_{t-1}) \{(\mu^* - \mu_{\phi_{t-1}})(\bm{x})\} dH(\bm{x}),
    \end{equation}
    in which $\tau_n(g)$ is the total number of times ($\leqslant n$) that $\phi$ samples from arm $g \in \mathcal{G}$ and is analogous to $\tau_n(j)$ in \eqref{3.1}. If leading arm transitions do not occur over $B$, $R_{n, \phi_{opt}^\infty}(B) = (\log n)^{1 + o(1)}$ as $n \rightarrow \infty$. On the other hand, if $B$ contains leading arm transitions, then $R_{n, \phi_{opt}^\infty}(B) = (\log n)^{2 + o(1)}$.
\end{theorem}
\textbf{Proof.} We combine the ideas in Section 2.3 for establishing the optimal asymptotic rate of the risk of $\phi_{opt}$ with those in Section IVB of \cite{LaiYakowitz} to tackle a non-denumerable set of stationary control laws that we have reviewed in the first paragraph of Section 3.2. First it follows from (C4) that
\begin{equation}\tag{A.2}\label{A2}
    \inf_{\bm{x} \in B} \log(1 - \nu_{i, \bm{x}}) \approx -v_i, \text{ where } \nu_{i, \bm{x}} = \nu\Big( \big\{g \in L_{\delta_i, \bm{x}}\big\} \Big| \bm{x}\Big),
\end{equation}
in which $\approx$ denotes the same order of magnitude (i.e., there exist positive constants $c_p$ and $\tilde{c}_p$ such that $c_p \delta_i^p \leqslant v_i \leqslant \tilde{c}_p \delta_i^p$) and for which we note that $L_{\delta_i, \bm{x}} \subset L_{\delta, \bm{x}}$ (since $\delta_i < \delta$) and $\delta_t^2 \sim (2 \log t) /t = o(1)$ for $t = \{n_i, \dots, 2n_i\}$. Moreover, if $\mu^*_{t-1} - \mu_{t-1, g}$ were known for $g \in K_t \setminus J_t$ and $n_i \leqslant t \leqslant 2n_i$, then
\begin{equation}\tag{A.3}\label{A3}
\begin{split}
    & \pp(\text{no } g \in K_t \setminus J_t \text{ satisfies} \eqref{3.13} \text{ with } \hat{\mu}^*_{t-1} - \hat{\mu}_{g, t-1} \text{ replaced by } \mu_{t-1}^* - \mu_{t-1, g}, \\
    & \text{ at times } t = n_i, n_i + 1, \dots, 2n_i)\\
    \leqslant & \pp\Bigg(\bigcap_{1 \leqslant m \leqslant \lfloor n_i / \ell_i \rfloor} \Big\{g_{i, t} \not\in L_{\delta_i, \bm{x}_t} \text{ for } n_i + (m-1) \ell_i \leqslant t < n_i + m \ell_i\Big\} \Bigg)
\end{split}
\end{equation}
in which $g_{i, t}$ is the arm selected from the consecutive block of times $n_i + (m-1) \ell_i \leqslant t < n_i + m\ell_i$ to have the largest expected reward $\mu_{t-1, g}(\bm{x}_t)$ (with $\bm{x}_t \in B$) among all arms $g \in K_t$. Combining \eqref{A2} and \eqref{A3} yields that the logarithm of the probability on the left-hand side of \eqref{A3} is bounded by $-\beta_p n_i$ for some $\beta_p > 0$, since $\ell_i v_i \approx \delta^{-p} \delta_i^p = 1$ and $\lfloor n_i /\ell_i\rfloor \ell_i \sim n_i$. Lai and Yakowitz's key assumption \eqref{3.9} for context-free bandits in \cite{LaiYakowitz} to handle a countably infinite set of arms by using increasing sets (indexed by time) of arms so that an optimal arm belonging to $L$ in \eqref{3.9} is contained in these sets within $O(1)$ time. To extend this idea to a non-denumerable set of arms, we replace $L$ by $L_{\delta_i, \bm{x}_t}$ for $t = n_i, n_i + 1, \dots, n_i + \lfloor n_i /\ell_i\rfloor \ell_i$ (with $n_i \sim a^i$) and let
\begin{equation}\tag{A.4}\label{A4}
    \tau = \inf \big\{n_i : \exists t \in \{n_i, n_i + 1, \cdots, n_i + \lfloor n_i /\ell_i\rfloor\ell_i\} \text{ and arm } g \in L_{\delta_i, \bm{x}_t}\big\}.
\end{equation}
Then $\E \tau \leqslant \sum_{i=1}^\infty \pp(\tau > n_i) \leqslant \sum_{i=1}^\infty e^{-\beta_p n_i} = O(1)$, since $\log \pp(\tau > n_i) \leqslant -\beta_p n_i$ as shown above.

However, $\mu^*_{t-1} - \hat{\mu}_{t-1, g}$ is actually unknown for $g\in K_t \setminus J_t$ and has to be estimated by $\hat{\mu}^*_{t-1} - \hat{\mu}_{t-1,g}$ using local linear regression, which is much more challenging than that in the second and third paragraphs of Section 2.3 because of the high-dimensional bandit process that involves a non-denumerable set (instead of a finite number $k$) of arms. In Section 3 and Supplement S2 of \cite{KimLaiXu}, high-dimensional covariates are considered and \cite{yang2015minimax}, \cite{dai2019smooth}, \cite{shen1994} and \cite{yang1999information} are referenced for the methodology and results. In particular, \cite{yang2015minimax} and the subsequent paper \cite{yang2016} by Yun Yang and coauthors are especially effective not only for high covariate dimension $p_n$ (with a single arm) but also for our setting of $k_i \approx a^i$ arms (with fixed covariate dimension $p$). Specifically, the pivotal assumption Q to handle $p_n$-dimensional covariates in Section 3.2 of \cite{yang2015minimax} can be restated in our notation as $H$ having compact support on $\mathbb{R}^{p_n}$ and a bounded density function (with respect to Lebesgue measure) that is bounded away from 0. Our assumption (C4), which is its counterpart for $k_i \approx a^i$ arms (with fixed covariate dimension $p$), can be used to establish consistency and ``minimax-optimality" of $(\hat{\mu}_{t-1}^* - \hat{\mu}_{t-1, g})(\bm{x}_t)$ with $\bm{x}_t \in B \subset \mathbb{R}^p$ by arguments similar to those of \cite{yang2015minimax} and \cite{yang2016} (which elucidates the role of assumption Q or (C4) in approximating a high-dimensional nonparametric regression problem by one with support near a fixed-dimensional manifold). In view of the consistency property, the argument in the preceding paragraph that assumes $\mu^*_{t-1} - \mu_{t-1, g}$ to be known is still applicable when $\mu_{t-1}^* - \mu_{t-1, g}$ is unknown and is substituted by $\hat{\mu}_{t-1}^* -\hat{\mu}_{t-1, g}$, thereby proving that the search procedure described in the last paragraph of Section 3.2 yields an arm $g \in L_{\delta_i, \bm{x}_t}$ within $O(1)$ expected time; see \eqref{A4} and the sentence below it.

Since an arm $g^*$ having expected reward $\mu_{t-1, g^*}(\bm{x}_t)$ with $\delta_i$ of $\mu^*(\bm{x}_t)$ is found by the search procedure with $O(1)$ expected time and is then added to the ``lumped" set \eqref{7} of leading arms, its status as a member of the lumped set of leading arms does not change with the covariate values, as explained in the first paragraph of Section 2.3. Hence the $\epsilon$-greedy randomization algorithm in Section 2.2 and the arm elimination procedure in Section 2.3 can work in the same way after $g^*$ is included in the set of leading arms, proving that $R_{n, \phi_{opt}^\infty} (B) = \big(\log n\big)^{1+o(1)}$ if $B$ contains no leading arm transition and that $R_{n, \phi_{opt}^\infty}(B) = \big(\log n\big) ^{2+o(1)}$ otherwise. It should be noted that the nonparametric contextual bandit theory in Section 2 assumes independent arms as in \cite{lai1985asymptotically}, \cite{lai1987} and \cite{KimLaiXu} whereas the arms we consider in Section 3 and in this theorem can be dependent. The decoupling inequalities reviewed in the first paragraph of Section 3.1, in particular \eqref{3.2} relating the tail probability of $\max_{1 \leqslant t \leqslant n} Y_t$ to that of $\max_{1 \leqslant t \leqslant n} \tilde{Y}_t$ for independent $\tilde{Y}_t$ such that $\tilde{Y}_t$ has the same distribution as $Y_t$, and \eqref{3.4} in which $\phi_t$ are bounded real-valued random variables such that $\phi_t$ is $\F_{t-1}$-measurable (as in adaptive allocation rules), show that the asymptotic rates of $R_{n, \phi_{opt}}^\infty (B)$ remain the same when the assumption of independence in these results is removed.

The following variant of Theorem 1 generalizes the non-denumerable ``multi-armed bandit problem" introduced by Shepp and coauthors in \cite{berry1997} from context-free Bernoulli bandits with a prior distribution of the arm parameters $\mu_g$ to nonparametric contextual bandits with a prior distribution $\nu$ on the arms.

\begin{theorem}
    Assume that $\nu$ satisfies \emph{(C4)} with finite $\mu^*$ and that \emph{(C1)}, \emph{(C2)} and \emph{(C3)} hold for covariate values in $B$, in which the regression function $m(\bm{x})$ and variance function $\sigma^2(\bm{x})$ in \emph{(C2)} refer to the Bayesian framework with $g$ generated by $\nu$ and then $m_g(\bm{x})$ and $\sigma_g^2(\bm{x})$ being the regression and variance functions. Then the risk $R_{n, \phi_{opt}^\infty}(B)$ over $B$ is of order $(\log n)^{1+o(1)}$ if $B$ does not contain leading arm transitions, and is of order $(\log n)^{2+o(1)}$ otherwise.
\end{theorem}

Instead of minimizing the asymptotic rate of the risk of adaptive allocation rules, defined by the regret \eqref{3.1} for context-free bandit problems, \cite{berry1997} follows Robbins' formulation in \cite{robbins1952} to maximize the long-run expected reward $\lim_{n \rightarrow \infty} n^{-1} \E\big(Y_1 + \cdots + Y_n\big)$ for multi-armed bandits with Bernoulli arms and a prior distribution on the means $\mu_g$ of the arms, in particular, the uniform distribution on $(0, 1)$ or $[a, b]$ with $0 < a < b < 1$ in Sections 2 and 3 of \cite{berry1997}. The posterior distribution of $\mu_g$ is beta or truncated beta with support $[a, b]$, with which \cite{berry1997} uses to analyze the long-run expected reward of several adaptive allocation strategies that they show to have considerably finite-sample performance than the ``nonrecalling strategy that uses arm $i$ until it gives $i$ failuers" and which has been shown by Herschkorn et al. \cite{herschkorn} to have $\mu^*$ as its long-run expected reward. They point out this strategy ``can spend inordinate amounts of time waiting for a long run of failures before dispensing with arms whose performances are clearly mediocre" and therefore ``has success proportion of only $0.79$ when $n = 500$", for which their strategies achieve ``success rates as high as $0.92$." Note that $\phi_{opt}^\infty$ preempts this difficulty due to a non-denumerable set of arms upfront by searching for $g^* \in L_{\delta_i, \bm{x}_t}$ first before applying the $\epsilon$-greedy randomization algorithm and arm elimination procedure, hence it is able to give minimax rate for the regret, which is a much sharper performance criterion than maximizing the long-run expected reward. Moreover, in contrast to our definitive theory in Theorem 2, although the case of a general continuous prior distribution on the success probabilities of the Bernoulli arms is considered in Section 4 of \cite{berry1997}, its treatment is brief and focuses on the expected failure proportions of three of the adaptive allocation strategies proposed in the paper.

We remark that \cite{herschkorn} references to both Mallows and Robbins \cite{mallows1964some} and Yakowitz and Lowe \cite{yakowitz1991nonparametric} whose ``policies return to previously observed arms infinitely often" and those in a preprint (which does not refer to \cite{herschkorn}) of \cite{berry1997} that has the feature of ``never returning to an arm once we have switched to a near one" and therefore the ``advantage that we need not retain any information about earlier arms." Yakowitz and Lowe \cite{yakowitz1991nonparametric} is the precursor of Lai and Yakowitz \cite{LaiYakowitz}, both of which give weaker results than Theorem 1 under stronger assumptions. Specifically, \cite{yakowitz1991nonparametric} considers independent arms, assumes finite absolute moment conditions on the rewards from the arms and derives polynomial growth results for the regret, whereas the Section IV of \cite{LaiYakowitz} assumes exponential bounds for the one-step reward of a controlled Markov chain with a non-denumerable set of stationary control laws. In contrast, Theorem 1 gives $(\log n)^{1 + o(1)}$ or $(\log n)^{2+o(1)}$ for the regret by making use of exponential bounds for self-normalized statistics \cite[Section 15.1]{pena2008self}. For a non-denumerable set of stationary control laws, \cite{LaiYakowitz} follows \cite{yakowtiz1992theory} to consider the weaker criterion of ``learning loss" than regret, defined as $\max_{g \in \mathcal{G}, \mu_g < \mu^* - \epsilon} \E_{\bm{x}} T_n(g)$, in which $\mathcal{G}$ is the metric space of stationary control laws, $T_n(g)$ is the number of times that the allocation strategy uses stationary control law $g$ up to time $n$, and $\E_{\bm{x}}$ denotes expectation when the initial state of the controlled Markov chain is $\bm{x}$.

\section{Summary and Discussion}

We have introduced herein a new approach to nonparametric multi-armed bandit theory involving both the bandit and the covariate processes. Following Shepp and his coauthors in a seminal paper in 1997, we assume a non-denumerable set of arms for the bandit process. The approach we develop herein can be readily extended to continuous-time processes, for which it bypasses the difficulties with index policies in continuous time by using $\epsilon$-greedy randomization and arm elimination instead of dynamic allocation indices. It also carries out a stochastic search with $O(1)$ expected time for a nearly optimal arm at covariate values in a given set $B$ before applying $\epsilon$-greedy randomization and arm elimination. The procedure is shown to attain the asymptotically minimal rates for the regret over $B$. We are further developing these results and methods to develop dynamic treatment regimes (DTRs) for learning health systems (LHS), which bear the responsibility for treating patients over time as their clinical status and learning needs evolve. Formalizing the notion of a complete care strategy, a DTR is a set of rules that dictates treatment decisions, given a patient's history of covariates and prior treatment \citep{lavori2004dynamic}. 
DTRs may be studied using a SMART, which begins with an initial treatment randomization and at each subsequent decision point, re-randomizes patients among further treatment options. A SMART culminates in an outcome $Y$ for each individual (which may be a function of $(x_1,\ldots,x_{I}))$, by which the treatments will be assessed. \cite{lavori2007improving,lavori2008adaptive} construct confidence intervals for comparing DTRs based on their expected outcomes. Key challenges in the design and analysis of SMARTs include incorporating patient covariates and handling estimations as the length of the decision tree grows, because the number of treatment strategies and possible patient histories explodes rapidly. Most SMARTS do not consider more than two decision points per patient. One approach for handling patient covariates is Q-learning \citep{sutton1998introduction,murphy2005experimental}. Q-learning seeks to model the patient's expected final outcome, conditional on taking action $a_i$ and given the history $(\tau_i,x_{1:i}, a_{1:i-1})$, and assuming optimal decision making thereafter. This model is thus a function, $E[Y | \tau_i,x_{1:i}, a_{1:i}] \sim Q(\tau_i,x_{1:i}, a_{1:i})$, called the {\it Q-function}.
In order to estimate a $Q$-function, Q-learning alternates between model estimation of expected values of states and backward induction to select optimal actions, using a modified version of Bellman's inequality. Q-learning is therefore an approximate dynamic programming technique. It has been combined with a variety of modeling approaches including linear models \citep{murphy2005experimental}, regression trees \citep*{ernst2005tree}, and kernels \citep{ormoneit2002kernel}. \cite{chakraborty2014dynamic} discuss nonregular asymptotics for Q-learning with the linear model.

Intuitions and approaches from MAB and CMAB theory, including \cite{lai1985asymptotically}, are proving fruitful in constructing algorithms with theoretical guarantees and good performance for analyzing DTRs. 
In the general DTR setting, \cite{zhang2019near} use the UCB approach to motivate a reinforcement learning algorithm and derive regret guarantees. 
Following the techniques of \cite{lai1985asymptotically} and \cite*{auer2009near}, at each time $t$, they construct a uniform confidence set $\mathbb{M}_t$ with two-sided bounds on the final payouts and transitions, and then use the Bellman equation recursively to find an optimal DTR in $\mathbb{M}_t$. Note that this approach also permits confidence bands for the values of individual DTRs. Zhang and Bareinboim derive regret guarantees, and further show that weak evidence from observational data collection can be used to narrow the range of possible transitions, thus narrowing $\mathbb{M}_t$ and improving performance.
\cite{hu2020dtr} analyze a two-stage DTR model. Assuming  a linear model for Q-functions, they extend the contextual bandit approaches of \cite{goldenshluger2013linear} and \cite*{bastani2020mostly} to the two-stage two-treatment DTR setting, using a combination of unbiased estimates from a small sample and biased estimates from the full sample. They derive regret bounds under several margin conditions on the Q-functions, notably showing under a sharp margin condition a regret bound of order $O(d(\log d)^{2/3}\log T + (d \log d)^2)$.
They demonstrate that their bounds have optimal dependence on $T$ by applying lower bounds from contextual bandits.
\cite{wang2016optimal} demonstrate an important connection between DTRs and contextual bandits in a Bayesian framework. They model binary outcomes using Bayesian generalized linear models and handle posterior computations using Laplace approximations. With quick recursive computation of the value function, they show how to collapse the DTR problem into a CMAB problem, where each decision point becomes a bandit sample, and payoffs are given by the change in the posterior expected value. This formulation naturally permits them to use Bayesian CMAB algorithms, including the knowledge gradient, Thompson sampling, and greedy Bayes algorithms, for learning and executing DTRs. 

The work in this chapter opens up new avenues for a definitive treatment of DTRs, which we are currently developing

\chapter{A Rigorous Framework for Complex Trial Design}

\section{Introduction}

In 2019, FDA began the Complex Innovative Trial Design (CID) Pilot Meeting Program to ``support the goal of facilitating and advancing the use of complex adaptive, Bayesian, and other novel clinical trial designs." \citep{food2018complex}
In this program, ``priority will be given to trial designs for which analytically derived properties (e.g., Type I error) may not be feasible and simulations are necessary to determine operating characteristics." In a public meeting on March 20, 2018, FDA officials highlighted key open questions for the use of complex simulation-based designs, including:

\begin{itemize}
    \item What should be the scope of simulations? [That is, which nuisance parameters can or should be considered? Should all null hypothesis combinations be considered? What about accrual rates?]
    \item At how fine of a grid?
    \item How to handle multiple hypotheses and multiple testing?
\end{itemize}

In this chapter, we shall lay a foundation for designers and regulators to computationally verify Type I Error, Family-Wise Error Rate (FWER), and other metrics over continuous hypothesis spaces. Our approach will combine Monte Carlo simulation at a grid of parameter values with technical arguments to extend estimates of Type I Error to the interstitial space. We require a well-specified data generating model which can be simulated or approximated as an exponential family process. This category includes most of the named distributional families for patient responses, including Bernoulli, Gaussian with known variance, Gaussian response with unknown variance, Poisson, and Negative binomial. Our approach is also compatible with censored survival data, and so can apply to Exponential, Gamma, and certain Weibull models. The methods can be further extended to Brownian motion, and thus asymptotically can be applied to describe many common estimators such as Cox regression under an assumption of proportional hazards. 

With increasing computation, our approach aims to approximate the Type I Error rate function over a complete region of the hypothesis space, a goal which previous approaches do not attempt. To the regulator, our method allays concern that an optimized method could be shifting Type I Error to an unseen area. To the designer, it permits a foundation for optimization of trial designs of nearly arbitrary complexity. In practice, we expect the key bottleneck of our approach will be rapid simulation over a fine grid of the null hypothesis space when it is high-dimensional.

In Section 5.2, we give background material on exponential families and how exponential family processes can be used to model adaptive clinical trials.
In Section 5.3, we discuss composite null hypothesis spaces and regulatory concerns for simulation.
In Section 5.4, we discuss Monte Carlo simulation for Type I Error estimation at grid points.
In Section 5.5, we show how to achieve upper bounds on the Type I Error over continuous hypothesis spaces.
In Section 5.6, we discuss examples of trials to which the method can be directly applied.
In Section 5.7, we discuss new opportunities unlocked by this methodology, including unplanned adaptations and Type I Error calibration for optimized designs. In the following subsection 5.1.1, we demonstrate our method by example, on an adaptive Bayesian trial.

\subsection{Example: A Two-Arm Bayesian Trial with Thompson Sampling}

To demonstrate the method by example, we show its application on a toy Bayesian trial. This trial will enroll a total of 100 patients, and has two treatment options to distribute between the patients. The treatment outcome of each patient is either success $(y_i = 1)$, or failure $(y_i = 0)$, observed immediately after each patient is treated. Thus we make may each treatment decision with the advantage of previously collected data.

We seek a response rate of above $60\%$ for each treatment. We have an unknown Bernoulli parameter for each arm, $\theta_1$ and $\theta_2$, and two null hypotheses: 

$$H_1: \theta_1 \leq .6$$ 

$$H_2: \theta_2 \leq .6$$

\begin{figure}
\begin{center}
\includegraphics[scale = .5]{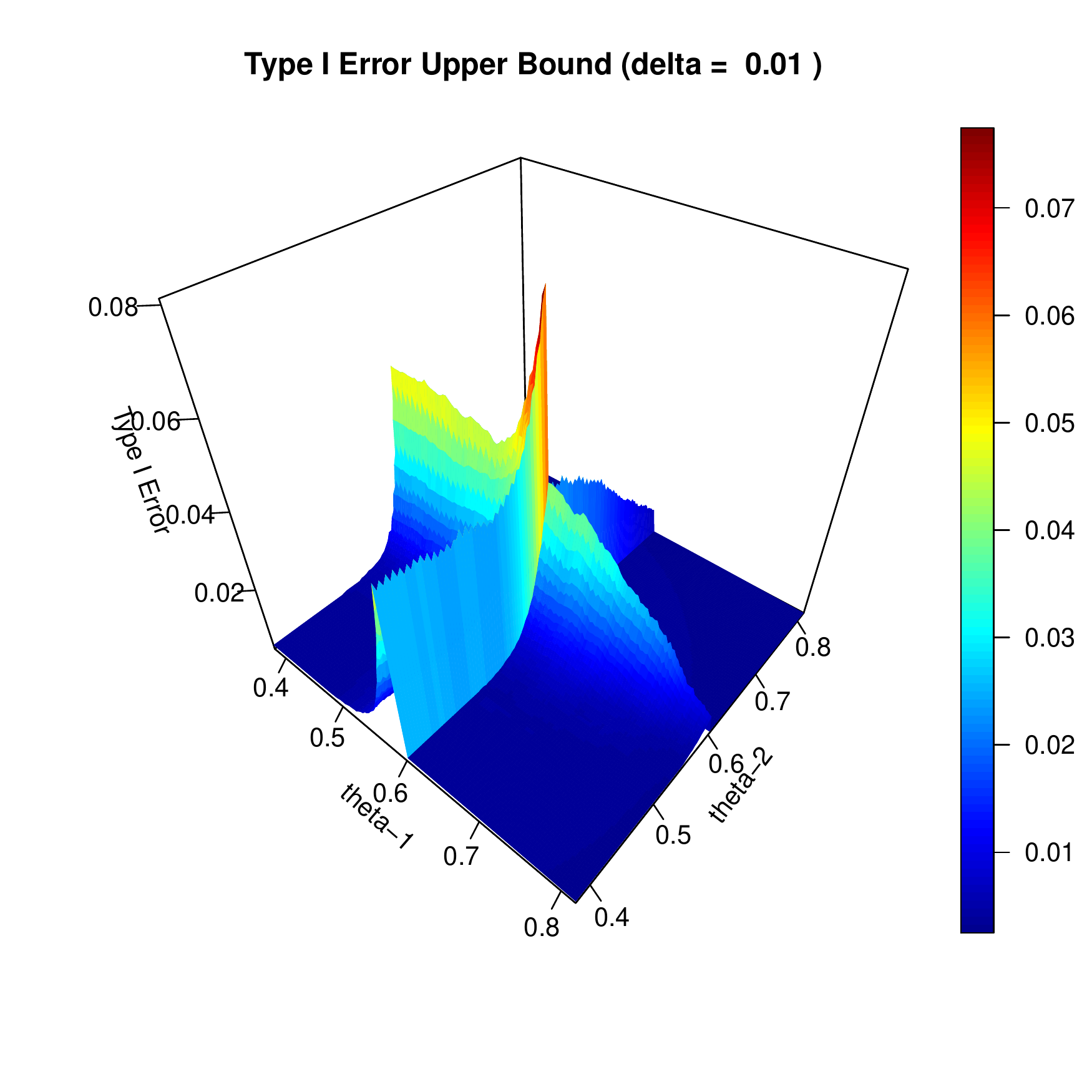}
\caption{Estimated Upper Bound on the Type I Error. Valid with pointwise confidence $1 - \delta$}
\label{ThetaExample}
\end{center}
\end{figure}

We will take a Bayesian approach, assuming independent uniform priors on $\theta_1$ and $\theta_2$. Allocation of patients to the two arms will proceed by the Bayesian adaptive algorithm known as Thompson Sampling, as described in \cite{russo2017tutorial}. At the conclusion of the trial, we will reject arm $i$ if its posterior credible probability that $\theta_i > 0.6$ is at least $95\%$.
For reasons that will become clear in section 5.5.3, we choose not to grid the parameter space evenly in terms of $\theta$, but instead in terms of the natural parameter $\eta$ of the binomial distribution
$$\eta_i = \log \frac{\theta_i}{1 - \theta_i}$$
Note that while $\theta_1$ and $\theta_2$ are bounded between $0$ and $1$, $\eta_1$ and $\eta_2$ range the whole real line. To keep our simulation area bounded in the space of $(\eta_1, \eta_2)$, we must have lower and upper bounds on $\theta_1$ and $\theta_2$ away from $0$ and $1$. Our grid will cover $\eta_i \in [-.5, 1.5]$ in 128 equal steps, for a total of $128^2 = 16384$ points (of which about 3/4 lie in the null hypothesis). This corresponds to a range of range of about $\theta_i \in [.38,.88]$
We perform >800,000 Monte Carlo simulations of the trial at each grid point. 
Our method yields the following upper bound to the Type I Error function, which holds over the continuous parameter space with $99\%$ pointwise confidence. That is, given a pair $(\theta_1, \theta_2)$, the corresponding height in Figure \ref{ThetaExample} is $\hat{\alpha}(\theta_1, \theta_2)$ which has the property

$$P(\hat{\alpha}(\theta_1, \theta_2) > \alpha(\theta_1, \theta_2)) \geq 1 - \delta.$$


\subsection{Existing Methods and Challenges for Type I Error Control}

For FDA submissions, a widely accepted standard is to bound the chance of coming to a false positive conclusion about a drug’s effects at less than 2.5\% \citep{food2017multiple}. In order to control this probability, a trial's design should be prospectively planned, including an analysis plan and statement of the hypotheses to be tested. When there are multiple hypotheses, the Family-Wise Error Rate (FWER), is the probability that one or more Type I Errors occur. The FDA standard thus implies a need for a FWER bound of 2.5\%.

Much work has gone into establishing Type I error (and FWER) control for adaptive and complex trial designs. In regulatory applications, the need for proof of Type I Error control often drives decisions for the trial design, analysis choice, or rules for adaptation. One commonly seen analytic approach is to use the approximate Gaussian or Brownian motion limits of estimators. For example, in group sequential settings, the central limit theorem can be applied to sufficient statistics for each group of patients between each interim analysis. For many adaptive trials, the Type I Error for the design can be evaluated at the borderline null hypothesis of $0$ treatment effect; type I error can be calculated with numerical integration using the Gaussian distribution, or with Monte Carlo methods.

In previous work on complex trials with multiple arms or subgroups, designs may be carefully constructed so that only a finite number of points require study - often, only the global null. A prominent example of this approach is the DEFUSE III trial, which uses the design of \cite{lai2014adaptive}. Yet, practitioners desire to use, and do use, designs where the Type I Error is not maximized at the global null. For example, a multi-arm Bayesian design which is tuned for FWER at level $\alpha$ under the global null, will usually have an inflated FWER $> \alpha$ when some treatment arms are null some have a strong positive effect.

\cite{ventz2015bayesian} discuss the problem of potential Type I error in inflation Bayesian designs. Their recommendation is to ask the regulator for a finite set of constraints, such as specific points in the null hypothesis space under which Type I Error can be estimated and controlled. This approach faces two key challenges: first, the dialogue process with regulators is costly for all parties. Second, it is not made clear how the regulator should decide on these constraints. It is difficult to have confidence that a small set of checks will be a sufficient guardrail for the overall space, when any sufficiently flexible optimization will, more or less by design, attempt to move the Type I Error to regions where it is less constrained. 

For multi-arm and adaptive trials, alternative analytic techniques exist, although they tend to be conservative. For example, the certainty equivalent technique of \cite{graves1996comparison} can be used, though it pays a large penalty for conservativeness. The multi-arm bandit literature contains various forms of concentration-style bounds, which we have discussed in Section 3.2.1. These approaches are typically loose by a constant or logarithmic factor of sample size efficiency.

Hybrid resampling, suggested by Lai and Bartroff, is inspired by the bootstrap, and has flavors of both analytic and simulation techniques. \cite{lai2006confidence} show that hybrid resampling can give efficient and accurate inference for a single parameter and demonstrate its use on Cox regression. However, hybrid resampling can be conservative in the presence of nuisance parameters. \cite{bartroff2013sequential} proposes to grid over the nuisance parameter to yield confidence bounds, but does not offer verification in between grid points. Thus, Type I Error control in complex trials remains a relevant challenge, and particularly so for designers aiming to pass muster of FDA pivotal trial requirements.

 
For estimating Type I Error with simulations, we will address two questions. First, how to show that our attention can be safely limited to a bounded region of the parameter space? We ultimately leave this question open, but suggest potential approaches in Section 5.3. Second: how to ensure that simulations have been performed on a sufficiently fine grid? This second question is our main target in Section 5.5. Our solution will apply to designs of nearly arbitrary complexity, as long as the data can be well-modeled and simulated at scale across a fine grid of the null hypothesis space, and a finite upper bound on the total sample size in each arm is available. In the next section, we introduce mathematical background for exponential family model, which is required for compatibility with our approach.


\section{Exponential Families}


Exponential families are a broad and powerful class of statistical models; they including most of distributions that have a commonly recognized name: Bernoulli, Binomial, Exponential, Gamma, Normal, Poisson, Negative Binomial, and some Weibull models. In Section 5.2.1 we define exponential families and introduce basic properties. In Section 5.2.2, we show how exponential family processes can be used to underpin simulations of adaptive trials.
In Section 5.2.3, we derive properties that will be used in later sections.

\subsection{Introduction to Exponential Families}

A family of distributions parametrized by an unknown parameter vector $\theta$ is an exponential family if the likelihood of the observed data $x$ given the parameter $\theta$ can be expressed as

$$f_{\theta}(x) = h(x)e^{\eta (\theta) \cdot T(x) - A(\theta)}$$

For example, a the family of Gaussian distributions with known variance equal to $1$ is an exponential family with a 1-dimensional unknown parameter vector $\mu$. The likelihood is

$$\frac{1}{\sqrt{2 \pi}} e^{-\frac{1}{2}(x^2 - 2 x \mu + \mu^2))} $$

which can be identified as an exponential family form with the substitutions

$$\eta(\mu) = \mu$$

$$T(x) = x$$

$$A(\theta) = \mu^2/2$$

$$h(x) = \frac{1}{\sqrt{2 \pi}}e^{-x^2/2}$$

The binomial distribution is also an exponential family. Its likelihood is

$${n \choose x} p^{x}(1 - p)^{n-x}$$

which can be put into exponential family form, via the substitutions:

$$\eta(p) = \log \frac{p}{1 - p}$$

$$T(x) = x$$

$$h(x) = {n \choose x}$$

$$A(p) = - n \log(1 - p)$$

We say that $\eta(\theta)$ is the natural parameter of the exponential family; one may transform the unknown parameter into $\theta' = \eta(\theta)$, in which case the likelihood may be written as

$$f_{\theta}(x) = h(x)e^{\theta' \cdot T(x) - A(\theta')}$$

This is called the canonical form of the exponential family.

Exponential families have many highly desirable properties for representing data outcomes of clinical trial patients. If we observe n independent outcomes from the same exponential family model, the resulting data remains in exponential family form, with likelihood:

\begin{equation} \label{combinedata} \tag{5.1}
    f_{\theta}(x_1, ... ,x_n) = \left[\prod \limits_{i = 1}^n h(x_i) \right] e^{ \eta(\theta) \cdot \left[ \sum \limits_{i = 1}^n  T(x_i) \right] - n A'(\theta)}
\end{equation}

We observe that $\sum \limits_{i = 1}^n  T(x_i)$ is a sufficient statistic for $\theta$.

We can also merge exponential families from separate data models together, yielding a joint exponential family in higher dimension. If we have $x_1$ from the exponential family model $\theta_1$ and $x_2$ from the exponential family model $\theta_2$, then we may join these into a new exponential family model for data $x' = (x_1, x_2)$ with unknown parameter vector $\theta' = (\theta_1, \theta_2)$ where

$$\eta(\theta') =
(\eta(x_1),\eta(x_2))$$

$$T(x') = (T(x_1),T(x_2)$$

$$A(\theta') = A(\theta_1) + A(\theta_2)$$

$$h(x') = h(x_1)h(x_2)$$

and similarly for more than $2$ models. Thus, if we have a separate exponential family for each arm of a trial, we may collect all of the observed data likelihood into a single exponential family representation.

\subsection{Describing Clinical Trials with Exponential Families}

For a clinical trial with pre-determined sample size where the outcomes are independent for each patient and follow an exponential family model, it is straightforward to combine these the data with the technique shown in \ref{combinedata}. However, adaptive clinical trial designs may, as data are obtained throughout the trial, make changes to the enrollment, halt the trial, or differently allocate patients between arms. We can conveniently use exponential family processes to handle many of these common types of decision-making. We will build up the model from the multi-arm bandit problem, as framed in \cite{lai1985asymptotically}.

Assume we have $K$ trial arms, with outcomes $x_{ik}$ corresponding to the $i$'th patient of arm $k$. Assume that for each $k$, $x_{ik}$ correspond to an i.i.d. draws from an exponential family with parameter $\theta_k$. We may start with an empty dataset, with likelihood 1 at time $t = 0$. At each timestep $t = 1, 2, ...$, we may either stop the trial according to a stopping time $\tau$, or select an arm from which to draw the next patient. That patient's outcome is drawn from their exponential family distribution, and added to the dataset. The decision rule can be random, with arm selection probability $p_{k,t}$. But, crucially, it must depend only on the so-far observed data, and cannot depend on the unknown parameter vector $\theta$. 

This can be defined more formally with the notation of nested sigma-algebras, as given in \cite{lai1985asymptotically}. They define an {\it adaptive allocation rule} $\phi$ as a sequence of random variables $\phi_{1}, \phi_{2}, \ldots$ in the set $\{1, ..., k \}$ such that the event $\{ \phi_{t} = k \}$ ("sample from arm k at step t") belongs to the $\sigma$-field $\mathcal{F}_{n-1}$ generated by the previous values $(\phi_1, x_1, \phi_2, x_2, \phi_{t-1}, x_{t-1})$. Since it is desirable to express randomness in the allocation rule, we may either achieve this by an appropriate random selection among deterministic allocation rules, or instead embed $\phi_t$ as a categorical draw from $\{1, \ldots, k \}$ with a probability vector $(p_1, \ldots , p_k)$ which lies in $\mathcal{F}_{n-1}$. Either way, the data likelihood at a stopping time $\tau$ has the exponential family form

\begin{equation} \label{adaptivelikelihood}\tag{5.2}
\left[ \prod \limits_{t = 1}^\tau p_{k^*_t,t} \right] \left[ \prod \limits _{k = 1}^{K} \prod \limits_{j = 1}^{n_k} h_k(x_j) \right] e^{ \left[ \sum \limits_{k=1}^K \left( \eta(\theta_k) \cdot \sum \limits_{j = 1}^{n_k}T_k(x_i) \right) \right] - \sum \limits_{k=1}^K n_{\tau,k} A(\theta_k)}
\end{equation}
where $k^*_t$ is the arm actually chosen at time $t$, and $n_{\tau,k}$ is the cumulative number of samples from arm $k$ at time $\tau$. 
Many common actions and strategies for clinical trial decision-making can be formulated within this probabilistic framework, including:

\begin{itemize}
    \item Rejections of hypotheses $H_m$ can correspond to stopping times $\tau_m$. We shall say hypothesis $H_m$ is rejected by time $\tau$ if $\tau_m \leq \tau$.
    \item Adaptive stopping of the trial, corresponding to a stopping time $\tau$ which is not constant
    \item Dropping an arm $k$ at time $\tau_d$, corresponding to a $\phi_t$ which allocates no chance to selecting arm $k$ for $t > \tau_d$
    \item Group sequential trials, corresponding to fixing probabilities in the arm decision rule $\phi$ at an interim $\tau_I$ and determining the next stopping rule for the subsequent interim.
    \item Adaptive randomization, e.g. Thompson sampling, corresponding to a certain choice of probabilities in the decision rule $\phi$
    \item Partial observation of survival outcomes due to independent censoring, lags before outcomes are observed, or random accrual times, corresponding to requiring decision rule probabilities $p_{k_t}$ and stopping times to be independent of the information that ``should be hidden" by censoring. In this case, a slight de-randomization of our method in Sections 5.5 is possible, if instead of the stopped exponential family likelihood one uses the appropriately censored likelihood.
\end{itemize}

To simulate the trial observation process, there are two ways, equivalent from a mathematical point of view: one may either take independent observations generated in sequence, or a sequence of pre-generated independent observations whose values are revealed in a monotonic fashion by the decision rule.

It bears mentioning that this exponential family approach can be extended to continuous time and Brownian motion, which has exponential family properties \citep{kuchler1989exponential}. Brownian motion is a common asymptotic limit for the evolution of model fit parameters throughout a trial, such as for Cox regression under suitable proportional hazard assumptions. Thus, the tools in this work can also be used to understand the asymptotic behavior of many models beyond exponential families. But we note that in these cases, additional questions may arise, such as whether the approximation is accurate in finite sample sizes, or whether model-misspecification such as non-proportional hazards in Cox regression is a possibility. In this case, we suggest the possibility of running simulations on both the Brownian limit model, to which the approaches in this work can be applied, and on other realistic data-generating models (which may or may not be exponential families to which our results apply) in an attempt to verify the accuracy and robustness of the approximation.

\subsection{Bounds for Exponential Family Processes}

In this subsection, we establish bounds for the processes described in section 5.2.2, when we have an upper bound maximum sample sizes given by $\tau_{max}$. That is, for trial stopping times $\tau$, we have

$$0 \leq n_{\tau,k} \leq n_{\tau_{max,k}}$$ 

We may introduce the notation $A_{\tau}$ in accordance with the exponent in equation (\ref{adaptivelikelihood}) to be

$$ A_{\tau}(\theta) := \sum \limits_{k = 1}^K n_{\tau,k} A(\theta_k)$$

And similarly define 

$$ A_{\tau_{max}}(\theta) := \sum \limits_{k = 1}^K n_{\tau_{max},k} A(\theta_k)$$

Then, we must have

\begin{equation} 
\label{hessian_increases} \tag{5.3}
0 \preccurlyeq \nabla^2 A_{\tau}(\theta) \preccurlyeq \nabla^2 A_{\tau_{max}}(\theta) 
\end{equation}

This equation follows directly from the fact that 

$$0 \preccurlyeq \nabla^2 A(\theta_k),$$

which is a consequence of the covariance identity for exponential families,

$$ \nabla^2 A(\theta_k) = Var(T_k(x_k))$$

Next, we shall prove an inequality that bounds the Hessian of the probability of sets. That is, for an exponential family process in canonical parametrization with parameter, if we consider a function $f$ which is of the form

$$f(\theta) = \E_{\theta} [\mathbbm{1}_{X_{\tau} \in C}]$$

For some set $C$ in $\mathcal{F}_{\tau}$, we prove the following bounds on the Hessian of $f$:
\begin{equation}\label{covineq}\tag{5.4}
-Cov_{\theta}(T(X_{\tau_{max}})) \preccurlyeq \nabla^2 f \preccurlyeq Cov_{\theta}(T(X_{\tau_{max}}))
\end{equation}

Within the natural parameter space,
the exponential family process likelihood has derivatives of all orders and is sufficiently well-behaved that we can interchange integration and derivation. 
Thus,
\begin{equation} \label{setinfo}\tag{5.5}
\nabla^2 f(\theta) = \nabla^2 E_{\theta} [\mathbbm{1}_{X_{\tau} \in C}] = \nabla^2 \int \limits_C P_{\theta}(X) dX =  \int \limits_C \nabla^2 P_{\theta}(X) dX
\end{equation}

We may think of this equation as the curvature or statistical information of the set $C$ with respect to $\theta$. Let us evaluate it, by taking two gradients of the exponential family density with respect to $\theta$: the first derivative yields

$$\nabla P_{\theta}(X) = (T(X) - \nabla A_{\tau}(\theta) )  P_{\theta} (X).$$

A further derivative yields

\begin{equation}
\label{2deriv}\tag{5.6}
\nabla^2 P_{\theta}(X) = \left[ (T(X) - \nabla A_{\tau}(\theta))(T(X) - \nabla A_{\tau}(\theta))^T - \nabla^2 A_{\tau}(\theta) \right] P_\theta (X).
\end{equation}

By the first half of inequality (\ref{hessian_increases}), we may eliminate $\nabla^2 A_{\tau}$ from (\ref{2deriv}) to yield the semidefinite inequality:

$$\nabla^2 P_{\theta}(X) \preccurlyeq (T(X) - \nabla A_{\tau}(\theta)(T(X) - \nabla A_{\tau}(\theta))^T P_{\theta}(X)$$

Now, substituting this inequality into  (\ref{setinfo}) yields

$$\nabla^2 f(\theta) = \int \limits_F \nabla^2 P_{\theta}(X) dX \preccurlyeq 
 \int \limits_F
(T(X) - \nabla A_{\tau}(\theta)(T(X) - \nabla A_{\tau}(\theta))^T P_{\theta}(X) dX 
$$

Since the integrand is always positive definite, we have

$$  \int \limits_F
(T(X) - \nabla A_{\tau}(\theta)(T(X) - \nabla A_{\tau}(\theta))^T P_{\theta}(X) dX 
\preccurlyeq \int \limits_{\Omega}
(T(X) - \nabla A_{\tau}(\theta)(T(X) - \nabla A_{\tau}(\theta))^T P_{\theta}(X) dX$$

Next, consider the linearly evaluated score

$$M_t = v^T \left( T_t(X) - \nabla A_t(\theta) \right)$$

$M_t$ is a martingale in $t$. The trial will finish at a stopping time $\tau$, at which point

$$ \int \limits_{\Omega}
v^T (T(X) - \nabla A_{\tau}(\theta)(T(X) - \nabla A_{\tau}(\theta))^T  v P_{\theta}(X_{\tau}) dX_{\tau} = Var(M_\tau)$$

But one can also consider the hypothetical of sampling from the exponential family process until reaching sample sizes indicated by $\tau_{max}$. Doing so would yield

$$ Var(M_\tau) \leq Var(M_{\tau_{max}}) = \int \limits_{\Omega}
v^T (T(X_{\tau_{max}}) - \nabla A_{\tau_{max}}(\theta)(T(X_{\tau_{max}}) - \nabla A_{\tau_{max}}(\theta))^T v P_{\theta}(X_{\tau_{max}}) dX_{\tau_{max}}$$

where $X_{\tau_{max}}$ is a dataset consisting of $\tau_{max,k}$ i.i.d. samples from each arm $k$. Since $v \in R^d$ is arbitrary, by an application of the variance identity for exponential families we arrive at

$$ \nabla^2 f \preccurlyeq Cov_{\theta}(T(X_{\tau_{max}}))$$



which is half of equation (\ref{covineq}).
We can achieve the corresponding lower bound by removing the term

$$(T(X) - \nabla A_{\tau}(\theta)(T(X) - \nabla A_{\tau}(\theta))^T$$

in equation (\ref{2deriv}). We are left with

$$\nabla^2 P_{\theta}(X) \succcurlyeq -  \nabla^2 A_{\tau}(\theta) P_{\theta}(X) $$

And therefore

$$\nabla^2 f 
\succcurlyeq - \int \limits_F \nabla^2 A_{\tau}(\theta) P_{\theta}(X)dX 
\succcurlyeq - \int \limits_\Omega \nabla^2 A_{\tau}(\theta) P_{\theta}(X)dX$$

By (\ref{hessian_increases}), we have

$$\nabla^2 f  \succcurlyeq - \int \limits_\Omega \nabla^2 A_{\tau_{max}}(\theta) P_{\theta}(X_{\tau_{max}})]dX_{\tau_{max}} = -Cov_{\theta}(T(X_{\tau_{max}}))$$

Thus, combining these bounds, we have shown both parts of (\ref{covineq}).

\section{Hypotheses, Rejections, and Design Characteristics}

We will assume that when the trial stops, decisions for hypothesis rejection are made as a function of the data up to the time of stopping. 
The null hypotheses chosen will typically correspond to a region of the parameter space $\Theta_0 \subset \Theta$. As an toy example, we may consider a trial comparing two treatments $T_1$ and $T_2$ against control $C$, where the distribution of patient responses are i.i.d Gaussian with a common known variance $\sigma^2$. In this case, the unknown model parameter vector is $$\theta := (\mu_{1}, \mu_{2}, \mu_C).$$ We may have null hypotheses

$$H_1: \mu_{1} \leq \mu_{C}$$
$$H_2: \mu_{2} \leq \mu_{C}.$$

One might also wish to compare $\mu_1$ and $\mu_2$, in which case we could ask about both directional null hypotheses:

$$H_3: \mu_1 \leq \mu_2$$
$$H_4: \mu_1 \geq \mu_2.$$

In such a setup, it may be the case that design and rejection rule of the trial are constructed to be invariant to additive change in each mean parameter - for example, if decisions are functions of two-group t-statistics. In this case, we might focus our attention on a lower-dimensional slice of the space. One simple way to achieve this is to limit our view to the slice $$\Theta' = \Theta \cap \{\mu_C = c\}$$ for some constant $c$, since we are assured that our results can be extended back to $\Theta$ via translation. 

Another potential trick for dimension reduction is re-scaling by the true standard error in the Gaussian model - although for this to be viable, sample sizing (or information-stopping) decisions would have to be expressible in terms of normalized statistics. Or, some designs may have a monotonicity property of the Type I Error function which can allow attention to be restricted to the boundary of the null hypothesis space. (For example, the DEFUSE 3 trial mentioned in Section 2.2.2; or, a Bayesian-decision-theoretic design which takes $K$ independently run trials and makes rejections based on the posterior credible interval of a Bayesian multi-level model, where the multi-level model has a fixed parameter for between-groups variance/heterogeneity.)

Because our main tool in this chapter is simulation, we must further restrict our attention to a bounded region $\Theta_0 \in \Theta$. In a regulatory setting, we leave justification for this as an open question, but mention here a few possibilities:

\begin{itemize}
    \item Argument via scientific bounds on the effect sizes
    \item Limiting the area under study to a 99.9\% confidence region around prior estimates from a pilot study
    \item A fallback aspect of the procedure, designed to occur with high probability when $\theta \not\in \Theta_0$, and constructed so that other methods (either analytic or simulation) may be used to bound Type I Error over $\Theta \backslash \Theta_0$. One design idea for achieving this is to limit the impact of extreme parameters from outside of the simulation region by constraining trial decisions to depend only on $(g_1, \ldots, g_d)$ where
    $$g_i := \min \left( \max ((\hat{\theta}_i, l_i ), u_i \right),$$
    where $\hat{\theta}_i$ is an estimate of the parameter $\theta_i$, and $l_i$ and $u_i$ are lower and upper bounds. For instance, under sufficiently large or small values of $\theta_i$ the boundary may be hit with high probability by the time of the first interim analysis, triggering a simplification such as removing the offending arm from the trial. It would then suffice to prove that this occurs with probability at least $1 - \epsilon_i$, alongside an analysis that Type I Error for the remaining arms is thereafter bounded by $\alpha - \epsilon_i$.
    
    \item Very extreme parameter values may already change the nature of the experiment to such a degree that the statistical model is irrelevant outside of $\Theta_0$. If patient outcomes are sufficiently poor in a subgroup, perhaps the trial should be stopped altogether. Alternatively, a stunning success in some arm could precipitate a change to the standard of care, thus requiring a stop or crossover of the control arm, or might address the concerns which initially motivated the statistical process in the first place.
\end{itemize}

Consider any exhaustive selection of null and alternative hypotheses of the form $$B_{p_1, p_2, ... p_M} := \bigcap \limits_{ m = 1}^{M} H_{p_m} \bigcap \limits_{p_s \not \in \{p_m \}_{m=1}^M} H_{p_s}^C$$


As a consequence of boundedness of our set of attention $\Theta_0$, its intersection with $B_{p_1, p_2, ... p_M}$, which we shall denote
$$A_{p_1, \ldots , p_M} := B_{p_1, p_2, ... p_M} \cap \Theta_0,$$
can be successfully approximated by an $\epsilon$-net of $\Theta_0$. For our $\epsilon$-net, we will use a grid denoted $\{ \theta_j\}_{j =1}^J$.
The subset of grid points used to approximate $A_{p_1, \ldots , p_M}$ does not have to be contained within $A_{p_1, \ldots , p_M}$. Thus, we do not require any special features of the null hypothesis regions: merely that false rejection and family-wise error are indicator functions of the parameter $\theta$ and the data, so that Type I Error rate and FWER are of the form

$$E_\theta[\mathbbm{1}_{X_\tau}\in F_{A_{p_1, \ldots , p_M}}]$$

Where the set $F_{A_{p_1, \ldots , p_M}}$ corresponds to the occurrence of a false rejection when the true parameter $\theta$ is in $A_{p_1, \ldots , p_M}$. Consequently, the techniques which follow can also be made to apply to other expectations of similar indicator functions of the data and parameters, such as power. In fact, the results can also be further extended to bounded metrics, such as the FDR.

Because we use Monte Carlo simulation, the guarantees that we give will be probabilistic in nature. Our method will accept an adaptive allocation rule for the trial design, an exponential family process model for the data with parameter $\theta \in \Theta$, a confidence parameter $\delta$, and a parameter region $\Theta_0 \subset \Theta$. We wish to bound an error metric $f(\theta)$. Our method will return an upper bound function $g(\theta)$ with a probabilistic guarantee of the form

$$\forall \theta \in \Theta_0, P(f(\theta) > g(\theta)) \leq \delta$$

Note that this statement is a pointwise guarantee, not uniform, over $\Theta_0$. For purposes of medical regulation, this is perfectly acceptable; there is only one treatment effect vector, and nature will not change it in response to our computer simulation. (However, to avoid risk of gaming the process by re-running the simulation, it may be necessary for simulation RNG seeds to be selected by a regulator or 3rd party after the design has been fixed). Section 5.7.2 will discuss a calibration method to help achieve pre-specified upper bounds on $g(\theta)$, such as a FWER bound of 2.5\%, so that a probabilistic chance of failing to meet the bound can be avoided.

\section{Type I Error Bounds from Monte Carlo Simulation}

Monte Carlo simulation is a robust tool for evaluating Type I Error when analytic tools are not available. Consider a fixed parameter value $\theta \in \Theta$. We will perform independent simulation replications $i = 1, \ldots, I$ (for example $I = 100,000$). Let $F_i(\Theta)$ be the event that a Type I (Family-Wise) Error occurs. The Monte Carlo estimator of Type I (Family-Wise) Error Rate is then

$$ \frac{\sum \limits_{i = 1}^N \mathbbm{1}_{F_i(\theta)}}{N}$$

This is an unbiased estimator of $f(\theta)$. Furthermore, $\sum \limits_{i = 1}^N \mathbbm{1}_{F_i(\theta)}$ follows a binomial distribution with probability $f(\theta)$. For approximate accuracy, one may build a $1 - \delta$ confidence interval for $f(\theta)$ by using the standard normal approximation to the binomial. [This approximation may be sufficient in most non-regulatory applications; the error is quite small as the number of simulations increases.] Or, for a rigorous and conservative guarantee, the upper Clopper-Pearson interval can be used.











\subsubsection{Identities for Fast Numerical Simulation}

Our methods in Section 5.5 will demand rapid and large-scale simulation. We have found it advantageous to re-use random number generation when possible. Here, we list some identities that may be useful for speeding up computation. These have the added benefit inducing correlation between the simulations for nearby values of $\theta$.

\begin{itemize}
    \item  To simulate a vector of Gaussians with parameters $(\mu_j, \sigma^2_j)$, one may draw

$$Z \sim N(0,1)$$

and compute with vectorized addition and multiplication,

$$\mu + \sigma Z.$$

\item To simulate a vector of Bernoulli random variables with parameters $p_j$, one may draw

$$u_j \sim U[0,1]$$

and compute

$p_j < u_j$

\item To draw from gamma distributions with parameters $a_j, \lambda = 1$, where $a_j$ is in ascending order, one may simulate independent gamma R.V's with parameters $a_j - a_{j-1}$ (with $a_0 = 0$) and take a cumulative sum.

\item To draw from beta distributions with parameters $\alpha_j, \beta_j$, one may use simulations of 

$$\beta(\alpha_j, \beta_j) \sim \frac{\Gamma(\alpha_j)}{\Gamma(\alpha_j) + \Gamma(\beta_j)}$$

\end{itemize}

These and similar identities can be especially useful for Bayesian methods with conjugate priors.

\section{Extending Monte Carlo Simulation to Continuous Space}

We will use Taylor's Theorem to extend the Monte Carlo simulations of the previous section to nearby continuous regions of space. In section 5.3, we discussed that hypothesis intersection sets $A_{p_1, p_2, ... p_M}$ can be $\epsilon$-covered with a collection $\{\theta\}_{j=1}^J$. For computational simplicity, we will now assume that $\{\theta\}_{j=1}^J$ is a grid, and that for each $A_{p_1, p_2, ... p_M}$ we may cover it with a collection of small rectangles $\{R_j\}$ centered at $\{\theta_j\}$. We aim to compute a $1 - \delta$ confidence bound for the Type I Error valid for each $R_j$. Let $U_p$ correspond to the event that hypothesis $p$ is rejected. Then, for our dataset $X_i$ when the trial stops, we may write the false rejection function $F(\theta,X_i)$ as

$$F_i(\theta,X_i) = \mathbbm{1}\{ \cup_p \left[ \{X_i \in U_p \} \cap \{\theta \in H_p\}\right]\}$$

and the Type I Error rate function as

$$f(\theta) = \E[F(\theta, X_i)]$$

\subsection{A Taylor Expansion}

Because as mentioned in Section 5.2.2 the likelihood of $X_i$ has derivatives of all orders in $\theta$, $f$ has derivatives of all orders in $\theta$ in the interior of $A_{p_1, p_2, ... p_M}$ (where the null status of the hypotheses remains constant). If $\theta \in R_j \subset A_{p_1, p_2, ... p_M}$, then the line connecting $\theta$ and $\theta_j$ is contained in $A_{p_1, p_2, ... p_M}$, so taking a Taylor expansion of $f$ in remainder form yields

\begin{equation}\label{taylor}\tag{5.7}
f(\theta) = f(\theta_j) + \nabla f(\theta_j)^T (\theta - \theta_j) + \int \limits_{a = 0}^{1} (1 - \alpha) S(\theta_j + \alpha v) d\alpha
\end{equation}

where 

$$v := \theta - \theta_j$$

$$S(\theta) := v^T \nabla^2 f(\theta) v$$

[If $R_j$ is not fully contained within a single $A_{p_1, p_2, ... p_M}$ but hits multiple hypothesis intersection sets, then a similar Taylor expansion holds for extensions of the Type I Error function from each set $A_{p_1, p_2, ... p_M}$ defined by

$$f^{A_{p_1, p_2, ... p_M}}(\theta') := \E_{\theta'} \left[ \mathbbm{1}\left\{ \bigcup \limits_{p \in \{p_1, \ldots, p_M\}}  \{X_i \in U_p \} 
\right\} \right]$$

so that $f^{A_{p_1, p_2, ... p_M}}$ agrees with $f$ on the set $A_{p_1, p_2, ... p_M}$. Therefore, to achieve our confidence interval for $R_j$, we can build confidence intervals for each $f^{A_{p_1, p_2, ... p_M}}$ such that ${A_{p_1, p_2, ... p_M}}$ intersects $R_j$, and then report the union of these confidence intervals (which will be the maximum of our upper bounds for $R_j$). To simplify notation and intuition, we will continue to refer to $f$, but intend $f^{A_{p_1, p_2, ... p_M}}$ in the case that $f$ has a discontinuity within $R_j$. Note that using a choice of rectangles $R_j$ that aligns with the (typically rectangular) boundaries of the intersection null hypothesis sets can essentially avoid this issue entirely.]

To construct a confidence interval for $f$ (or $f^{A_{p_1, p_2, ... p_M}}$), we can conservatively estimate the right hand side of equation (\ref{taylor}) over the rectangle $R_j$, with an upper bound that is valid with probability at least $1 - \delta$. The degree of conservativeness in our estimate will depend on both the number of replications performed, and the fineness of the grid; the larger the dimensions of $R_j$, the more loss we will incur from Taylor expansion. We shall separate and maximize over $R_j$ the three terms on the right hand side of equation (\ref{taylor}) to consider the following quantities: 
$$\delta_I := f(\theta_j)$$

$$\delta_{II} := \sup \limits_{\theta \in R_j} \nabla f(\theta_j)^T (\theta - \theta_j)$$ 

$$\delta_{III} := \sup \limits_{\theta \in R_j} \int \limits_{a = 0}^{1} (1 - \alpha) S(\theta_j + \alpha v) d\alpha$$

In Section 5.4, we showed how it is possible to upper bound $\delta_1$ via Monte Carlo with a $1 - \delta/2$ confidence interval. With $n_j$ replications, the confidence width shrinks around the true value of $f$ at a rate proportional to $n_j^{-1/2}$.
In sections 5.5.2, we will show how to construct a $1 - \delta/2$ upper bound on $\delta_{II}$. As one shrinks the grid distances at rate proportional to $\epsilon \to 0$ in every coordinate and increases $n_j$, this term decreases at the rate $\epsilon n_j^{-1/2}$. 
In section 5.5.3, we discuss approaches for bounding $\delta_{III}$. As one shrinks the grid distances at rate proportional to $\epsilon \to 0$ in every coordinate, this bound decreases at the rate $\epsilon^2$. 

The upper bound for each $R_j$ is then expressed as a sum of $\delta_{I} + \delta_{II} + \delta_{III}$. Finally, we may stitch together our confidence upper bounds for each $R_j$ to yield an upper bound function $g$ defined across all of $\theta_0$.

\subsection{Upper Bounds on $\delta_{II}$}

\subsubsection{Unbiased Estimation of $\nabla f$}

 Let the set $F^{A_{p_1, p_2, ... p_M}}(X)$, correspond to a false rejection under data $X$ as if the true intersection null hypothesis were $A_{p_1, p_2, ... p_M}$; as in the previous section, we will suppress its notation and denote it as $F$. Then, given a simulation from true parameter $\theta_0$ we could generally say.

$$\nabla f(\cdot)= \nabla P_{(\cdot)}(F) = \nabla \int \limits_F \left( \frac{dP_{(\cdot)}}{dP_{\theta_0}} (X)\right) dP_{\theta_0}(X) = \int \limits_F \left( \nabla \frac{dP_{(\cdot)}}{dP_{\theta_0}} (X) \right) dP_{\theta_0}(X)$$

The gradient passes through for likelihoods which are smooth in $\theta$ and 
have sufficient tail decay in $X$. 
With $\theta_0 \neq \theta_j$, this representation generally yields an importance sampling estimator - an approach we leave to future work. But, for simplicity, a natural and easy choice is to use only simulations where $\theta_0= \theta_j$, so that we are simply estimating the gradient at $\theta_j$ using Monte Carlo. 
Since $P_{\theta_0}$ is a probability distribution, this representation leads directly to an unbiased estimate of $\nabla f$. 

For each $l = 1,...,L$, set $y_l$ to be 0 if Monte Carlo sample $l$ has no false rejections, and $y_l = \widehat{\nabla}f_{\theta_j}(X_l) = \nabla_{\theta} \frac{dP_{\theta}}{dP_{\theta_j}} |_{\theta = \theta_j}$ 
if there is a false rejection. Then our natural Monte Carlo estimate is

$$\widehat{\nabla} f_{\theta_j} = \frac{1}{L}\sum \limits_{l=1}^L y_l = \frac{1}{L}\sum \limits_{l=1}^L \nabla_{\theta} \frac{dP_{\theta}}{dP_{\theta_j}} |_{\theta = \theta_j}(X_l) F_l$$ 

Where $F_l$ is an indicator for false rejection of simulation $l$. This estimate is very similar in form to the estimate in Section 5.4, and is asymptotically multivariate normal. However, handling the supremum in term $\delta_{II}$ will require slightly more work.

\subsubsection{Confidence Intervals for $\delta_{II}$ in Exponential Families with Monte Carlo}

We first compute the unbiased Monte Carlo score estimator for canonical exponential families. The likelihood of observing a trial run-out stopping at time $\tau$ with dataset $X$ and parameter $\theta$ is:

$$ \left[\prod \limits_{t = 1}^\tau p^*_t \right] \left[\prod \limits_{t = 1}^\tau h_t(x_t) \right] e^{ \theta^T T(X) - A_\tau(\theta)}$$

where $T(X) = \sum \limits_{t = 1}^\tau  T_t(x_i)$ 
and $A_\tau = \sum \limits_{t = 1}^\tau  A_t(\theta)$

We shall denote the occurrence of a Type I Error with the indicator function $\mathbbm{1}_{X \in F}$.

Thus, the Type I Error Rate at $\theta$ is

$$f(\theta) = E_{p(x,\theta)}[\mathbbm{1}_{X \in F}] = \int \mathbbm{1}_{X \in F} 
\left[\prod \limits_{t = 1}^\tau p^*_t \right] \left[\prod \limits_{t = 1}^\tau h_t(x_t) \right]
e^{ \theta^T T(X) - A_\tau(\theta)} dX$$


Taking a derivative with respect to $\theta$,

$$ \nabla_{\theta} f(\theta) = \nabla_{\theta} \int \mathbbm{1}_{X \in F} \left[\prod \limits_{t = 1}^\tau p^*_t \right] \left[\prod \limits_{t = 1}^\tau h_t(x_t) \right] e^{ \theta^T T(X) - A_\tau(\theta)} dX$$

Under sufficient regularity, which holds for the exponential families we consider, the derivative and integral can be interchanged, yielding:

$$= \int \mathbbm{1}_{X \in F} \left[\prod \limits_{t = 1}^\tau p^*_t \right] \left[\prod \limits_{t = 1}^\tau h_t(x_t) \right] \nabla_{\theta} e^{ \theta^T T(X) - A_\tau(\theta)} dX$$

$$= \int  \left[\prod \limits_{t = 1}^\tau p^*_t \right] \left[\prod \limits_{t = 1}^\tau h_t(x_t) \right] e^{ \theta ^T T(X) - A_\tau(\theta)} \mathbbm{1}_{X \in F} \left( T(X)  - \nabla_{\theta} A_\tau(\theta) \right) dX$$

Note that this differs from an integral of the sampling likelihood, by a factor of $\mathbbm{1}_{X \in F} \left( T(X)  - \nabla_{\theta} A_\tau(\theta) \right)$.
Thus, similar to the previous Monte Carlo estimate in section 5.4, an unbiased estimator of $\nabla f$ is

$$\widehat{\nabla f(\theta_j)} = \frac{1}{n_j} \sum \limits_{i = 1}^{n_j} \widehat{\nabla f(\theta_j)}_i = \frac{1}{n_j} \sum \limits_{i = 1}^{n_j} \mathbbm{1}_{X_i \in F} \left( T_{\tau_i}(X_i) - \nabla_{\theta_j} A_{\tau_{i} }(\theta_j) \right) $$

where $X_i$ are i.i.d. Monte Carlo samples from the trial simulation under $\theta_j$, and $\tau_i$ are the relevant stopping times. 




\subsubsection{A Variance Bound on the Gradient Estimate}

We can also compute a bound on $Var(\widehat{\nabla f})$. Since if $d >1$ the variance matrix is of dimension $d \times d$, we shall use semidefinite matrix comparisons. We define for matrices $B$ and $C$ of dimension $m \times m$ by the relation

$$B \preccurlyeq C \iff \forall v \in R^{m}, v^T B v \leq v^T C v $$

Or, equivalently, if $C - B$ is positive semidefinite. 

Because the Monte Carlo estimate of $\widehat{\nabla f}$ is composed of i.i.d. unbiased summands, for $v \in R^d$ we have

$$ Var(v^T \widehat{\nabla f}) = \frac{1}{N} Var(v^T \widehat{\nabla f}_i)$$

Now, we may write each unbiased summand as follows:

$$v^T \widehat{\nabla f}_i = F_i v^T( T(X_i) - \nabla A_{\tau_i}(\theta_j))$$

Compare this to the following linear evaluation of the score function:

$$M_t := v^T( T(X_{t}) - \nabla A_{t}(\theta_j))$$

and note that

$$ v^T \widehat{\nabla f}_i = M_{\tau_i}F_i$$

$M_t$ is linear evaluation of the score function of the exponential family, and is therefore a martingale under $\theta_j$. In consequence, the stopped random variable $M_\tau$ has expected value $0$. Therefore,

$$Var(v^T \widehat{\nabla f}_i) = \inf \limits_{u \in R} E[(v^T \widehat{\nabla f}_i - u)^2] \leq E[ (v^T\widehat{\nabla f}_i)^2]$$ 

$$= E[(M_{\tau_i}F_i)^2] \leq E[M_{\tau_i}^2] = Var(M_{\tau}).$$

Because $M_t$ is a martingale, the variance of $M_\tau$ is upper bounded by the variance at the time upper-bound, that is,

$$Var(v^T(
T(X_{T_{max}}) - \nabla A_{T_{max}}(\theta_j))$$

which, noting that the $\nabla A$ term is constant, is equal to

$$v^T Cov(T(X_{T_{max}})) v = v^T \nabla^2 A_{\tau_{max}}(\theta_j) v.$$

Hence, we arrive at the upper bound,

$$ Var(v^T \widehat{\nabla f}_i) \leq v^T \nabla^2 A_{\tau_{max}}(\theta_j) v$$

Which, because $v^T$ was chosen arbitrarily, implies

$$ Cov(\widehat{\nabla f}_i) \preccurlyeq  \nabla^2 A_{\tau_{max}}(\theta_j)$$


And thus,

$$ Cov(\widehat{\nabla f}) \preccurlyeq \frac{1}{N} \nabla^2 A_{\tau_{max}}(\theta_j)$$

The existence of this variance bound and i.i.d. sampling suggests the possibility of using a central limit theorem. But to avoid any discussion about the rate of distributional convergence, in the next section we will take a simpler approach to constructing confidence bounds, using Cantelli's Inequality which requires only a variance bound.

\subsubsection{A Cantelli Inequality to Upper Bound $\nabla f$}

There is an extra difficulty in creating a confidence bound for $\delta_{II}$ as opposed to $\delta_{I}$. $\nabla f(\theta_j)$ is a multi-dimensional quantity, and our desire is to bound
$$\sup \limits_{\theta \in R_{j}} \nabla f(\theta_j)^T (\theta - \theta_j)$$

A hyper-rectangle $R_j$ in dimension $d$ has $2^d$ choices of corner points. We shall index them as $\theta_{m}$ for $m \in 1,...2^d$ and consider $v_m = \theta_m - \theta_j$. For each $v_m$, we will construct a separate 1- $\delta/2$ confidence upper bound $c_m$, so that

$$P\left(v_m^T \nabla f(\theta_j) \leq c_m \right) \geq 1 - \delta/2.$$ 

Then, because the supremum of a linear function over a hyper-rectangle is achieved at a corner, taking an union of these confidence sets yields a conservative confidence bound over $R_j$. Thus, we have

$$P \left( \sup \limits_{v \in R_j} v^T \nabla f(\theta_j) \leq \max \limits_m c_m  \right) \geq 1 - \delta/2$$

Which is of the desired form. We show how to determine $c_m$ in the next subsection.


\subsubsection{Construction of $c_m$ with Cantelli's Inequality}

Cantelli's inequality, for a R.V. $Y$ with variance $\sigma^2$ and any positive number $\lambda > 0$, states:

$$P\left( Y + \lambda \leq E[Y] \right) \leq \frac{\sigma^2}{\sigma^2 + \lambda^2} $$

We shall take the random variable $Y = v_m^T \widehat{\nabla f(\theta_j)} = \frac{1}{N} \sum v_m^T \widehat{\nabla f(\theta_j)}_i$, for which we have an variance upper bound from Section 5.5.3. We may combine this with knowledge of maximum samples size per arm $T_{max}$, yielding

$$\sigma^2 \leq \frac{1}{n_j} v_m^T \nabla^2_{\theta_j} A_{\tau_max}(\theta_j) v = \frac{1}{n_j} v_m^T Cov_{\theta_j}(T (X_{\tau_{max}})) v_m^T$$ 

To ensure a confidence of $1 - \delta/2$, we seek $\lambda$ such that

$$ \delta/2 = \frac{\sigma^2}{\sigma^2 + \lambda^2} = \frac{v_m^T Cov(T(X_{\tau_{max}})) v_m}{v_m^T Cov_{\theta_j}(T(X_{\tau_{max}})) v_m + \lambda^2 n_j}$$

Solving for $\lambda$ yields

$$\lambda := \sqrt{\frac{v_m^T Cov_{\theta_j}(T(X_{\tau_{max}})) v_m}{n_j} \left(\frac{1}{\delta/2} -1\right)} = \sqrt{\frac{v_m^T \nabla^2_{\theta_j} A(\theta_j) v_m}{n_j} \left(\frac{1}{\delta/2} -1\right)}$$
  
Then, our upper confidence bound is simply $$c_m = Y + \lambda = \frac{1}{N}\sum v_m^T \widehat{\nabla f(\theta_j)}_i + \sqrt{\frac{v_m^T \nabla^2_{\theta_j} A(\theta_j) v_m}{n_j} \left(\frac{1}{\delta/2} -1\right)}.$$

In some problems, distributional information can offer a deterministic upper bounds $d_m$. For example, a multivariate Gaussian distribution with covariance matrix equality to $I_{d \times d}$ has 

$$\|\nabla f\|_1 \leq \frac{d}{\sqrt{2\pi}}.$$

If one can generate a deterministic upper bound 

$$v_m^T \nabla f(\theta_j) \leq d_m $$

one may use it to refine the bound, to 

$$c'_m = \min (c_m, d_m)$$

which may help for small Monte Carlo replications $n_j$.

\subsection{A Bound for $\delta_{III}$}

We will attempt to bound the final term $\delta_{III}$ by bounding

$$\delta_{III} \leq \sup \limits_{\theta \in R_j} \int \limits_{a = 0}^{1} (1 - \alpha) v^T \nabla^2 f(\theta_j + \alpha v ) v d\alpha$$

where $$v = \theta - \theta_j.$$

We can achieve this by use of equation (\ref{covineq}). Note that (\ref{covineq}) is very similar to the bound in Section 5.5.3. Besides its simplicity and form, there is a striking feature to note: we are being charged no penalty for stopping the trial in a complex fashion, as compared to a trial that samples all the way forward to $\tau_{max}$. As $\tau_{max}$ upper bounds the information available in the trial, it is effectively a general constraint on the curvature of the Type I Error function even under adaptive stopping.

Let us consider the case of a multi-arm trial where each arm has Gaussian outcomes with known variance $\sigma^2_i$. Conveniently, after we have chosen a sampling upper bound, $\tau_{max}$, the bound we get for $\delta_{III}$ {\it does not vary}, regardless of $\theta$ or the trial stopping rule. The matrix bound is

$$\nabla^2 f \preccurlyeq diag(\sigma^{-2}_k \tau_{max,k})$$ 

This leads to the straightforward quadratic bound on $\delta_{III}$:

$$\delta_{III} \leq \sup \limits_{v + \theta_j \in R_j} \frac{1}{2} v^T diag(\sigma_k^{-2} \tau_{max,k}) v.$$

If $R_j$ is a (hyper-)rectangle centered at $\theta_j$, the sup is achieved by taking $v$ to be any corner point.
For other exponential family distributions besides the Gaussian, the matrix bound is typically possible to maximize conservatively on a case-by-case basis. For example, in a 1-dimensional binomial model as discussed in section 5.1.1, $T(x) = x$, 

\begin{equation}
    Cov_{\theta}(T(X_{\tau_{max}})) = \tau_{max} p(1 - p) \leq \tau_{max} / 4 \tag{5.8}\label{binomvar}
\end{equation}

and so we may also use a constant upper bound, similar to the Gaussian. Or, we could use the corners of each $R_j$ to inform a tighter bound. We emphasize that for these results to apply, it is necessary for the exponential family to be in canonical form; and thus for the binomial this Hessian bound applies in the transformed space of $\log(p/1 - p)$, not the original space of $p$. This necessity becomes especially clear when one considers what ought to happen if, incorrectly, $\theta_j$ were to be gridded as $p \in [0,1]$. The curvature will rises extremely fast near the boundary points, so that $\delta_{III}$ and the standard error contribution to $\delta_{II}$ are very large. We interpret the the mild behavior of the curvature in (\ref{binomvar}) as due to the canonical transformation stretching the original parameter space to have a more even information metric.

\section{Examples}

In this section we demonstrate our computational results for two basic examples: a trivial trial with Gaussian outcomes in 5.6.1, and a Thompson Sampling example with Bernoulli outcomes in section 5.6.2. We also derive the relevant exponential family quantities for the case of Gaussian outcomes with an unknown variance parameter in section 5.6.3.

\subsection{Example: A Trivial Trial with Gaussian Outcomes}

Consider the simplest example of a trial in one dimension: that would be $n$ patients with Gaussian outcomes, 
$$y_i \sim N(\mu,\sigma^2),$$ with a known standard error $\sigma$, and testing whether the unknown mean $\mu$ is significantly greater than $\mu_0$.

The textbook approach is to reject when 

$$\frac{\bar{y} - \mu_0}{\sigma/\sqrt{n}} > z_{1-\alpha},$$

where $\alpha$ is a one-sided confidence level, often taken to be 2.5\%. In this case, the Type I error or power function denoted $f_1(\mu)$ is given by the simple formula,


$$\Phi\left(z_{\alpha} + \frac{\mu - \mu_0}{\sigma / \sqrt{N}}\right)$$

To further simplify, we may assume $\mu_0 = 0$ and $\sigma = 1$, $n = 10$.

\begin{figure}
\begin{center}
\includegraphics[scale = .4]{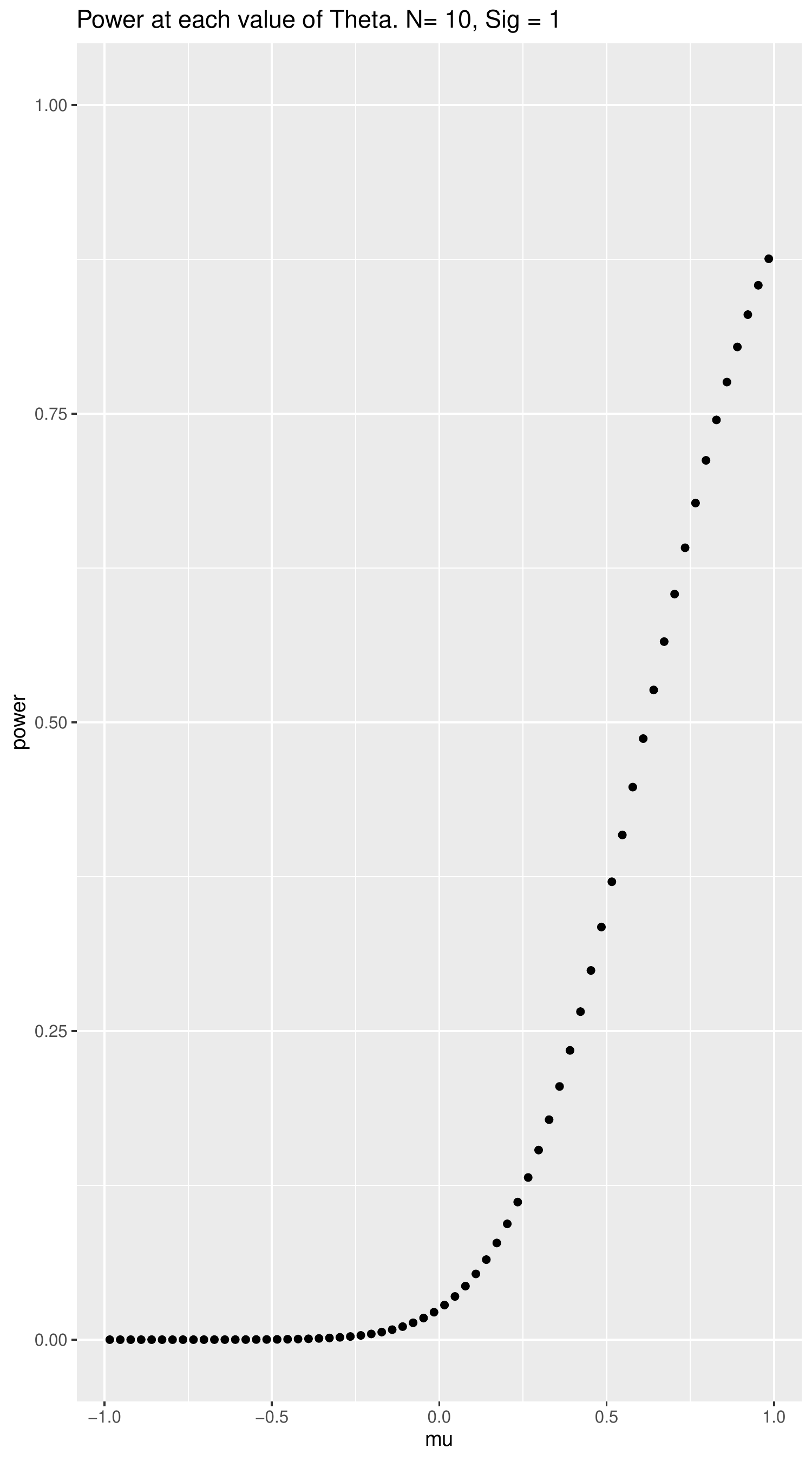}
\caption{The power function for a simple 1-parameter Gaussian trial}
\label{1Dpower}
\end{center}
\end{figure}

\begin{figure}
\begin{center}
\includegraphics[scale = .5]{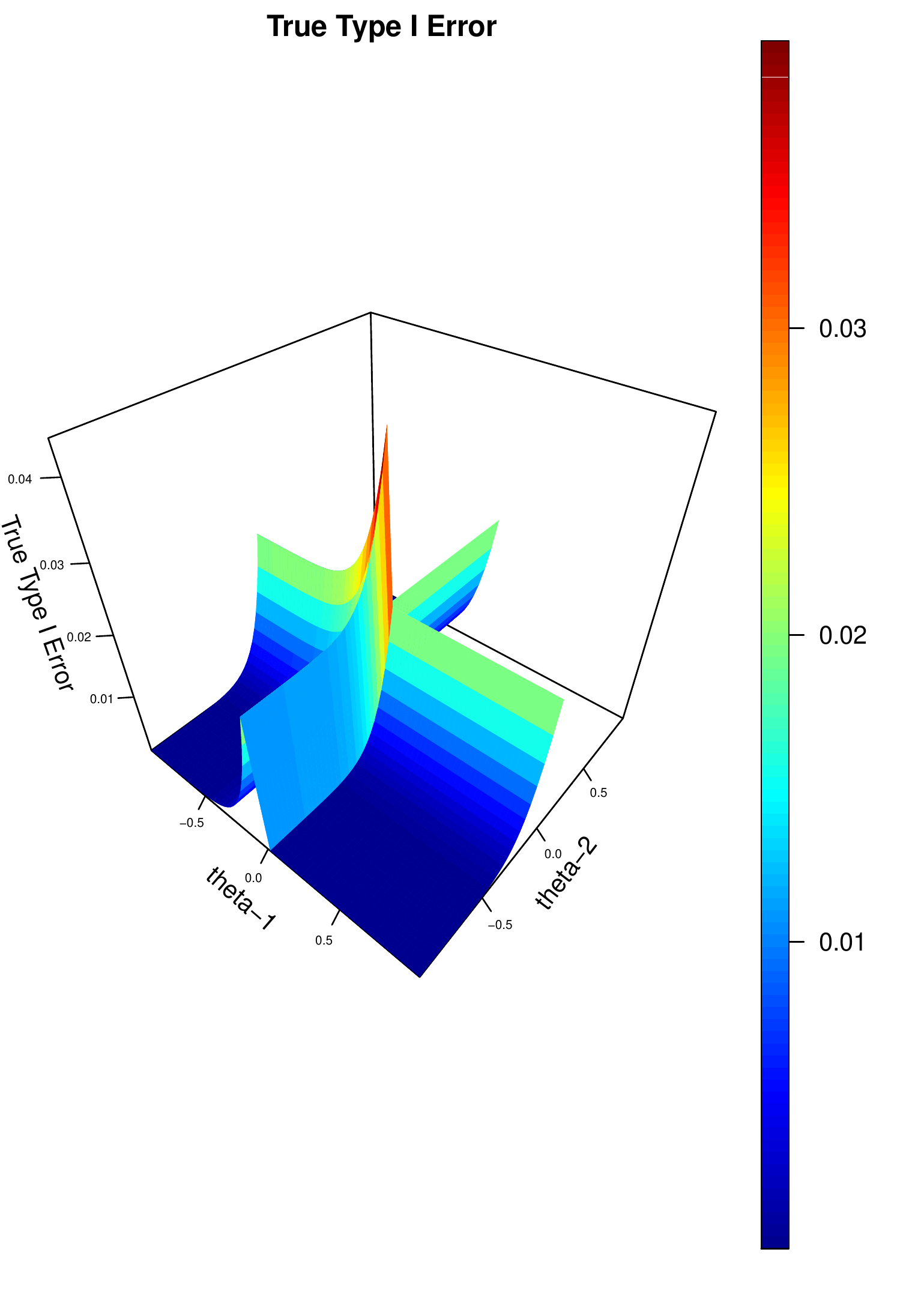}
\caption{The exact Type I Error function of running two parallel Gaussian trials}
\label{SimpleTrueTypeI}
\end{center}
\end{figure}

\begin{figure}
\begin{center}
\includegraphics[scale = .5]{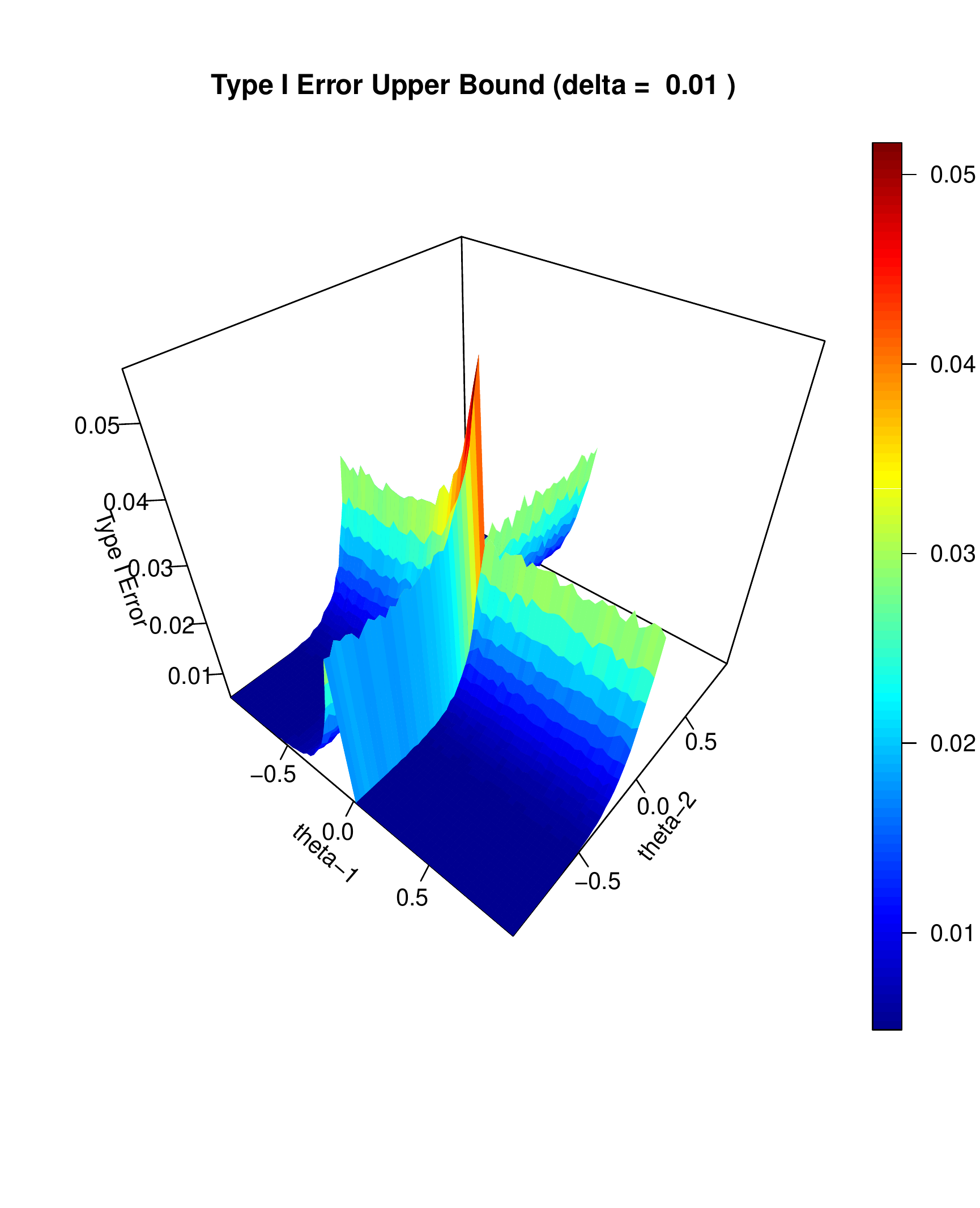}
\caption{Our upper bound on the Type I Error function of running two parallel Gaussian trials, with 99\% pointwise confidence}
\label{ApproxTypeI}
\end{center}
\end{figure}

Figure \ref{1Dpower} shows the familiar power plot for this example.
To slightly increase the difficulty, we will now introduce a second dimension, by running two parallel, independent copies of the previous trial with different parameters $(\theta_1, \theta_2)$. The resulting Type I error function is 
$$f_2(\theta_1,\theta_2) = 1 - (1- f_1(\theta_1))(1 - f_1(\theta_2))$$
Figure \ref{SimpleTrueTypeI} is a plot of the Type I error function.
Note that we have excluded the positive quadrant entirely, for reasons of presentation, to focus on the null hypothesis space. Our method yields the following approximate upper bound to this $f$, shown in Figure \ref{ApproxTypeI} using 4096 gridpoints at distances 1/32 and 50,000 Monte Carlo draws at each point. In this case, our upper bound has the right shape but is slack by about 1\%, due to a combination of the gridding gaps and limited amount of Monte Carlo replications. Both of these can be improved with more computational scale - see next example.

\subsection{Example: Thompson Sampling}

\begin{figure}
\begin{center}
\includegraphics[scale = 0.5]{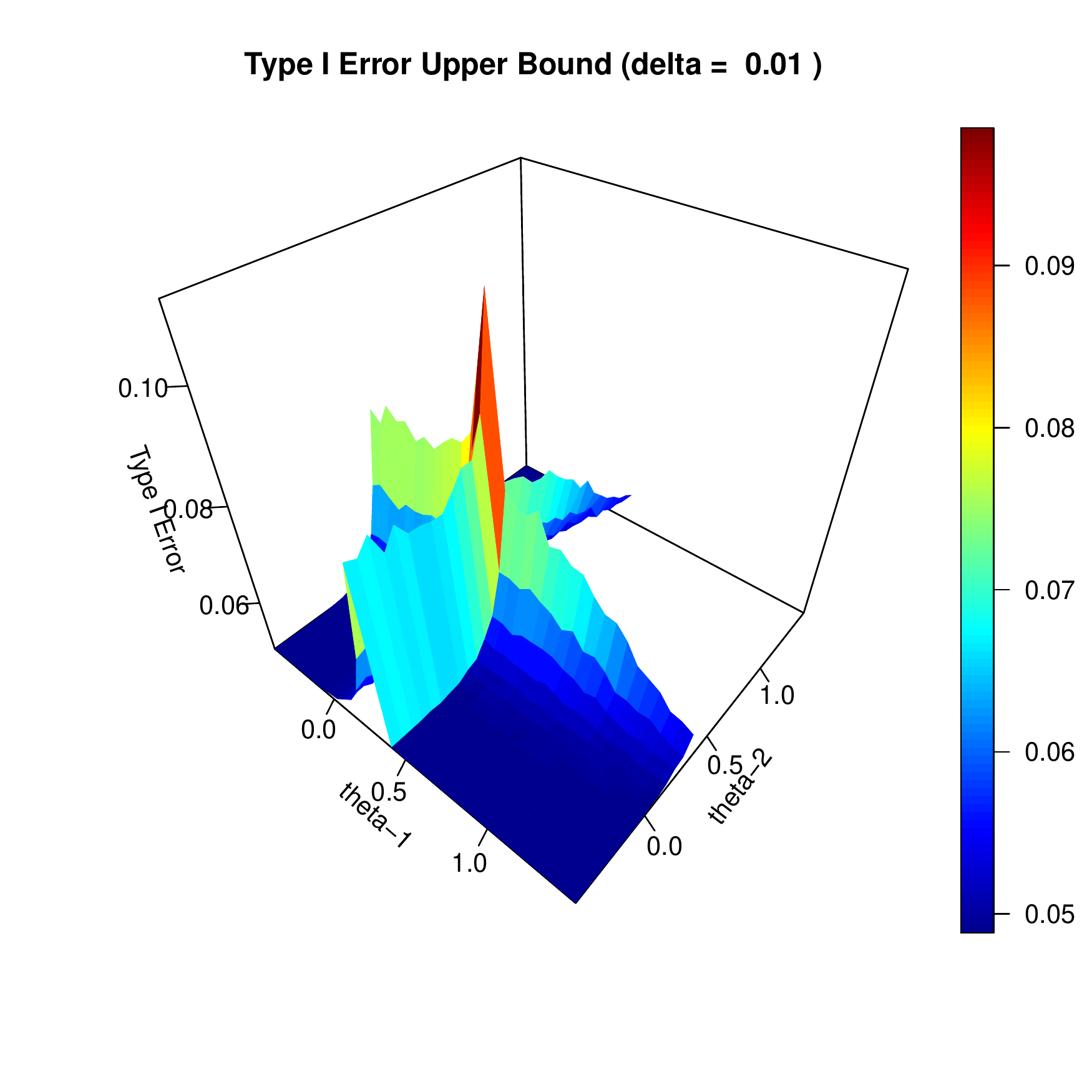}
\caption{An upper bound on the Type I Error of the Bayesian trial, with a 99\% pointwise guarantee. 1024 grid-points with 25,000 Monte Carlo replications each. The x and y axes are log-odds of the arm probabilities. Note the poor resolution and conservativeness due to the large gridding gaps and small number of Monte Carlo simulations.}
\label{fig:smallthompson}
\end{center}
\end{figure}

Here we show more details of the example in section 5.1.2. This trial emulates an uncontrolled two-arm phase II trial. We will use a total of n = 100 patients, split adaptively between the arms by Thompson sampling. 
Figure \ref{fig:smallthompson} shows a simulation of the Type I error shown in terms of the canonical-link (log-odds) space given by $\theta = \log(p/1-p)$, with simulations supported at 1024 points, and taking K = 25000 Monte Carlo samples at each point.

\begin{figure}
\begin{center}
\includegraphics[scale = 0.5]{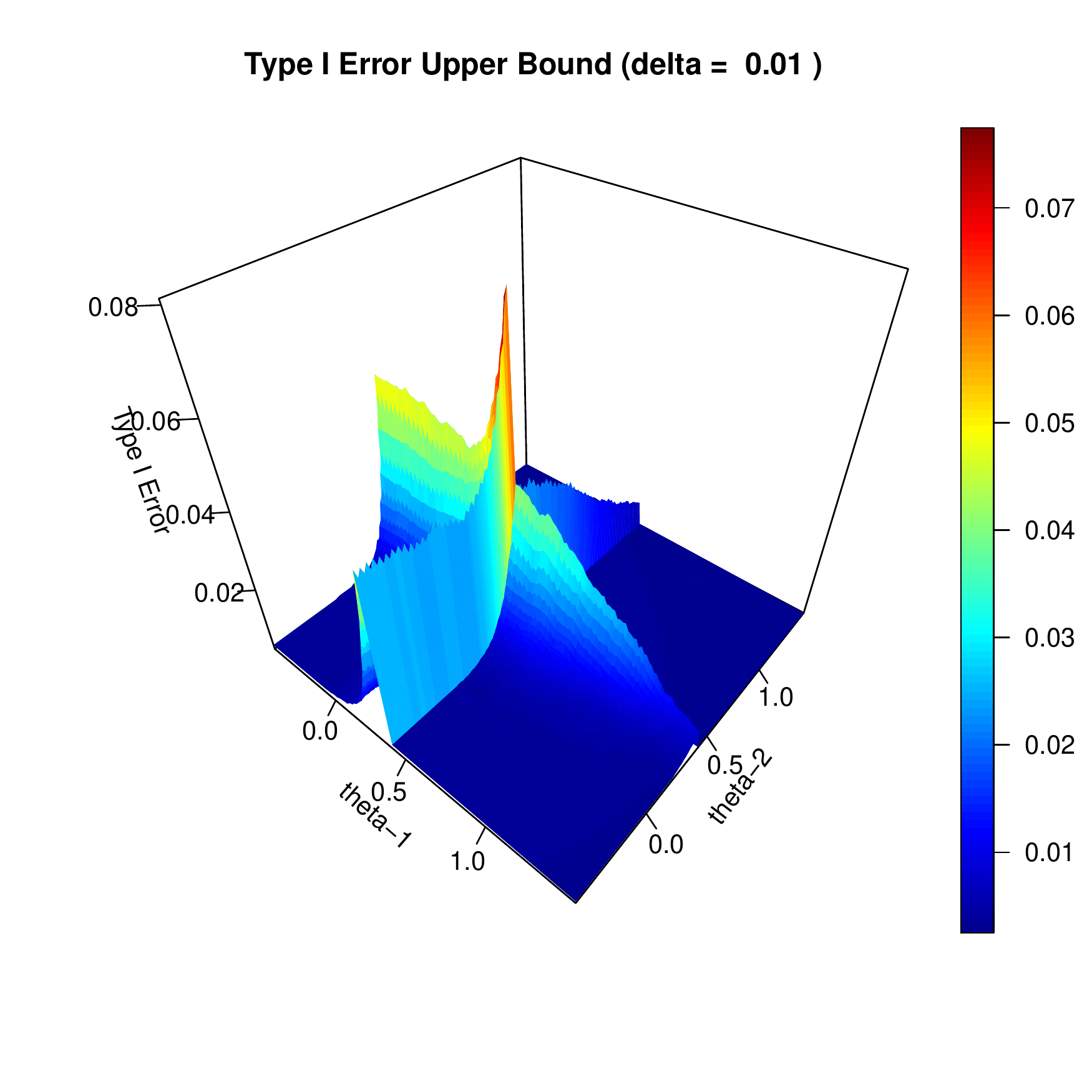}
\caption{An upper bound on the Type I Error of the Bayesian trial, with a 99\% pointwise guarantee. The x and y axes are log-odds of the arm probabilities. 16384 grid-points with 838,000 Monte Carlo replications per point. Resolution and conservativeness are improved.}
\label{fig:bigthompson}
\end{center}
\end{figure}

With this relatively small computation, we are experiencing both noise and significant costs for the gridding. If we instead use a finer grid, and perform many more simulations at each point, our confidence upper bound differs from the unbiased Monte Carlo simulation estimate by under $1 \%$. At the point where they differ the most, roughly 10\% of the difference is standard errors from estimating $f$, 20\% is standard errors from $\nabla f$, and 60-70\% is (approximately correct) inflation based on the estimated gradient and the second order bound. See Figure \ref{fig:bigthompson}.

\subsection{Example: Gaussian Nuisance Parameter}

The Gaussian likelihood with both unknown mean and variance is an exponential family:

$$ \frac{1}{2\pi\sigma^2} e^{(x - \mu)^2 / 2\sigma^2} =  \frac{1}{\sqrt{2\pi}} e^{\frac{x^2}{2\sigma^2} + \frac{\mu x}{\sigma} - \frac{\mu^2}{2\sigma^2} - log(\sigma) }$$

This gives us the natural parameter $\eta = (\eta_1, \eta_2)$, where

$$ \eta_1 = \mu/\sigma^2$$

$$ \eta_2 = -1/2\sigma^2$$

The sufficient statistics are 

$$T(x) = (x, x^2).$$

Then, the log-partition function is

$$A(\eta) = \frac{-\eta_1^2}{4\eta_2} - \frac{1}{2}\log(-2 \eta_2)$$

We can then evaluate,

$$\nabla A(\eta) = \left(  \frac{-\eta_1}{2\eta_2} , \frac{\eta_1^2}{4\eta_2^2} - \frac{1}{2 \eta_2}
\right)$$

$$\nabla^2_{11} A(\eta) = \frac{-1}{2\eta_2}$$

$$\nabla^2_{12} A(\eta) = \nabla^2_{21} A(\eta) = \frac{\eta_1}{2 \eta_2^2} $$

$$\nabla^2_{22} A(\eta) = \frac{-\eta_1^2}{2\eta_2^3} + \frac{1}{2\eta_2^2}$$

As we accumulate samples, the log-partition function scales linearly:

$$\Psi_t(\eta) = t \Psi(\eta).$$

This time, however, $\nabla^2_{22} A(\eta)$ is not diagonal, so to compute $\delta_{III}$ conservatively some care must be taken. One approach is to take the maximum and minimum over $R_j$ of the contribution due to each sub-component $\nabla^2_{ij} A(\eta)$.



\section{Further Capabilities}

With sufficient computational scale, the rigorous control these methods offer may enable subsequent paradigm shifts. In Section 5.7.1, we discuss capability for highly flexible adaptation for complex designs such as platform trials. 
In Section 5.7.2, we discuss calibration of Type I Error.

\subsection{Unplanned Changes to Complex Trials with Minimal Loss}

With sharp estimation and bounding of Type I Error over the null hypothesis space, new opportunities arise for finely controlled adaptation and re-design. In a standard adaptive trial with only one unknown parameter, unplanned adaptation is typically managed using the conditional error rate principle; i.e., at a stopping time $\tau$, one may compute the remaining conditional Type I Error at the boundary null hypothesis point $\theta_0$:
$$\alpha^1_{\theta_0}(\tau) := \mathbbm{E}(\mathbbm{1}_{F} | \mathcal{F}_\tau)$$
where $F$ is the set of false rejections under $\theta_0$. Then, one may re-design the remainder for the trial as long as the new conditional Type I Error under the re-design, $\alpha^2_{\theta_0}(\tau)$, is not greater than $\alpha^1_{\theta_0}(\tau)$. This ensures that the conditional Type I Error rate is a supermartingale, and consequently the overall Type I Error rate is bounded by $$\alpha^1_{\theta_0}(0) = \alpha.$$

In a well-modeled complex trial, the same principles can hold when we widen our view of the Type I Error rate to be a function over $\Theta_0$. At an interim point $\tau$, we could compute the remaining conditional Type I Error rate function, $f_{1}(\theta)$, and introduce a new design with Type I Error function $f_{2}$ such that

$$ f_2(\theta) \leq f_1(\theta), \forall \theta \in \Theta_0$$

Using simulation, we may estimate a lower bound with pointwise confidence $1 - \delta$ for $f_1$, which we will denote $g_1^-$. (This can be done with a small change to the methods of Section 5.5: Instead of upper bounding the probability of the set $F$, which corresponds to Type I Error, one upper bounds the probability of the set $F^C$ and subtracts the resulting estimate from 1.) Similarly, if we use Section 5.5 to estimate an upper bound for $f_2$, which we shall denote $g_2^+$, holding with point-wise confidence $1 - \delta_2$, if we are assured that

$$g_2^+(\theta) \leq g_1^-(\theta), \forall \theta \in \Theta_0$$
then the overall procedure would stay within the initial Type I Error budget over $\Theta_0$, with pointwise confidence at least $1 - \delta_1 - \delta_2$. Note, however, that we have not yet shown how to achieve this without risking $g_2^+$ exceeding $g_1^-$ randomly due to simulation error - this will be the subject of Section 5.7.2.

But first, we mention that there is often additional improvement possible: the Type I Error rate function of the initial design, $f_0(\theta)$, will typically have a positive amount of slack, $\alpha - f_0$, at most of the volume of the null hypothesis space. The trial designer could, as part of their design plan, pre-specify what is to happen with this extra slack. For example, at each null point $\theta ' \in \Theta_0$, one could pre-specify a ``shadow rejection rule" specific to $\theta'$ with total budget $\alpha$ so that in the event of a design change, the Conditional Type I Error for the shadow rejection rule can be used. This approach is essentially equivalent to declaring under each $\theta \in \theta_0$ a different supermartingale bounded in [0,1], which matches the evolution of the conditional Type I Error budget for the shadow rule.

However, a more likely scenario is that a designer will leave this slack un-specified. Without a declaration of how to invest it, we suggest this slack should simply remain static, but able to be re-allocated in an interim design. Thus, in the same setting as before, if the initial design has an upper bounding function $g_0^+$ for the overall Type I Error rate  with confidence $1 - \delta_0$, if the new design satisfies

\begin{equation}
    \label{redesign}\tag{5.9}
    g_2^+(\theta) \leq g_1^-(\theta) + \alpha - g_0^+(\theta)
\end{equation} 
then the adaptation retains an overall Type I Error guarantee with confidence $$1 - \delta_0 - \delta_1 - \delta_2.$$

It is often noted that unplanned adaptations are typically inefficient relative to a well-pre-specified plan. Thus, it is generally recommended to pre-specify an efficient initial adaptative design, and use unplanned re-design to adapt when new information arrives or circumstances change.
Within this framework, it is possible to make nearly arbitrary adaptations, including adding new unanticipated arms to the trial. To add new arms or parameters to the mix, we must only change how we define the parameter space: instead of the initial parameter space $\Theta$, we must place the trial as having been within a larger space $\Theta'$ all along - increased by the necessary dimension(s) to accommodate the new parameter(s). We may trivially extend $g_0$ and $g_1$ into functions $g_0'$ and $g_1'$ on $\Theta'$ which are constant over the newly added dimensions. Then, the designer must seek to find a $g_2$ obeying (\ref{redesign}). If they succeed, they can be satisfied that FWER will be controlled by the overall re-design process.

\subsection{Safe Calibration}

For regulatory purposes, a designer would typically like to hit 2.5\% Type I Error probability guarantee on the nose; with our methods, this is not obvious how to achieve given the randomness of the Monte Carlo upper bound. Here we present a way to ensure that the upper bound for at least one point in the null hypothesis space $\theta_0$ does sharply hit a pre-specified bound $\alpha(\theta)$ (such as a constant .025  - $\delta$ bound for an initial design, or the right hand side of \ref{redesign}), with the rest of $\Theta_0$ remaining within budget.

We begin with a near-optimized design $D$. We shall keep its sampling decisions fixed, but place its rejection rule in a family of rules $D_{\lambda}$ indexed by a 1-dimensional parameter $\lambda$.
We require that $D_{\lambda}$ obeys a monotonicity property in $\lambda$, so that if $\lambda_1 < \lambda_2$, a rejection under $D_{\lambda_1}$ implies a rejection under $D_{\lambda_2}$.
For example, if $D$ uses a p-value threshold based rejection rule, $D_\lambda$ can be parametrized by multiplying or adding $\lambda$ to the p-values used by $D$. [Or, if using a Bayesian thresholding rule involving posterior probabilities, $\lambda$ may tune the threshold.] We may assume without loss of generality that all possible values of $f_{\lambda (\theta)}$ between 0 and 1 are achieved, as this is always possible with the use of randomization.

Then, generate a fixed base $B$ of Monte Carlo samples of $X_{\tau}$ under the design $D$. For a sequence of $\lambda$'s we use the methods in this chapter on $B$ to compute $1 - \delta$-confidence upper bound functions
$$g_\lambda(\theta).$$
for each rejection rule implied by the values of $\lambda$ considered. Then, we we may select as our final rejection rule any $D_{\lambda^{'}}$ such that


$$ \lambda' \leq \sup \{\lambda : \forall \lambda'' \leq \lambda; \forall \theta \in \Theta_0, g_{\lambda''}(\theta) \leq \alpha  \} $$

To the extent that $g_{\lambda}(\theta) - \alpha(\theta)$ is approximately continuous in $\lambda$, we should be able to find a $\lambda^{'}$ which is close to the desired budget.

\subsubsection{Proof}

Fix any $\theta^*$ in the null hypothesis space.
In the parametrized family of tests $D_\lambda$,
we may define 
$$\lambda^* := \inf \{\lambda : f_\lambda(\theta*) >  \alpha(\theta*) \}$$
where $f_\lambda(\theta*)$ is the Type I Error rate function for $D_\lambda$ under parameter $\theta^*$. 
Then with probability at least $1 - \delta$, by the simulation method's guarantee, $$g_{\lambda^*}(\theta^*)\geq \alpha(\theta^*).$$
Since the $\lambda'$ which our procedure chooses will always have
$$g_{\lambda'}(\theta^*) \leq \alpha(\theta^*).$$
We conclude that $g_{\lambda'}(\theta^*) \leq g_{\lambda^*}(\theta^*)$, and therefore $\lambda' \leq \lambda^*$, with probability at least $1 - \delta$. By monotonicity of $f_\lambda$ and the definition of $\lambda^*$, our procedure is under budget with pointwise confidence $1 - \delta$ at $\theta^*$, and thus for all $\theta \in \Theta_0$.

\section{Discussion and Conclusion}
 
As FDA seeks to establish new standards for complex trial design, we hope the methods in this chapter can support their needs and mission. By enabling technical control of highly complex designs, we aim to spur further automation of the simulation design and verification process, reducing negotiation and labor. In principle, these techniques could be used as an inner loop of a constrained optimization (to compute the constraint); one wonders whether this approach could permit highly sophisticated optimizations such as neural networks. But ultimately, the utility of this framework is likely to be determined by the power and availability of simulation software. Thus, we invite talented computer scientists to help further accelerate this rapidly developing field.

This past year, the world has been engulfed in a global pandemic, and many of the ideas reviewed in this thesis have taken on greater significance. Governments are learning how to control the outbreak using a combination of interventions, including masks and “social distancing,” rapid learning by intensivists faced with a virus having pleiotropic clinical effects, development of new therapeutics and the repurposing of existing drugs, and ambitious programs of vaccine development. They are also dealing with the economic disruption, both directly from the pandemic and indirectly from the control efforts. 

The limitations of observational, non-experimental approaches and conventional randomized clinical trials have been cast in sharp relief. Each scientific specialty has begun to propose ways to make its responses better and faster ``next time around.” Virologists propose beginning therapeutic drug and vaccine development, even in advance of knowing the identity of the new pandemic agent. Ecologists and wildlife conservationists urge a greatly expanded global project to survey likely animal sources of the next spillover event, and to target those agents that are likely to pose a substantial global threat. Clinical scientists and trialists seek to create pre-formed platforms for rapid testing of the drugs, non-pharmacologic interventions, and vaccines that will be proposed. In this area, innovative experimental design will be critical, and as statisticians are recruited to help prepare for the next emergency, they will find, as we have, that recent advances in adaptive trial design will provide a sturdy, flexible, and reliable framework for their efforts.


\bibliography{bibliography}

\end{document}